\documentclass[iop,apj,numberedappendix]{emulateapj}
\usepackage{amssymb,latexsym,amsmath}
\usepackage{apjfonts}
\usepackage{graphicx}

\usepackage{appendix}
\usepackage{natbib}
\usepackage{bbm}
\def \nh {N${\rm _H}$}

\def \arcmin {\hbox{$^\prime$}}
\def \arcsec {\hbox{$^{\prime\prime}$}}

\def\spose#1{\hbox to 0pt{#1\hss}}
\def\ltsim{$\mathrel{\spose{\lower 3pt\hbox{$\sim$}}
        \raise 2.0pt\hbox{$<$}}$\thinspace}
\def\gtsim{$\mathrel{\spose{\lower 3pt\hbox{$\sim$}}
        \raise 2.0pt\hbox{$>$}}$\thinspace}
\def \msun {${\rm M_\odot}$}

\def \nh {$N_{\rm H}$}
\def \deg {$^\circ$}

\newcommand{\source}{\mbox{IC\,1860}}

\newcommand{\apec}{APEC}
\newcommand{\mekal}{MEKAL}
\newcommand\emin{\hbox{{$E_{\rm min}$}}}

\newcommand{\zfe }{${\rm Z_{Fe}}$}

\newcommand{\chandra }{{\em Chandra}}

\newcommand{\xspec }{{\em Xspec}}
\newcommand{\acis }{{\em ACIS}}
\newcommand{\ciao }{{\em CIAO}}

\newcommand{\xmm }{{\em XMM}}
\newcommand{\asca }{{\em ASCA}}
\newcommand{\rosat }{{\em ROSAT}}

\newcommand{\gmrt }{{\em GMRT}}

\newcommand\omegam{\hbox{{$\Omega_{\rm m}$}}}
\newcommand\omegalambda{\hbox{{$\Omega_{\Lambda}$}}}
\newcommand\kmsmpc{{\rm km s$^{-1}$ Mpc$^{-1}$}}
\newcommand\ho{\hbox{{$H_{0}$}}}

\newcommand\kms{{\,km\,s$^{-1}$}}

\newcommand{\Kms}{\,\mathrm{km}\,\mathrm{s}^{-1}}
\slugcomment{Accepted by the Astrophysical Journal}
\shorttitle{Sloshing cold fronts in the IC1860 group}
\shortauthors{Gastaldello et~al.}
\begin{document}
\title{Sloshing cold fronts in galaxy groups and their perturbing disk galaxies: 
an X-ray, Optical and Radio Case Study.}
\author {Fabio Gastaldello\altaffilmark{1,2},
         Laura Di Gesu\altaffilmark{1,3},
         Simona Ghizzardi\altaffilmark{1},
         Simona Giacintucci\altaffilmark{4,5},
         Marisa Girardi\altaffilmark{6,7},
         Elke Roediger\altaffilmark{8,9}, 
         Mariachiara Rossetti\altaffilmark{3,1},
         Fabrizio Brighenti\altaffilmark{10,11},
         David A. Buote\altaffilmark{2},
         Dominique Eckert\altaffilmark{12},
         Stefano Ettori\altaffilmark{13,14},
         Philip J. Humphrey\altaffilmark{2},
         and William G. Mathews\altaffilmark{11}
}
\altaffiltext{1}{IASF-Milano, INAF, via Bassini 15, Milan 20133, Italy}
\altaffiltext{2}{Department of Physics and Astronomy, University of
California at Irvine, 4129 Frederick Reines Hall, Irvine, CA 92697-4575, USA}
\altaffiltext{3}{Dipartimento di Fisica, Universit\`a degli Studi di Milano, via Celoria 16, 20133, Milan, Italy}
\altaffiltext{4}{Department of Astronomy, University of Maryland, College Park, MD 20742-2421, USA}
\altaffiltext{5}{Joint Space-Science Institute, University of Maryland, College Park, MD, 20742-2421, USA}
\altaffiltext{6}{Dipartimento di Fisica, Universit\`a degli Studi di Trieste, Sezione di Astronomia, via Tiepolo 11, 3
4133, Trieste, Italy}
\altaffiltext{7}{INAF, Osservatorio Astronomico di Trieste, via Tiepolo 11, 34133 Trieste, Italy}
\altaffiltext{8}{Jacobs University Bremen, PO Box 750 561, 28725 Bremen, Germany}
\altaffiltext{9}{Hamburger Sternwarte, Universitaet Hamburg, Gojensbergsweg 112, D-21029 Hamburg, Germany}
\altaffiltext{10}{Dipartimento di Astronomia, Universit\`a di Bologna, via
Ranzani 1, Bologna 40127, Italy}
\altaffiltext{11}{UCO/Lick Observatory, University of California at Santa Cruz,
 1156 High Street, Santa Cruz, CA 95064, USA}
\altaffiltext{12}{ISDC Data Centre for Astrophysics, Geneva Observatory, ch. d'Ecogia, 16, 1290 Versoix, Switzerland
}
\altaffiltext{13}{INAF, Osservatorio Astronomico di Bologna, via Ranzani 1, Bologna 40127, Italy}
\altaffiltext{14}{INFN, Sezione di Bologna, viale Berti Pichat 6/2, I-40127 Bologna, Italy}

\begin{abstract}
We present a combined X-ray, optical, and radio analysis of the galaxy group IC 1860 using
the currently available \chandra\ and \xmm\ data, literature multi-object spectroscopy 
data and \gmrt\ data. The \chandra\ and \xmm\ imaging and spectroscopy reveal two surface brightness discontinuities
at 45 and 76 kpc shown to be consistent with a pair of cold fronts. These features are interpreted as due to sloshing 
of the central gas induced by an off-axis minor merger with a perturber. This scenario is further supported by the presence of a peculiar
velocity of the central galaxy IC 1860 and the identification of a possible perturber in the optically disturbed spiral galaxy IC 1859.
The identification of the perturber is consistent with the comparison with numerical simulations of sloshing.
The \gmrt\ observation at 325 MHz shows faint, extended radio emission contained within the inner cold front, as seen in some galaxy clusters
hosting diffuse radio mini-halos. However, unlike mini-halos, no particle reacceleration is needed to explain the extended radio emission, 
which is consistent with aged radio plasma redistributed by the sloshing.
There is strong analogy of the X-ray and optical phenomenology of the IC 1860 
group with two other groups, NGC 5044 and NGC 5846, showing cold fronts.
The evidence presented in this paper is among the strongest supporting the currently favored model of cold-front formation in relaxed 
objects and establishes the group scale as a chief environment to study this phenomenon.
\end{abstract}
%
%=====================================================================
% INTRODUCTION  INTRODUCTION   INTRODUCTION   INTRODUCTION 
%=====================================================================
%
\section{Introduction} 
\label{Introduction} 
The current generation of X-ray telescopes (\xmm\ and to a greater degree \chandra) with their angular
resolution and sensitivity has revealed a wealth of small scale features in the intra-cluster 
(ICM) and intra-group medium (IGM). One particularly interesting is the phenomenon of the "cold fronts", i. e. sharp surface 
brightness discontinuities, interpreted as contact edges between regions of gas with different 
entropies \citep[see the review by][]{Markevitch.ea:07}. Cold fronts appear to be almost
ubiquitous in galaxy clusters \citep{Markevitch.ea:03*1,Ghizzardi.ea:10,Lagana.ea:10} and they are found both in 
dynamically active objects and in cool core relaxed clusters. 
In merging clusters cold fronts arise during merger events through 
ram-pressure stripping mechanisms that induce the discontinuity among the merging dense subcluster and 
the less dense surrounding ICM \citep[e.g.][]{Vikhlinin.ea:01}. In relaxed clusters 
cold fronts are most likely induced by minor 
mergers that produce a disturbance on the gas in the core of the main cluster, displace it from the 
center of the potential well, and decouple it from the underlying dark matter halo through ram 
pressure \citep{Markevitch.ea:01,Churazov.ea:03,Ascasibar.ea:06}. 
The oscillation of the gas of the core around the minimum of the potential generates
a succession of radially propagating cold fronts, appearing as concentric edges in the surface 
brightness distribution of the cluster. These fronts may form a spiral structure when the sloshing 
direction is near the plane of the sky. When the
sloshing direction is not in the plane of the sky concentric arcs are observed.
The sequence of events is described in great detail in the simulations presented in 
\citet{Ascasibar.ea:06}.

Cold fronts are unique tools to understand the physical properties of the ICM 
\citep[e.g.][]{Markevitch.ea:07}. They can also be in principle used as a gauge of the
merger activity \citep{Owers.ea:09*2}, in particular of the more frequent minor merging of smaller 
subsystems compared to the less frequent and more spectacular major mergers.
Cold fronts do not need a favorable geometry for being detected as the more elusive shock fronts
\citep[e.g.][]{Markevitch:10} and they may provide the only indication at X-ray wavelengths  
of a minor merger in a seemingly relaxed cluster. Cold fronts can be used, with
the help of simulations, to infer many merger characteristics, as for example
from the presence and size of a spiral pattern in the surface brightness distribution
the direction and time of peri-center passage of the perturber 
\citep{Ascasibar.ea:06,Johnson.ea:10,Roediger.ea:11}.

Recently, gas sloshing in the core has also been invoked to explain 
the formation of diffuse radio minihalos in relaxed, cool-core clusters. These faint,
steep-spectrum radio sources are relatively rare, with only few clusters
with confirmed detections \citep{Feretti.ea:12}. A spatial correlation between 
the minihalo emission and cold fronts has been observed in a few clusters 
\citep{Mazzotta.ea:08,Giacintucci.ea:13}, with the
minihalos contained within the region confined by the cold fronts,
suggesting a tight connection between sloshing motions and the origin of 
minihalos. Recently, \citet{ZuHone.ea:13} showed with high resolution MHD simulations
that gas sloshing can lead to turbulent reacceleration of relativistic 
electron seeds (e.g., from past AGN activity) and produce radio 
diffuse radio emission within the envelope of the sloshing cold fronts.

Another complementary path to the study of presence and dynamics of cluster mergers
uses optical data to perform a spatial and kinematical analysis of member
galaxies. These types of studies allow us to reveal and measure the amount of substructure,
and to detect and analyze possible pre-merging clumps
or merger remnants \citep[e.g.,][]{Girardi.ea:02}. This is certainly interesting
for major merger clusters which are very rich in optical substructures \citep[see for example the 
DARC - Dynamical Analysis of Radio Clusters - project,][]{Girardi.ea:07} but even more in the
case of the minor merger scenario for sloshing. The assumption of a single Gaussian distribution
is a good description for the galaxy velocities of the main cluster and  the recently merged perturber
may be identified by means of optical/dynamical substructure search \citep[e.g.,][]{Owers.ea:11*1}.
Optical studies can also investigate the presence of peculiar velocities of the bright central 
galaxy (BCG). In fact one of the observable effects of the merger scenario discussed by \citet{Ascasibar.ea:06} 
is the fact that, if the BCG sits at the peak of the dark matter distribution, it is expected to start
oscillating along with it after each subcluster flyby. Gas sloshing and BCG peculiar velocities 
are caused by the same minor mergers. Given the increasing number of objects showing
sloshing cold fronts this can explain also the large and apparently puzzling 
number of systems showing peculiar velocities of their BCGs 
\citep[e.g.,][]{vandenBosch.ea:05,Coziol.ea:09}. As a matter of fact \citet{Miralda-Escude:95}
looking at the early evidence of BCGs spatially coincident with the centers of dark matter halos
producing lensing features, commented that ``the fact that cD galaxies often have large peculiar velocities relative 
to the average of the cluster galaxies has been used as an argument against their identification as cluster centers: 
however, clusters are continuously merging, and their density peaks do not need to coincide with their centers of mass. 
Substructure will cause the density peaks to move, in response to the gravitational forces of the 
in-falling material.''

It is yet another manifestation of the hierarchical nature of 
structure formation the fact that sloshing cold fronts have been
detected also at the smaller mass scales of poor clusters
\citep[e.g., Virgo,][]{Simionescu.ea:10} and of groups of galaxies 
\citep{Gastaldello.ea:09,Randall.ea:09,Machacek.ea:11}. At this mass scale
even a single massive galaxy can be the responsible perturber and the X-ray signature
of sloshing cold front is a signature of the gravitational interaction between the galaxy
(and its halo) and its host group. 

The aim of this paper is to investigate sloshing features, such as cold fronts and spirals in the 
surface brightness distribution, in the nearby, X-ray bright galaxy group IC 1860.
We will look for other observable properties of the sloshing scenario such as a peculiar velocity 
of the BCG and extended radio emission possibly related to the presence of sloshing. 
We present in detail the results of the available
\xmm\ and \chandra\ data of the object, the available optical
data in the literature and new Giant Metrewave Radio Telescope (\gmrt) data. 
We will then compare the observed features in IC 1860 with two other groups
hosting cold fronts, NGC 5044 and NGC 5846.

The group IC 1860 has been originally classified as a Dressler cluster
\citep[DC-0247-31,][]{Dressler:80} and it is also known as Abell S301
\citep{Abell.ea:89}. It appears then in many other catalogues, in
particular the catalogue of \citet{Maia.ea:89}.
It is a nearby group of galaxies at z=0.022 (at a distance of 97 Mpc
with the cosmology adopted in this paper) and its optical structure/membership
has been studied in \citet{Dressler.ea:88} and in particular, given the many additional redshifts
provided by the 2dF redshift survey, by \citet{Burgett.ea:04}. It has been discovered
in the X-ray band by the Einstein X-ray Observatory \citep{Burstein.ea:97} and ROSAT showed the presence of 
extended and diffuse group X-ray emission at a temperature of $\sim$1 keV \citep{Mulchaey.ea:03}.
\chandra\ imaging showed no evidence of AGN induced activity in the form of cavities in the core \citep{Dunn.ea:10}.
In our previous \xmm\ analysis \citep{Gastaldello.ea:07*1} we mainly focused on the azimuthally averaged radial
properties of the system for the purpose of mass analysis, however noticing the presence of
a possible disturbance in the surface brightness distribution. 

The paper is organized as follows: in \S\ref{obs} we summarize the \xmm\ and \chandra\ data reduction and analysis;
in \S\ref{images} we analyze the X-ray images and surface brightness profiles finding evidence of a set
of two surface brightness discontinuities and an excess brightness spiral feature; in \S\ref{spc} we perform a spectral
analysis of the temperature structure of the group finding evidence that the surface brightness discontinuities are cold fronts;
in  \S\ref{fit_prof} we model the pressure jumps between the cold fronts and in  \S\ref{syst} we discuss the possible
systematics of our X-ray analysis; in \S\ref{optical} and in \S\ref{radio} we present the analysis of the available optical and \gmrt\ data, 
respectively. In \S\ref{simulations} we use simulations to constrain the merger geometry of IC 1860 and in \S\ref{discussion} we discuss our results, making 
comparisons with the groups NGC 5044 and NGC 5846. Additional analysis
performed on these groups is presented in appendix \ref{comparison}; in \S\ref{summary} we summarize our results.
We also report on the narrow angle tail radio galaxy IC 1858 in appendix \ref{sec.appendixic1858}.

The cosmology adopted in this paper assumes a flat universe with \ho = 70 \kmsmpc,
\omegam = 0.3 and \omegalambda = 0.7. All the errors quoted are at the 68\% confidence
limit. At the distance of IC 1860 1\arcmin\ corresponds to 26.7 kpc.
%
%=====================================================================
% DATA REDUCTION 
%=====================================================================
 %%%%%%%%%%%%%%%%%%%%%%%%%%%%%%%%%%%%%%%%%%%
\section{Observations and data preparation}
\label{obs}
%%%%%%%%%%%%%%%%%%%%%%%%%%%%%%%%%%%%%%%%%%%
%
%%%%%%%%%%%%%%%%%%%%%%%%%%%%%%%%%%%%%%%%%%%%%%%%%%%%%%%%%%%%%%%%%%%%%%%%%%%%%%%%&&&&&&&&&&&&&&&&&&&&&&&&&&&&
\begin{figure}[th]
\centerline{
\includegraphics[height=0.27\textheight]{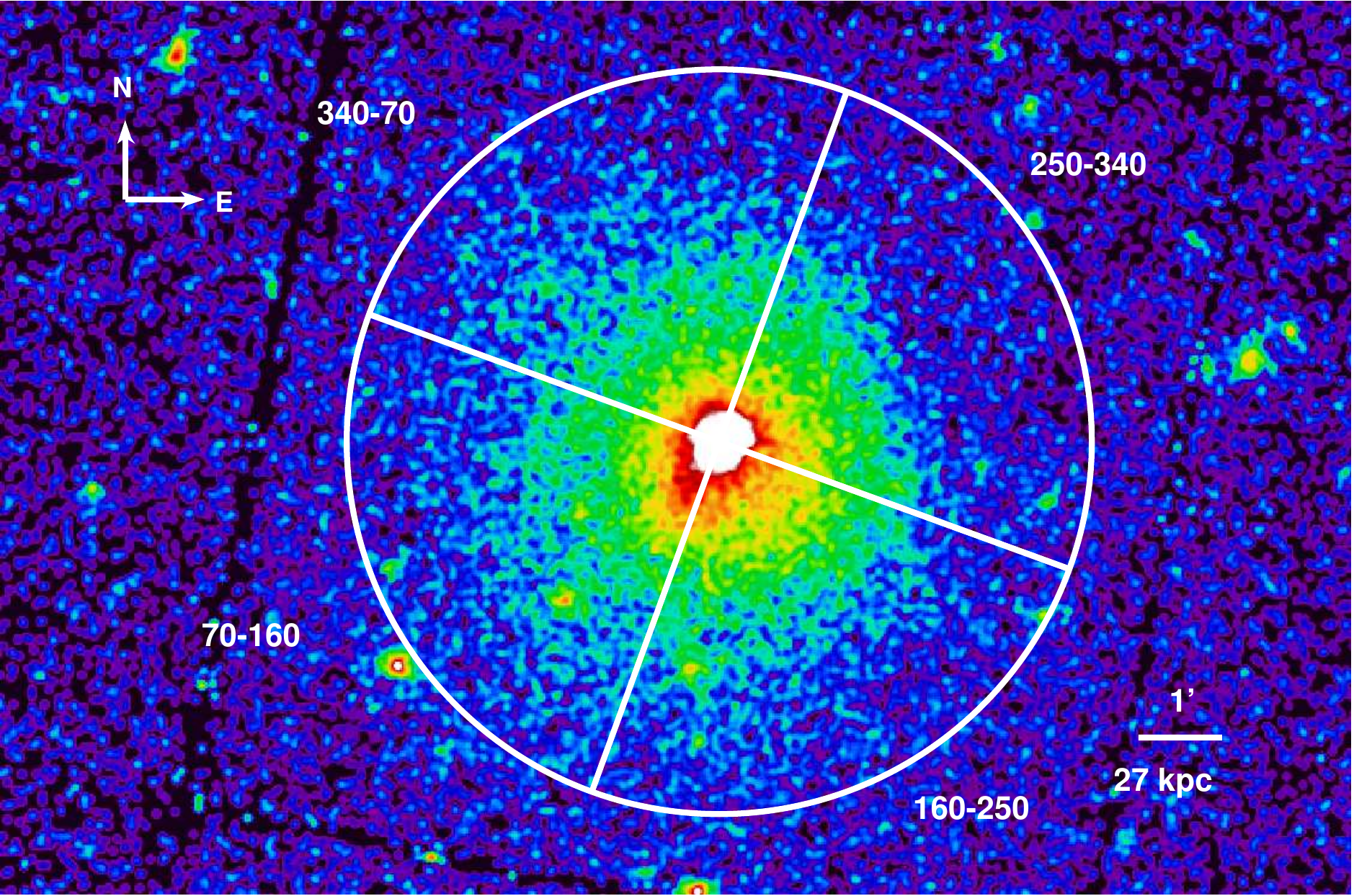}}  
\caption{\label{xmm.image} \footnotesize Exposure-corrected mosaic of the \xmm\ MOS1 and MOS2 
images Gaussian smoothed on a 3\arcsec\ scale. Superposed over the image
are the angular sectors used for the surface brightness profiles of Fig. \ref{xmm.prof} 
and discussed in the text. 
%The dashed curves highlights the position of the 
%brightness discontinuity we detected in the angular sector 250\deg-340\deg.
 }
\end{figure}

%%%%%%%%%%%%%%%%%%%%%%%%%%%%%%%%%%%%%%%%%%%%%%%%%%%%%%%%%%%%%%%%%%%%%%%%%%&&&&&&&&&&&&&&&&&&&&&&&&&&&&&&&&&&
\begin{figure*}[th]
%%\vspace{-0.5cm}
\centerline{
\parbox{0.5\textwidth}{
\includegraphics[height=0.28\textheight]{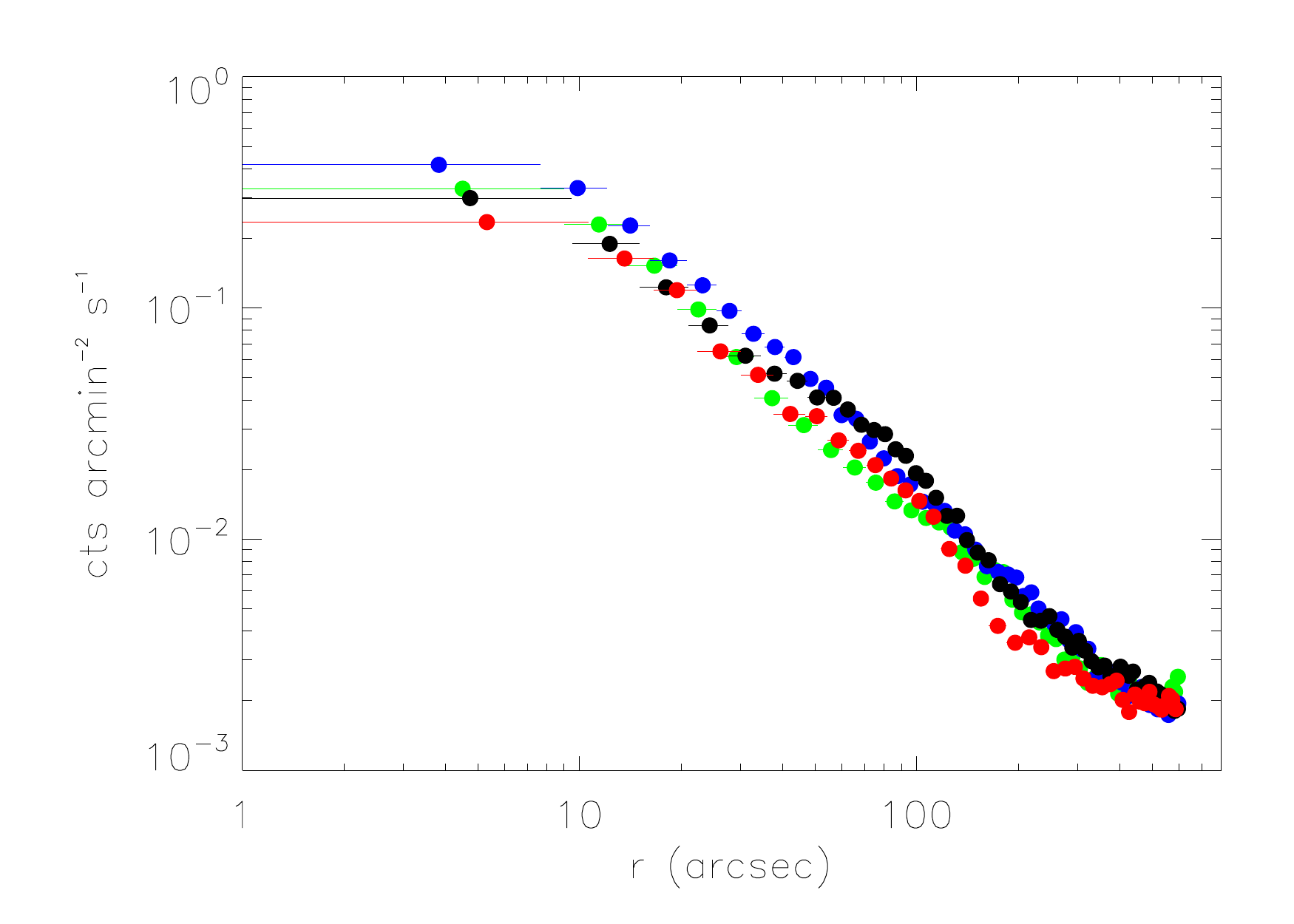}}  
\parbox{0.5\textwidth}{
\includegraphics[height=0.28\textheight]{fig2a}}
}  
\caption{\label{xmm.prof} \footnotesize
\emph{Left panel:} \xmm\ surface brightness profiles in the selected angular sectors discussed in the text 
(section \S\ref{subsection.cf.ext}) and shown in Fig. \ref{xmm.image}. Green circles: sector P.A. 340\deg-70\deg. 
Blue circles: Sector P.A. 70\deg-160\deg. Black circles: sector P.A. 160\deg-250\deg. Red circles: sector P.A. 250\deg-340\deg. 
The outer edge can be easily seen in the sector P.A. 250\deg-340\deg. 
\emph{Right panel:} Zoom over the interesting radial region for the outer edge. The slope of  the source profile 
is steeper in the sector P.A. 250\deg-340\deg\ than in the control sector P.A. 340\deg-70\deg.
}
\end{figure*}
%%%%%%%%%%%%%%%%%%%%%%%%%%%%%%%%%%%%%%%%%%%%%%%%%%%%%%%%%%%%%%%%%%%%%%%%%%%%&&&&&&&&&&&&&&&&&&&&&&&&&&&&&&&&

%
%%%%%%%%%%%%%%%%%%%%%%%%%%%%%%%%%%%%%%%%%%%%%%%%%%%%%%%%%%%%%%%%%%%%%%%%%%%%%%%%%%%%%%%%%%%%%%%%%%%%%%%%%%%%
\begin{figure}[h]
\centerline{
\includegraphics[height=0.25\textheight]{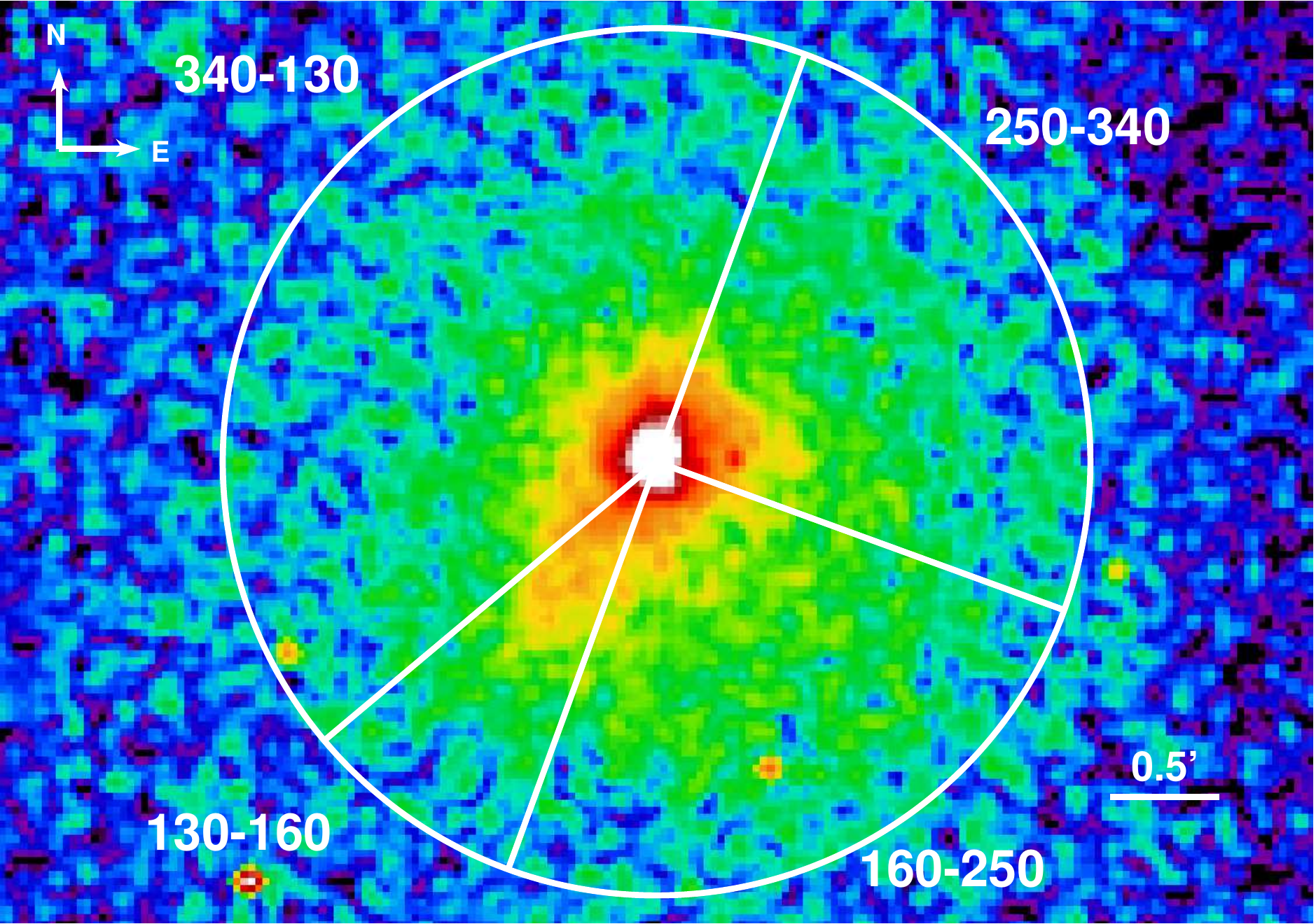}
}  
\caption{\label{chandra.image} \footnotesize 
\chandra\ 0.7-2.0 keV image Gaussian smoothed on a 3\arcsec\ scale and corrected for exposure map.
Superposed over the image are the angular sectors used for the surface brightness profiles of Fig. \ref{chandra.prof} and discussed in the 
text.
%The dashed curves highlights the position of the brightness discontinuity
%we detected in the angular sector 220\deg-250\deg.
 }
\end{figure}
%
%%%%%%%%%%%%%%%%%%%%%%%%%%%%%%%%%%%%%%%%%%%%%%%%%%%%%%%%%%%%%%%%%%%%%%%%%%%%%%%%%%%%%%%%%%%%%%%%%%%%%%%%%%%%%
%
%%%%%%%%%%%%%%%%%%%%%%%%%%%%%%%%%%%%%%%%%%%%%%%%%%%%%%%%%%%%%%%%%%%%%%%%%%%%%%%%%%%%%%%%%%%%%%%%%%%%%%%%%%%%%
\begin{figure*}[th]
%%\vspace{-0.5cm}
\centerline{
\parbox{0.5\textwidth}{
\includegraphics[height=0.28\textheight]{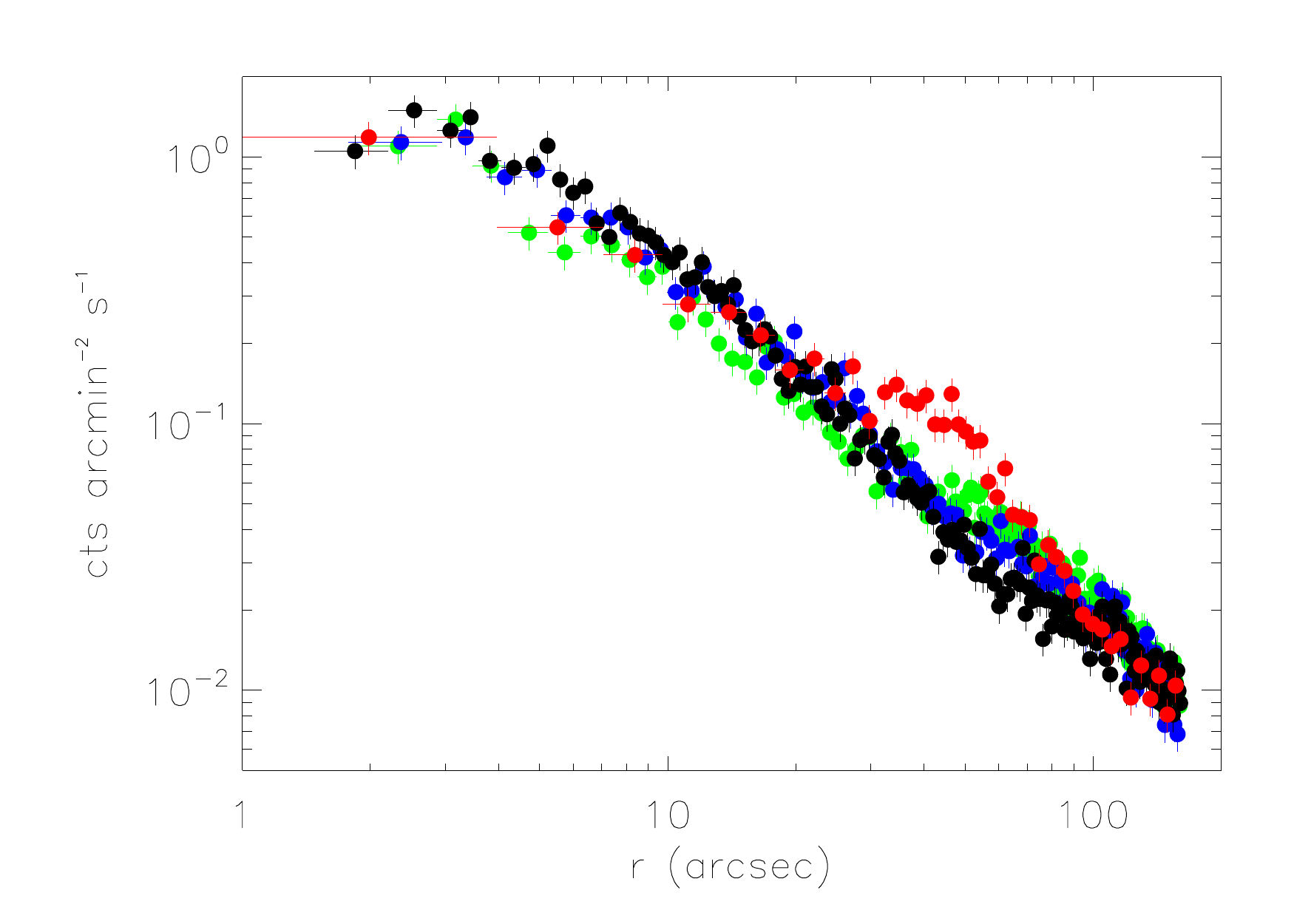}}
\parbox{0.5\textwidth}{
\includegraphics[height=0.28\textheight]{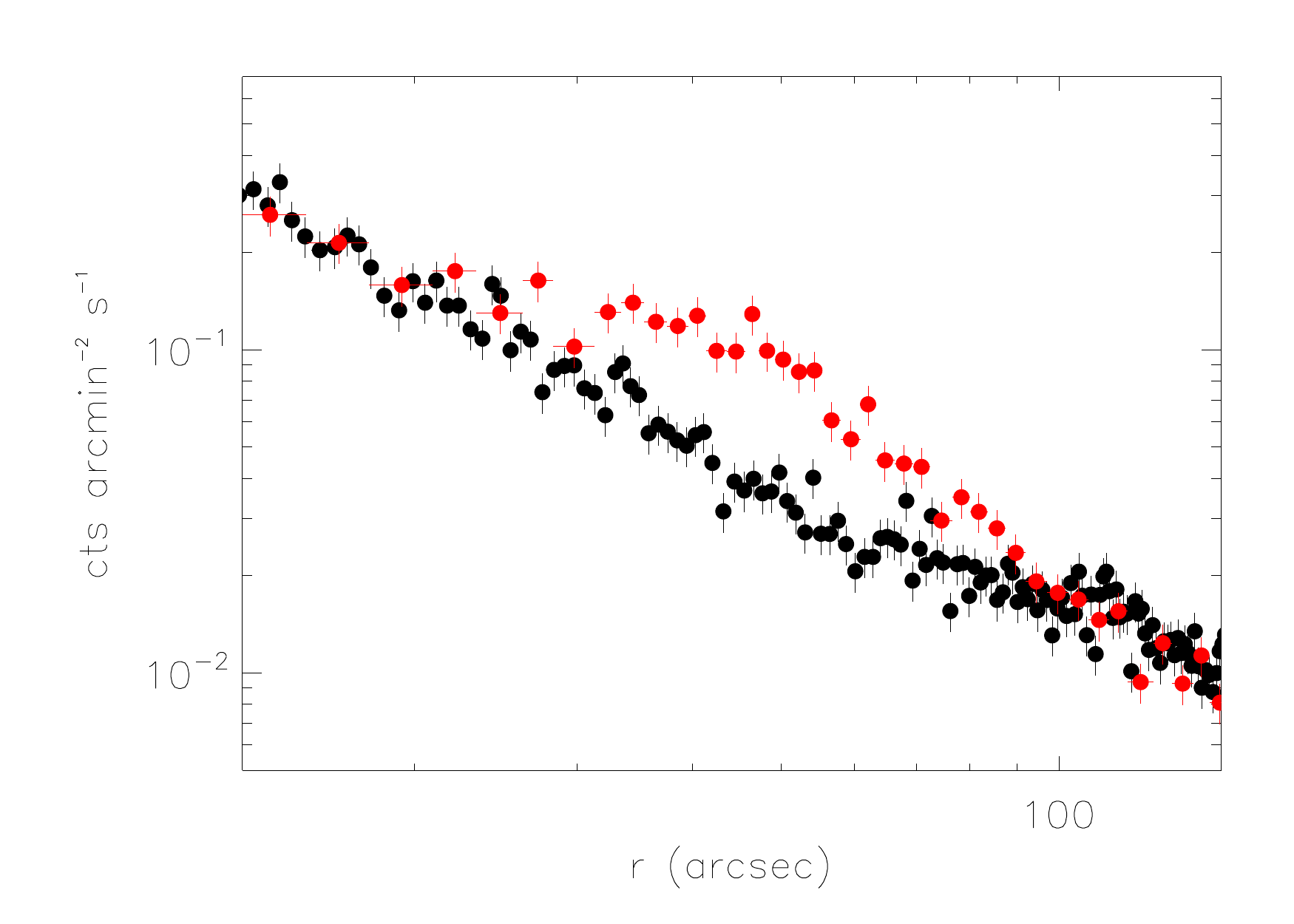}}
}
\caption{\label{chandra.prof} \footnotesize
\emph{Left panel:} \chandra\ surface brightness profiles in the selected angular sectors discussed in the text 
(section \S\ref{subsection.cf.int}) and shown in Fig. \ref{chandra.image}. 
Black circles: sector PA 340\deg-130\deg. Red circles: sector P.A. 130\deg-160\deg. Green circles: sector PA 160\deg-250\deg. 
Blue circles: sector 250\deg-340\deg.
The inner edge and enhancement of the surface brightness profile are evident in the sector P.A. 130\deg-160\deg. A steepening of the sector  
P.A. 250\deg-340\deg\ profile can also be seen at the edge of \chandra\ field of view, consistent with the \xmm\ feature of 
Fig.\ref{xmm.prof}.
\emph{Right panel:} Zoom over the interesting radial region for the inner edge. The slope of  the source profile is 
rapidly changing in the sector  P.A. 130\deg-160\deg\ than in the control sector P.A. 340\deg-130\deg.
}
\end{figure*}
%%%%%%%%%%%%%%%%%%%%%%%%%%%%%%%%%%%%%%%%%%%%%%%%%%%%%%%%%%%%%%%%%%%%%%%%%%%%%%%%%%%%%%%%%%%%%%%%%%%%%%%%%%%%%
%
%
%
\source\ has been observed by \xmm\ on February 04th 2003 (obsID
0146510401) with the EPIC MOS and pn cameras for 39 and 38 ks respectively
and by \chandra\ with the ACIS-S configuration on 
September 2009 (obsID 10537) for 40 ks.
We used both \chandra\ and \xmm\ available archival data and we took advantage 
of the different characteristic of the two satellites. We exploited the wider field of view 
provided by XMM to analyze the source structure at larger radii while the higher resolution of Chandra
allowed us to explore the details of the source in the inner region.
We processed the data following the standard SAS
\footnote{http://xmm.esa.int/sas/current/documentation/threads/} 
and CIAO  \footnote{http://cxc.harvard.edu/ciao/threads/index.html}
 threads; in the following subsections we briefly describe the data preparation.

\subsection{XMM}
Starting from the Observation Data Files retrieved from the \xmm-Newton archive,
we generated calibrated event files with SAS, version 11.0, using the tasks
{\em emchain} and {\em epchain}. We removed proton flares both from the hard and
the soft light-curve, using respectively a fixed threshold criterium and a $\sigma$-clipping technique,
as described in \citet{Rossetti.ea:10}. We also checked for residual contamination by a 
quiescent soft-protons component, evaluating the in-over-out ratio as defined in \citet{De-Luca.ea:04}. 
Net exposure time after protons flare correction is 34.5 ks for MOS-1, 34.8 ks for MOS-2 and 
29.4 ks for pn.
We checked the observation for contamination by solar wind
charge exchange: the ACE (Advanced Composition Explorer) SWEPAM\footnote{the data available at http://www.srl.caltech.edu/ACE/ACS/}
proton flux was less than $4\times10^{8}$ protons s$^{-1}$ cm$^{-2}$ and the SWICS O$^{+7}$/O$^{+6}$
ratio was less than 0.3, values which are typical of the quiescent Sun \citep{Snowden.ea:04}.
The spectra of the out-of-field-of-view events of CCD 4 of MOS1 showed an anomalously high flux in the soft band \citep[see][]{Kuntz.ea:08}
and it was therefore excluded from our analysis.

We then filtered the events according to standard pattern and flag criteria and we performed
out-of-time correction for pn. For each detector we created images in the 0.5-2 keV
band and exposure maps. We detected point sources using the task {\em ewavelet} and
masked them using a circular region, centered at the source position and with a 25\arcsec\ radius.
Using the task {\em emosaic}, we combined the MOS images into a single exposure-corrected 
image shown Fig.\ref{xmm.image} and discussed in Sec \ref{subsection.cf.ext}.

\subsection{Chandra} 

In order to ensure the most up-to-date calibration,
we reprocessed all \chandra\ data starting from
level 1 event files, using  the X-ray analysis package \ciao\ 4.3
in conjunction with the \chandra\ calibration database (\emph{Caldb}) version 4.1.3. 
We took into account time-dependent drift in the detector gain and charge transfer inefficiency 
as implemented in the \ciao\ tools.
To clean the data from periods of enhanced background, we performed a $\sigma$-clipping
to the light-curve in the 0.5-7 keV band. The final exposure time after deflaring
is 35.7 ks. Point sources were detected with the the \ciao\ tool {\em wavdetect} and then excluded 
from all the following steps of the data analysis using an appropriate mask region.
We used the blank-sky dataset provided in the CALDB to perform background analysis, 
after properly re-processing and re-projecting it. 
To take into account possible temporal variations of instrumental background, 
we rescaled the background file for the the ratio between the count-rate of
the observation and the background. To calculate the proper rescaling
factor, we extracted spectra for both background and source files 
from an external region, not contaminated by source emission and we 
quantified the count-rate ratio in the hard band 9-12 keV.
We then generated an image the 0.7-2.0 keV band 
and corrected it for its exposure map. 
The analysis we performed on the Chandra
image is discussed in section \ref{subsection.cf.int}.

%=====================================================================
% IMAGING
%=====================================================================

\section{X-ray images and surface brightness profiles}

The sloshing scenario has been introduced to explain a number of features present in the
surface brightness distribution of relaxed clusters and groups, in particular the presence
of concentric sharp surface brightness edges which have a temperature jump consistent with being cold fronts and
the presence of an excess corresponding to spiral- or arc-shaped
brighter regions inside the actual fronts. The morphology depends mainly on the angle
between our line-of-sight (LOS) and the orbital plane of the subcluster: a spiral like
structure is seen if the interaction is seen face-on, if on the contrary the line of sight
is parallel to the orbital plane arcs are seen on alternating sides of the cluster core 
\citep{Ascasibar.ea:06,Roediger.ea:11}.
We will investigate in the following sections the surface brightness distribution of the X-ray emission of IC 1860 as 
revealed by the \xmm\ and \chandra\ data to look for such features. 
\label{images}
\subsection{The outer surface brightness edge}
\label{subsection.cf.ext}

Visual inspection of the mosaic image of the two MOS cameras (Fig.\ref{xmm.image})
clearly reveals the presence of a sharp brightness discontinuity in the north-west direction
(P.A. 250\deg-340\deg, position angles measured from the N direction).
In the opposite direction (P.A. 70\deg-160\deg), at large radii the group is more elongated 
and in the inner region a luminous plume is seen: this feature is sharper in the
\chandra\ image so we will focus on it in section \ref{subsection.cf.int}.
To confirm this qualitative impression, we compared the surface brightness profiles in
four angular sector 90\deg wide as depicted in Fig.\ref{xmm.image}.
An evident slope change is seen in the 250\deg-340\deg\ profile,
at a radius of $\sim 170$\arcsec\ (Fig. \ref{xmm.prof}); conversely the profiles 
of the other three control sectors are fairly smooth with a continuous derivative. 
In the right panel of Fig. \ref{xmm.prof} we provide a zoom over the
interesting radial region where the edge is seen. The feature is also detected 
in the pn and the \acis\ data (Fig.\ref{chandra.image}): however it appear 
less evident in the \chandra\ image, being close to the outer edge of the CCD.

\subsection{The inner surface brightness edge}
\label{subsection.cf.int}

The \chandra\ image of Fig.\ref{chandra.image} provides a more detailed
insight over the inner region of \source, confirming what already pointed out in section 
\ref{subsection.cf.ext}: in a narrow sector at P.A. 130\deg-160\deg\ the source is
elongated in a luminous tail ending in a surface brightness edge. In Fig.\ref{chandra.prof} we compare the profile of the source 
in this sector, with the profile of the source in other two control sectors as depicted in
Fig.\ref{chandra.prof}. The profile of the source in the sector at P.A.130\deg-160\deg 
is clearly different with respect to the other control profiles in the radial range 20\arcsec-100\arcsec; 
the feature is more evident in the zoom of the left panel of Fig.\ref{chandra.prof}: 
a peak correspondent to the luminous
tail is seen, then at a radius of $\sim100$\arcsec, the profile flattens out.
The feature is clearly seen also in the \xmm\ data.
%The outer surface brightness discontinuity we detected in the \xmm\ profile is also marginally visible 
%in the \chandra\ data; the profile at PA -20\deg-70\deg begins to steepen at a radius of $\sim$100\arcsec.
%
\subsection{The spiral feature}
\label{subsection.spiral}

%%%%%%%%%%%%%%%%%%%%%%%%%%%%%%%%%%%%%%%%%%%%%%%%%%%%%%%%%%%%%%%%%%%%%%%%%%%%%%%%
\begin{figure}[th]
\centerline{
\includegraphics[height=0.27\textheight]{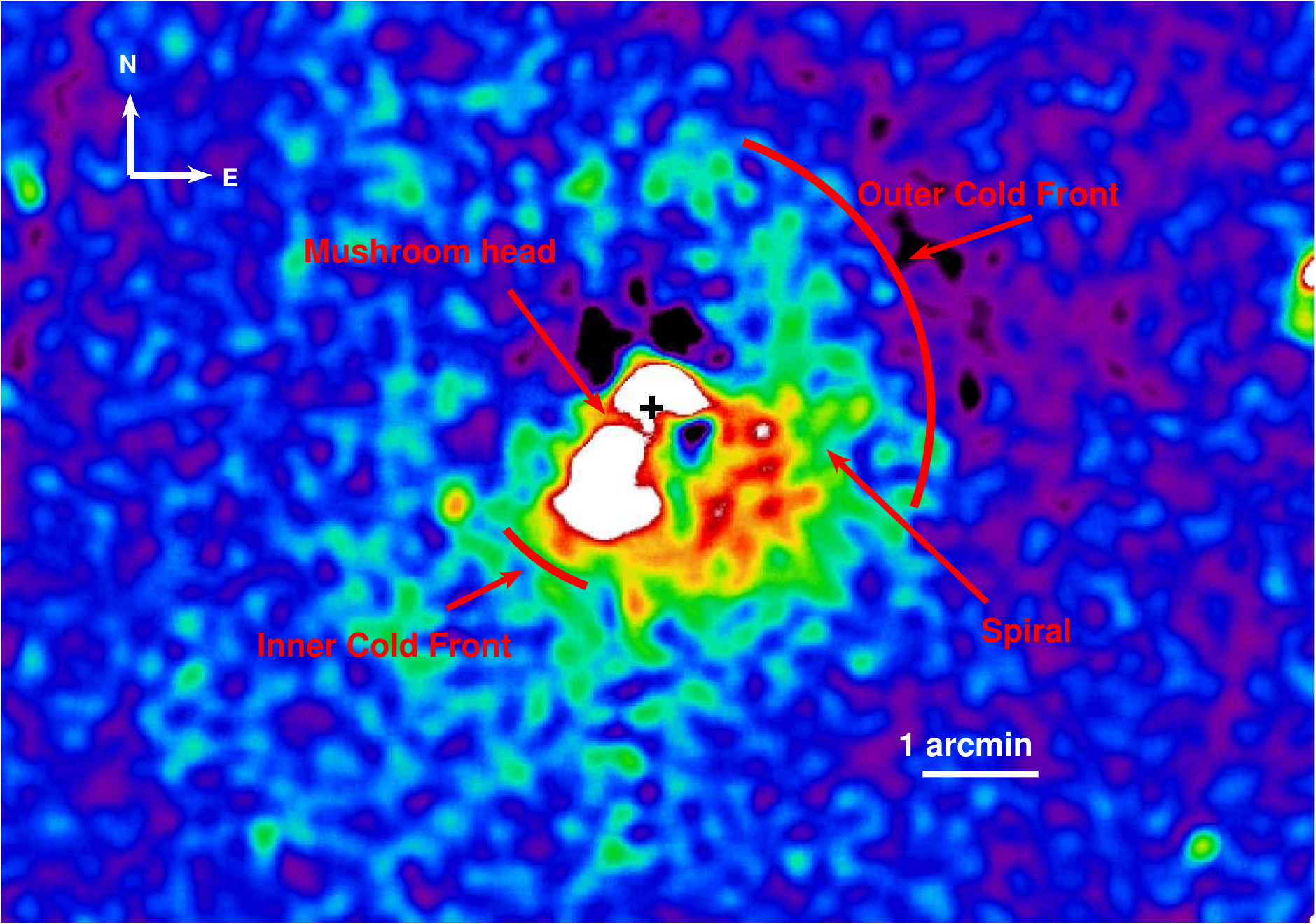}}
\caption{\label{fig.residuals} \footnotesize \xmm\ residual map, obtained by subtracting to the image of Fig.\ref{xmm.image}
the best fit beta model for the data. The center of IC 1860 is shown by the black cross and the position
and extent of the surface brightness edges by the red arcs. The ``mushroom head'' discussed in the text is the central 
region with white colour. The image has been processed to remove point sources.
}

\end{figure}
%%%%%%%%%%%%%%%%%%%%%%%%%%%%%%%%%%%%%%%%%%%%%%%%%%%%%%%%%%%%%%%%%%%%%%%%%%%%%

In Fig.\ref{fig.residuals} we present the surface brightness residual map between the
\xmm\ data and the best-fit radially symmetric two-dimensional $\beta$-model describing the surface
brightness distribution obtained with \emph{Sherpa}: the residual map highlights the appearance of sloshing features. 
A characteristic spiral pattern in surface brightness can be seen, with the tail being confined 
by the outer cold front. The qualitative impression is that the overall 
morphology is tilted with respect to the plane of the sky, with the mushroom structure formed by 
the coolest gas and seen in many hydrodynamic simulation \citep[e.g., the snapshot at 1.9 Gyr of Fig. 7 of][]{Ascasibar.ea:06} 
seen almost face on. The southern edge of this ``mushroom head'' coincides with the inner surface brightness discontinuity.

%
%
%=====================================================================
% SPECTRA
%=====================================================================
%
%%%%%%%%%%%%%%%%%%%%%%%%%%%%%%%%%%%%%%%%%%%%%%%%%%%%%%%%%%%%%%%%%%%%%%%%%%%%%%%%
\begin{figure*}[th]
%%\vspace{-0.5cm}
\centerline{
\parbox{0.5\textwidth}{
\includegraphics[height=0.28\textheight]{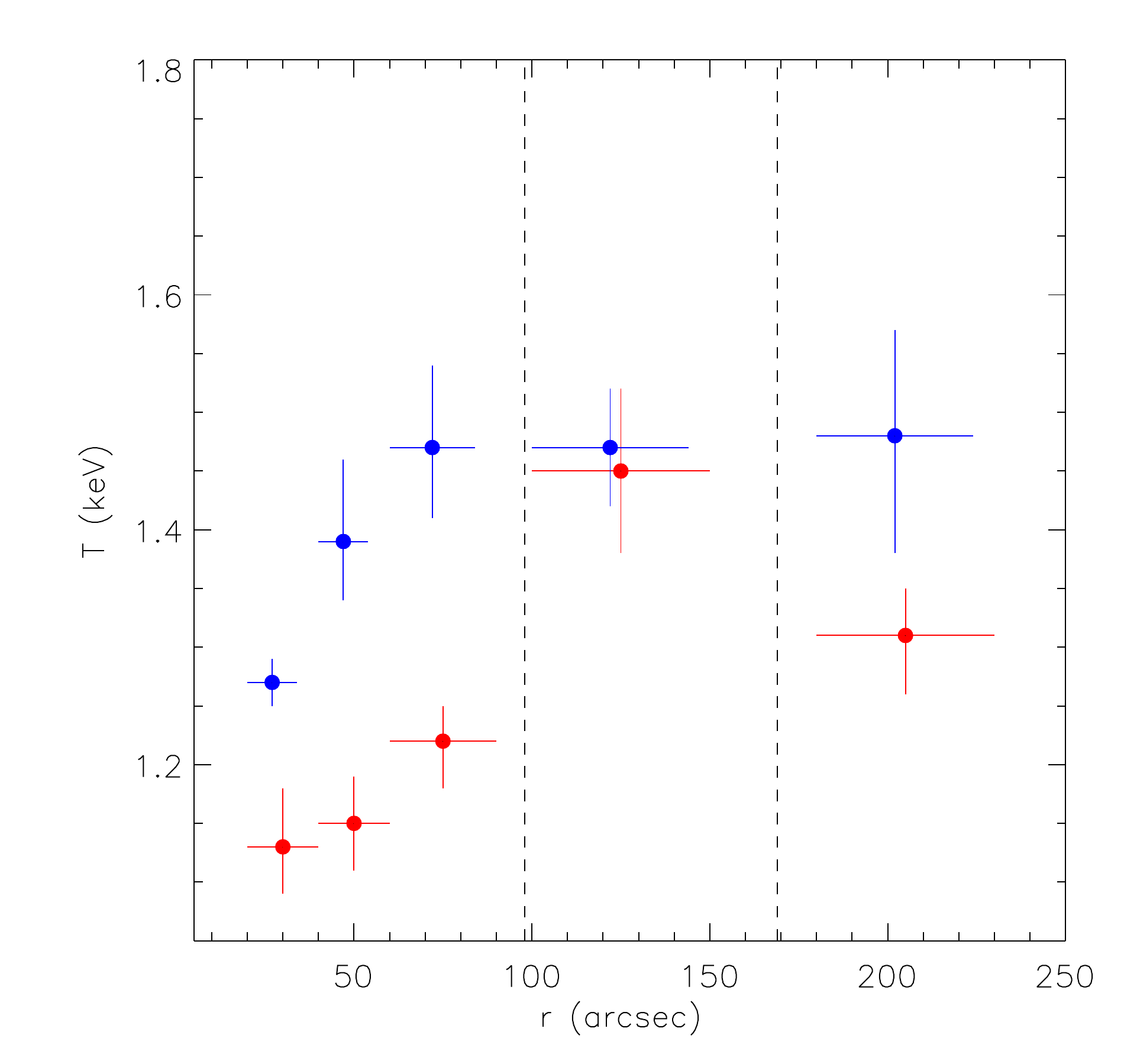}}
\parbox{0.5\textwidth}{
\includegraphics[height=0.28\textheight]{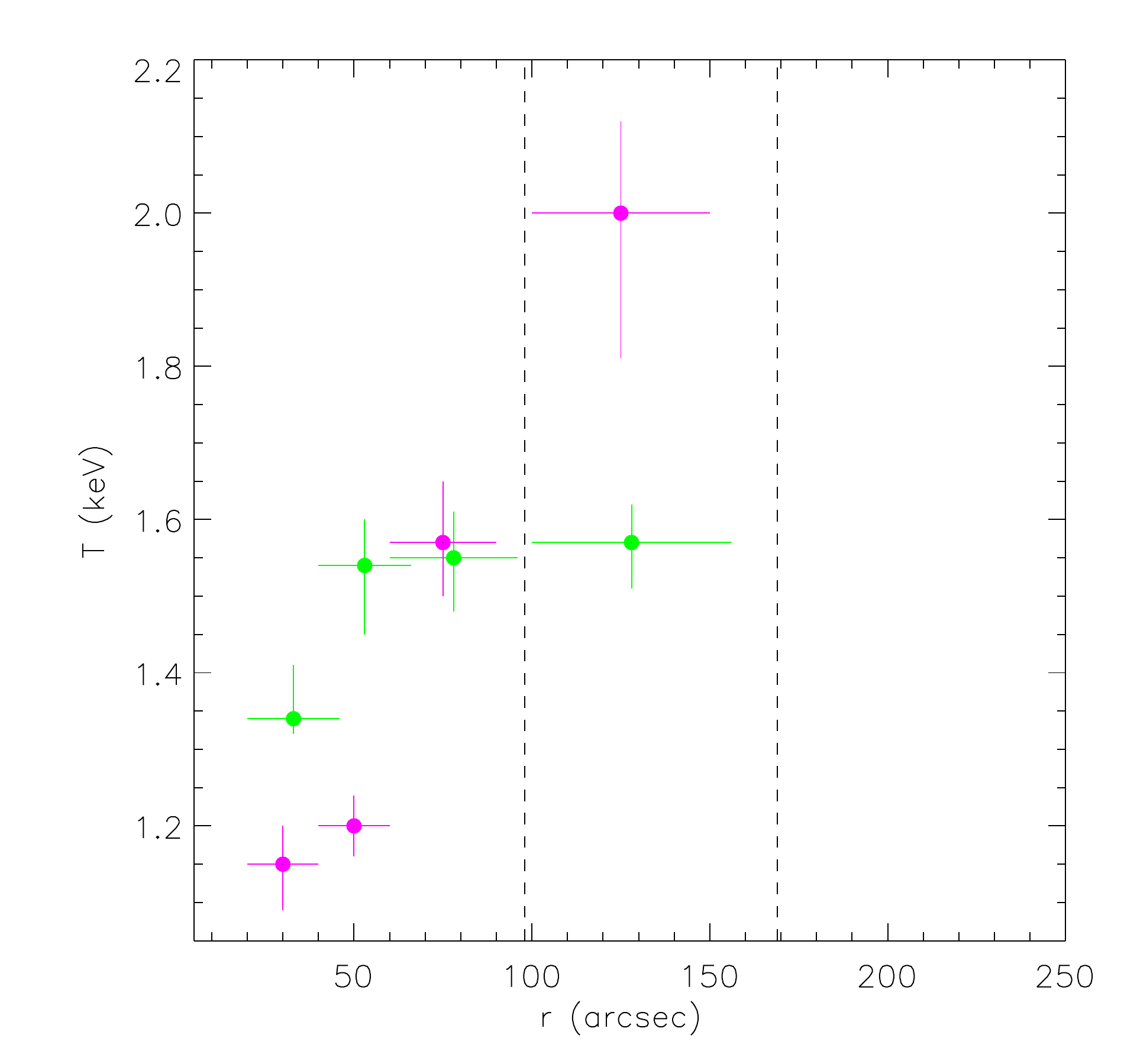}}
}
\caption{\label{prof.temp} \footnotesize
\emph{Left panel:}\xmm\ temperature profile along the sectors containing the surface brightness edge. Blue points: 
outer edge sector, P.A. 250\deg-340\deg. Red points: inner edge sector, P.A. 130\deg-160\deg. Horizontal error bars 
marks the radial extensions of the bins we used for the spectral analysis.The vertical dashed lines marks the radial position 
of inner and outer surface brightness edges.
\emph{Right panel:} \chandra\ temperature profile along the sectors containing the surface brightness edge. Green points: 
outer edge sector, P.A. 250\deg-340\deg. Magenta points: inner edge sector, P.A. 130\deg-160\deg. Horizontal error bars 
marks the radial extensions of the bin for the spectral analysis. The vertical dashed lines marks the radial position of 
the surface brightness edge as in the left panel: note that the more external radial bin of the \xmm\ temperature profile falls outside 
\chandra\ field of view and for clarity of the plot the temperature axes do not have the same extent in the two profiles. 
}
\end{figure*}
%%%%%%%%%%%%%%%%%%%%%%%%%%%%%%%%%%%%%%%%%%%%%%%%%%%%%%%%%%%%%%%%%%%%%%%%%%%%%%%%%%%%%%%%%%%%%%%%%%%%%
%
%%%%%%%%%%%%%%%%%%%%%%%%%%%%%%%%%%%%%%%%%%%%%%%%%%%%%%%%%%%%%%%%%%%%%%%%%%%%%%%%%%%%%%%%%%%%%%%%%%%%%
\begin{table*}[t] \footnotesize
\caption{Parameters from the spectral fits across the edges
\label{spectra.tab}}
\begin{center} \vskip -0.4cm
\begin{tabular}{ccccccc}
\tableline\tableline\\[-7pt]
& \multicolumn{3}{c}{XMM} & \multicolumn{3}{c}{CHANDRA}\\
& & $T$ & \zfe\ & & $T$ & \zfe\ \\
Radii PA & $\chi^2$/dof & (keV) & solar & $\chi^2$/dof & (keV) & solar  \\
\tableline \\[-7pt]
{\emph{Outer edge}} \\
100-150\arcsec\ 250\deg-340\deg &   98/105  & $1.47^{+0.05}_{-0.05}$ & $0.53^{+0.07}_{-0.07}$ &   
64/70  & $1.57^{+0.05}_{-0.06}$ & $0.52^{+0.10}_{-0.09}$  \\ 
180-230\arcsec\ 250\deg-340\deg &   55/68  & $1.48^{+0.09}_{-0.10}$ & $0.53^{+0.07}_{-0.07}$ &
 \nodata &  \nodata &  \nodata \\ 
{\emph{Control regions}} \\
100-150\arcsec\ 340\deg-70\deg &   84/122  & $1.59^{+0.04}_{-0.04}$ & $0.53^{+0.07}_{-0.07}$ & 
58/66  & $1.59^{+0.05}_{-0.06}$ & $0.57^{+0.11}_{-0.10}$ \\ 
180-230\arcsec\ 340\deg-70\deg &   73/91  & $1.42^{+0.08}_{-0.07}$ & $0.36^{+0.08}_{-0.07}$ &   
 \nodata &  \nodata &  \nodata  \\ 
100-150\arcsec\ 160\deg-250\deg &   139/150  & $1.48^{+0.04}_{-0.04}$ & $0.49^{+0.06}_{-0.06}$ &   
72/72  & $1.59^{+0.05}_{-0.06}$ & $0.55^{+0.10}_{-0.09}$ \\ 
180-230\arcsec\ 160\deg-250\deg &   94/113  & $1.58^{+0.05}_{-0.06}$ & $0.44^{+0.07}_{-0.07}$ &   
 \nodata &  \nodata &  \nodata \\ 
{\emph{Inner edge}} \\
60-90\arcsec\ 130\deg-160\deg &   48/54  & $1.22^{+0.04}_{-0.03}$ & $0.38^{+0.07}_{-0.06}$ &  
 24/21  & $1.57^{+0.08}_{-0.07}$ & $0.45^{+0.14}_{-0.12}$ \\ 
100-150\arcsec\ 130\deg-160\deg$^{\dagger}$ &   29/50  & $1.45^{+0.07}_{-0.07}$ & $0.41^{+0.09}_{-0.09}$ &    
39/47  & $2.00^{+0.12}_{-0.19}$ & $0.62^{+0.16}_{-0.18}$  \\ 
{\emph{Control regions}} \\
 60-90\arcsec\ 340\deg-70\deg&   63/81  & $1.68^{+0.04}_{-0.04}$ & $0.69^{+0.11}_{-0.09}$ &
 29/41  & $1.83^{+0.16}_{-0.14}$ & $0.53^{+0.17}_{-0.14}$ \\ 
100-150\arcsec\ 340\deg-70\deg &   84/122  & $1.59^{+0.04}_{-0.04}$ & $0.53^{+0.07}_{-0.07}$ &    
58/66  & $1.59^{+0.05}_{-0.06}$ & $0.57^{+0.11}_{-0.10}$ \\ 
 60-90\arcsec\ 160\deg-250\deg &   126/123  & $1.34^{+0.01}_{-0.01}$ & $0.50^{+0.05}_{-0.04}$ &
 60/55  & $1.52^{+0.05}_{-0.06}$ & $0.77^{+0.14}_{-0.14}$  \\ 
100-150\arcsec\ 160\deg-250\deg &   139/150  & $1.48^{+0.04}_{-0.04}$ & $0.49^{+0.06}_{-0.06}$ &
72/72  & $1.59^{+0.05}_{-0.06}$ & $0.55^{+0.10}_{-0.09}$\\ 
\tableline 
\vspace{0.5 cm}
\end{tabular}
\tablecomments{Results of \xmm\  and \chandra\ spectral fits, in the radial bins inside and outside the surface brightness 
edges, as discussed in section \S \ref{spc}. The first and the second columns refers to the radial and angular range of 
each bins. \xmm\ and \chandra\ are reported as labeled on the top of the table. No entry in the \chandra\ rows means 
that the referred bins falls out the available field of view.\\
 $^{\dagger}$ The \chandra\ data have been extracted in a 100\deg-160\deg\ PA region to increase the statistics}
\end{center}
\end{table*}
\vspace{0.5 cm}
%
%%%%%%%%%%%%%%%%%%%%%%%%%%%%%%%%%%%%%%%%%%%%%%%%%%%%%%%%%%%%%%%%%%%%%%%%%%%%%%%%%%%%%%%%%%%%%%%%%%%%%%%%%%%%%%%
%
%%%%%%%%%%%%%%%%%%%%%%%%%%%%%%%%%%%%%%%%%%%%%%%%%%%%%%%%%%%%%%%%%%%%%%%%%%%%%%%%%%%%%%%%%%%%%%
\begin{figure*}[th]
%%\vspace{-0.5cm}
\centerline{
\parbox{0.5\textwidth}{
\includegraphics[height=0.28\textheight]{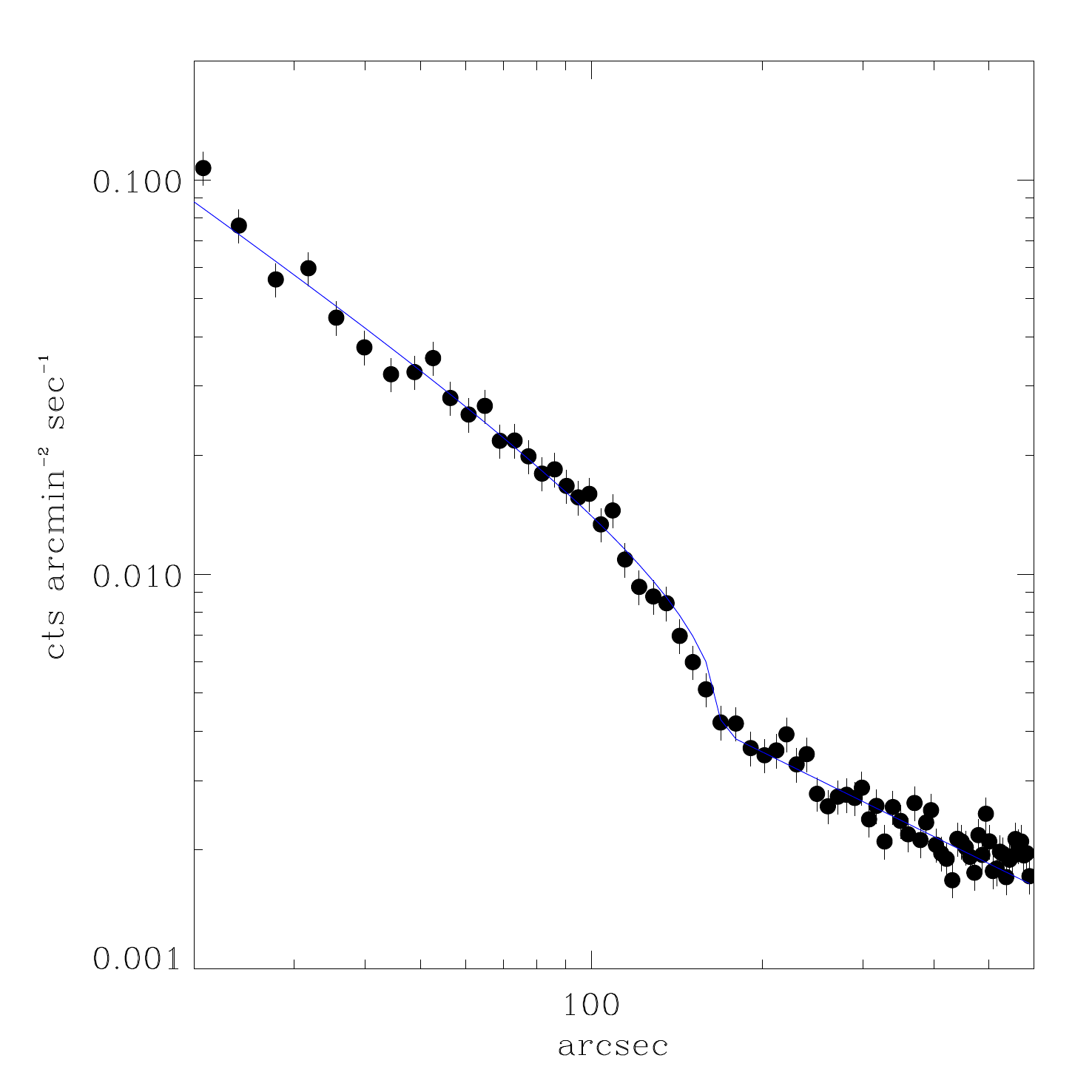}}
\parbox{0.5\textwidth}{
\includegraphics[height=0.28\textheight]{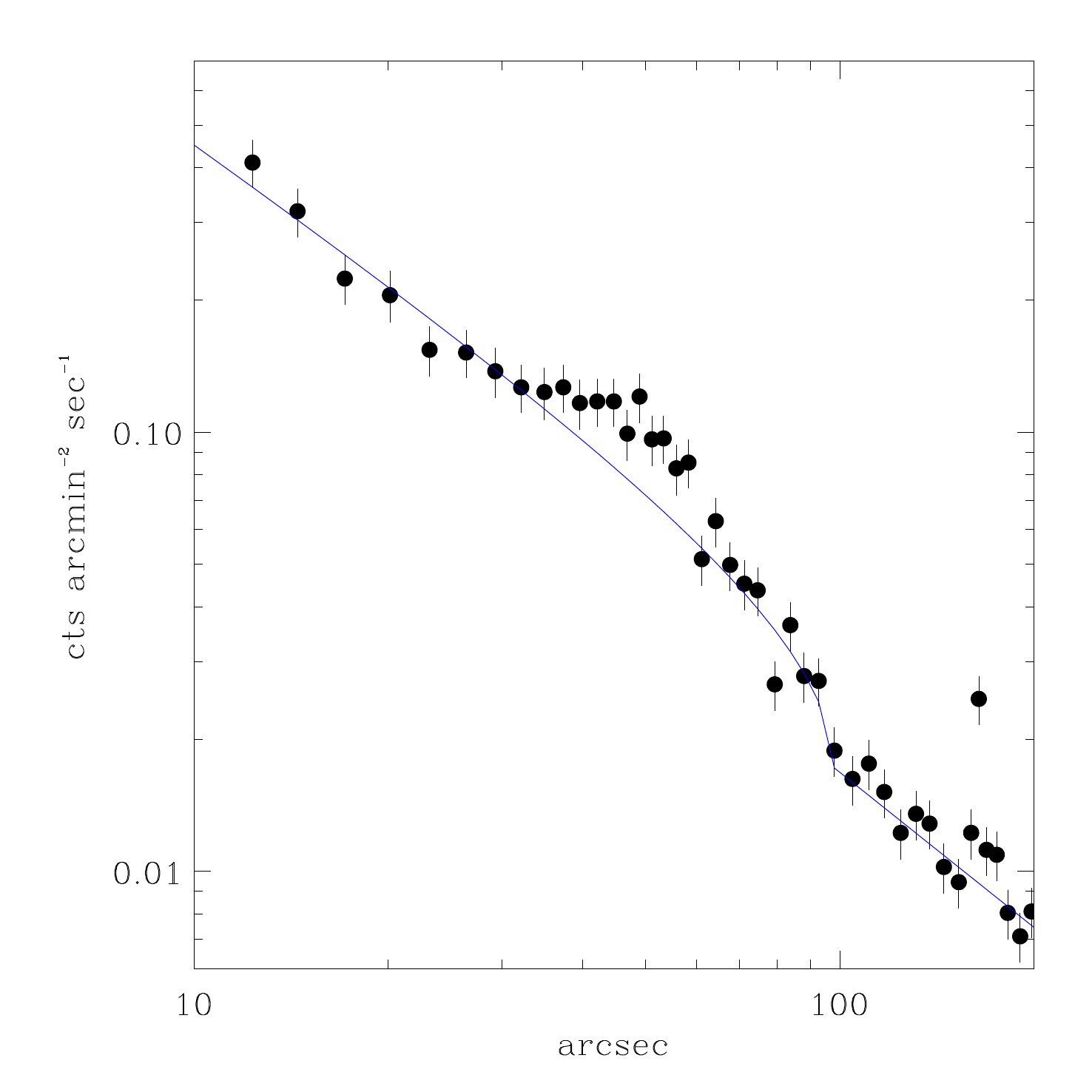}}
}
\caption{\label{fit_prof.plot} \footnotesize
\emph{Left panel:}Fit of the \xmm\ surface brightness profile along the outer edge. The superposed solid line represents the 
best fit  two power law model, as discussed in section \S\ref{fit_prof}.
\emph{Right panel:} Fit of the \chandra\ surface brightness profile along the outer edge. The superposed solid line 
represents the best fit two power law model. Note that, as discussed in section \S\ref{fit_prof}, the model does not fit 
the profile in the radial range $\sim$20\arcsec-$\sim$50\arcsec, where the source is elongated in a narrow luminous tail.
}
\end{figure*}
%
%%%%%%%%%%%%%%%%%%%%%%%%%%%%%%%%%%%%%%%%%%%%%%%%%%%%%%%%%%%%%%%%%%%%%%%%%%%%%%%%%%%%%%%%%%%%%%
%
\begin{table*}[th] \footnotesize
\caption{Parameters from the fits of the surface brightness profiles across the edges.
\label{fit_prof.tab}}
\begin{center} \vskip -0.4cm
\begin{tabular}{cccccccc}
\tableline\tableline\\[-7pt]
& r (kpc) & $\gamma_{in}$ & $\gamma_{out}$ &$\frac{n_{in}}{n_{out}}$ & $\frac{P_{in}}{P_{out}}$ & $M$ \\
\tableline \\[-7pt]
Outer edge P.A. 130\deg-160\deg &  76 & $0.88\pm0.09$ & $0.85\pm0.03$ & $1.28\pm0.03$ & $1.27\pm0.10$ &$0.55\pm0.08$ \\ 
Inner edge P.A. 250\deg-340\deg &  45  & $0.90\pm0.05$ & $1.00\pm0.06$ & $1.65\pm0.06$ & $1.30\pm0.13$ & $0.57^{+0.10}_{-0.14}$ \\ 
\tableline 
\vspace{0.5 cm}
\end{tabular}
\tablecomments{Results from the \xmm\ and \chandra\ profiles fits, along the outer and the inner edge respectively. The 
first two columns refer to the internal and external power-law index, as discussed in the text, section \S \ref{fit_prof}. 
We then report the derived density and pressure jumps and the Mach number $M$.}
\end{center}
\end{table*}
%%%%%%%%%%%%%%%%%%%%%%%%%%%%%%%%%%%%%%%%%%%%%%%%%%%%%%%%%%%%%%%%%%%%%%%%%%%%%%%%%%%%%%%%%%%%%%%%%%%%%%%
%
\section{X-ray spectral analysis}
\label{spc}

To further investigate the nature of the brightness discontinuities we
analyzed the temperature profiles across them.
For this purpose we extracted \xmm\  and \chandra\ spectra  
in several radial bins\footnote{The range of the radial bins used for the
spectral analysis are: 20\arcsec-40\arcsec;
40\arcsec-60\arcsec;
60\arcsec-90\arcsec;
100\arcsec-150\arcsec;
180\arcsec-230\arcsec;},
along the angular sectors containing the edges  (P.A. 250\deg-340\deg and P.A. 130\deg-160\deg)
and in other two control sector (P.A. 340\deg-70\deg; P.A. 160\deg-250\deg) in order to measure 
the temperature trend across the edges and check if it is actually different from the overall undisturbed 
temperature profile.

We extracted \xmm\ spectra generating a response file and an ancillary response file
using the standard SAS tasks {\em rmfgen} and {\em arfgen}  in extended source mode;
we extracted \chandra\ spectra generating count-weighted spectral response matrices 
appropriate for each region  using the task {\em specextract}.
For each region, we extracted also background spectra 
from the blank-field event files. All spectra were re-binned to ensure
a minimum 20 counts bin$^{-1}$ and fitted in the 0.5-5.0 keV band,
with an \apec\ \citep{Smith.ea:01} plasma model 
as implemented in \xspec\ version 11.0.
We take into account galactic absorption using a {\em phabs} model 
with the hydrogen column density frozen to the value provided in \citet{Kalberla.ea:05}. 
In section \ref{syst} we will explore the sensitivity of the results
to the above assumptions.

In Table \ref{spectra.tab} we present the results of the spectral fits
in the interesting radial bins inside and outside the edges and 
in the same radial bins in the control sectors. In Fig.\ref{prof.temp} we show 
the temperature profile along the sectors containing the edges 
as obtained with \xmm\ (left panel) and \chandra\ (right panel): 
the vertical dashed lines mark the radius of the two brightness discontinuity.
The region outside the outer edge is covered only
by the larger \xmm\ field of view; therefore the complete
temperature profile along the outer edge is available
only in the left panel of Fig.\ref{prof.temp}. 

For the inner edge both  \xmm\ and  \chandra\
temperature profiles show an enhancement of the gas temperature
across the discontinuity; the same behavior is not observed
in the control sectors. It is also clear that as expected there is asymmetry in the
temperature profile, in particular the part of the ``mushroom head'' of emission in the SE leading to the inner edge is 
cooler than the surroundings. 
The temperature trend observed
by the two instruments is the same, although in the outer radial bin 
the \chandra\ temperature is 38\% higher than the \xmm\ one.
We interpreted this discrepancy as due to a temperatures mixing in the \xmm\ spectrum, induced by
its larger PSF. To check this hypothesis we compared \xmm\ and \chandra\
temperatures in two wider (P.A: 100\deg-1600\deg) radial bins 
at the same radii where we observed the temperatures
discrepancy (100\arcsec-150\arcsec) and further away 
(150\arcsec-210\arcsec). Coherently with the given
interpretation, away from the disturbed region where
the edge is seen, the discrepancy between \chandra\
and \xmm\ temperature™es falls to the 12\% level. 

There is no statistically significant change in the temperature across the outer edge.
The quality of the data allows us only to measure projected temperatures meaning that the true de-projected difference of temperatures
across the front can be lower. Furthermore when contrasted with the declining trend of the
control regions at different azimuthal angles, the picture is consistent with cooler gas inside the 
edge. A similar behavior has been seen for the outer edge in the NGC 5044 \citep{Gastaldello.ea:09}
and NGC 5846 \citep{Machacek.ea:11} groups.
%
%=====================================================================
%FIT PROFILI FIT PROFILI  FIT PROFILI FIT PROFILI FIT PROFILI FIT PROFILI
%=====================================================================
%
\section{Fits of the surface brightness edges}
\label{fit_prof}
In order to characterize in a quantitative way the density jump across the cold front
we modeled the the surface brightness profile across the discontinuity following the
approach of \citet{Rossetti.ea:07} and \citet{Gastaldello.ea:10}. We assume that
the gas density profile is described by a power law on either side of the edge,

\begin{equation}
\label{eq.coldfront}
n=\left\{\begin{array} {ll}
n = n_{in} \left(\frac{r}{r_{cf}}\right)^{-\gamma_{in}} & r < r_{cf} \\
n = n_{out} \left(\frac{r}{r_{cf}}\right)^{-\gamma_{out}} & r > r_{cf} \end{array}\right.
\end{equation}
where $r_{cf}$ is the cold front radius and we derive the parameters of this model from the fit of the projected 
surface brightness profile, which can be expressed as the integral of the emissivity along the line of sight:
\begin{equation}
\Sigma=K\int_0^{+\infty}n^2(z) \Lambda(T,Z) dz,
\end{equation}
where K is a constant and we took into account the dependence of the emissivity of the temperature $T$ and 
metal abundance $Z$.

We fit the outer part of the surface brightness profile to set the external component parameters 
and we successively derive the best fit parameters of the innermost part.
We applied this method to the \xmm\ profile across the outer edge and 
to the \chandra\ profile across the inner edge. Our results are summarized
in Table \ref{fit_prof.tab} and in figure \ref{fit_prof.plot} we plot the surface brightness profile
and the best fit model respectively for the outer (left panel) and inner edge (right panel).

For the outer edge the agreement between the chosen model
and the data is good. We find a density jump of $n_{in}/n_{out} = 1.28 \pm 0.03$  
that, when combined with the temperature and abundance information in the regions inside and
outside the front reported in Table \ref{spectra.tab}, gives a pressure ratio of
$P_{in}/P_{out} = 1.27 \pm 0.10$. If this pressure ratio is interpreted as evidence for bulk motion 
of the cold front then following \citet{Vikhlinin.ea:01} we infer that
the front is moving subsonically with a Mach number $M = 0.55 \pm 0.08$.

In the profile of the inner edge a surface brightness bump
is seen between $\sim$20\arcsec and $\sim$50\arcsec, which
is not well fitted by the adopted two power law model. This
discrepancy was expected, given the peculiar shape of the
source in the sector containing the edge: the observed bump
corresponds roughly to the region where the source is elongated
in a narrow luminous tongue. Except for this peculiar structure,
the profile inside and outside the edge is well described by
the model and then internal and external power law indexes are
well determined. We find a density jump of $n_{in}/n_{out} = 1.65 \pm 0.06$
and using the \chandra\ temperature and abundance determination 
the pressure ratio is $P_{in}/P_{out} = 1.30 \pm 0.13$ corresponding to a Mach number 
$M = 0.55^{+0.10}_{-0.14}$. We will explore the robustness of this result with 
respect to various analysis choices in \S\ref{systcoldfront}.

%
%=====================================================================
% SYSTEMATICS  SYSTEMATICS  SYSTEMATICS  SYSTEMATICS  SYSTEMATICS   
%=====================================================================
%
%
\section{Systematic Errors}
\label{syst}
In this section we provide an investigation 
of  the possible systematic errors affecting the temperatures 
and abundances across the edges quoted in Table \ref{spectra.tab}. 
Readers who are uninterested in these technical details can safely skip
to the following section.
\subsection{Plasma Codes}
\label{plasma}
We compared the results obtained using the \apec\ code to those
obtained using the \mekal\ code \citep{Kaastra.ea:93} 
The different implementations of the atomic physics and
different emission line lists in the plasma codes produce 
no qualitative differences between the fits and no significative variations
of $\chi^2$. Variations of the  fitted \xmm\ temperatures are all within $\sim$5\% 
and well comprised into the statistical errors; all fitted \xmm\ abundance agrees 
within $\sim$9\%.

For the \chandra\ data the use of the different plasma code produces a 5\% 
variation of the inner spectral temperature and 4\% variation of the outer one.
\subsection{Bandwidth}
We explored the sensitivity of our results to the default lower limit of
the bandpass, $\emin=0.5$~keV. For comparison we performed spectral fits with 
$\emin=0.4$~keV and $\emin=0.7$~keV. The fitted temperatures are substantially
unchanged and consistent within the statistical errors, both for \xmm\ and for
\chandra; an analogous result holds for the iron abundances.\\
\subsection{Variable \nh}
\label{nh}
To check for possible deviations of \nh\ from the galactic value of 
\citet{Kalberla.ea:05}, we perform a spectral fits leaving \nh\ as free
parameters. For the \xmm\ spectra the fitted \nh\ values are higher by a factor $\sim$ 3
with respect to the Galactic value; the correspondent variations
of fitted temperatures are within $\sim$9\%; absolute variations 
of temperatures and abundances are however well accounted 
by the quoted statistical errors.\\
The parameters of the \chandra\ spectrum of the region inside the inner edge
are basically unchanged and the fitted \nh\ agrees with the Galactic value 
within 1$\sigma$. The spectral fits of the region outside the inner edge
gives instead a higher \nh\ by a factor $\sim$ 4; the correspondent
fitted temperature (T=$1.97^{+0.31}_{-0.47}$) however agrees with
the quoted one within the statistical errors and its lower limit is still
consistent with an enhancement of the temperature outside the inner edge.\\
\subsection{Background}
To account for systematic errors in background normalization, we performed
spectral fits allowing it to vary of $\pm$ 5\%. Temperatures and abundances
are insensitive to this variations in all the considered \xmm\ spectra; variations 
of \chandra\ temperatures are all within 10\%, while iron abundances are 
basically unchanged.\\
\subsection{Cold front modeling}
\label{systcoldfront}
The inner cold front is not perfectly modeled by the expression given in 
Eq.\ref{eq.coldfront} and there are discrepancies between the spectral parameters
at either sides of the front between the \chandra\ and \xmm\ values. We explore the sensitivity 
of the derived Mach number with respect to the choice of the fitting range of the surface brightness profiles
and the adopted temperature and abundance values. The fiducial measurement is derived by using the
\chandra\ spectral values and fitting the internal component in the range 7\arcsec-40\arcsec.
If we fit the internal component in the range 7\arcsec-30\arcsec\ we obtain a worse fit
and a Mach number $M=0.35^{+0.15}_{-0.35}$ and if we use instead a 60\arcsec-90\arcsec\ range
to better model the excess, with the problem of deriving a very high density in the inner regions, we
obtain a Mach number $M=0.48^{+0.12}_{-0.18}$. 
Using the \xmm\ spectral parameters instead of the \chandra\ values we obtain $M=0.67^{+0.06}_{-0.08}$.
We therefore conclude that our determination in \S\ref{fit_prof} is rather robust to possible systematic errors.

%
%=====================================================================
% OPTICAL ANALYSIS
%=====================================================================
%
\section{Optical Analysis}
\label{optical}

%%%%%%%%%%%%%%%%%%%%%%%%%%%%%%%%%%%%%%%%%%%%%%%%%%%%%%%%%%%%%%%%%%%%%%%%%%%%%%%%
\begin{figure*}[t]
\centerline
{\includegraphics[width=0.9\textwidth]{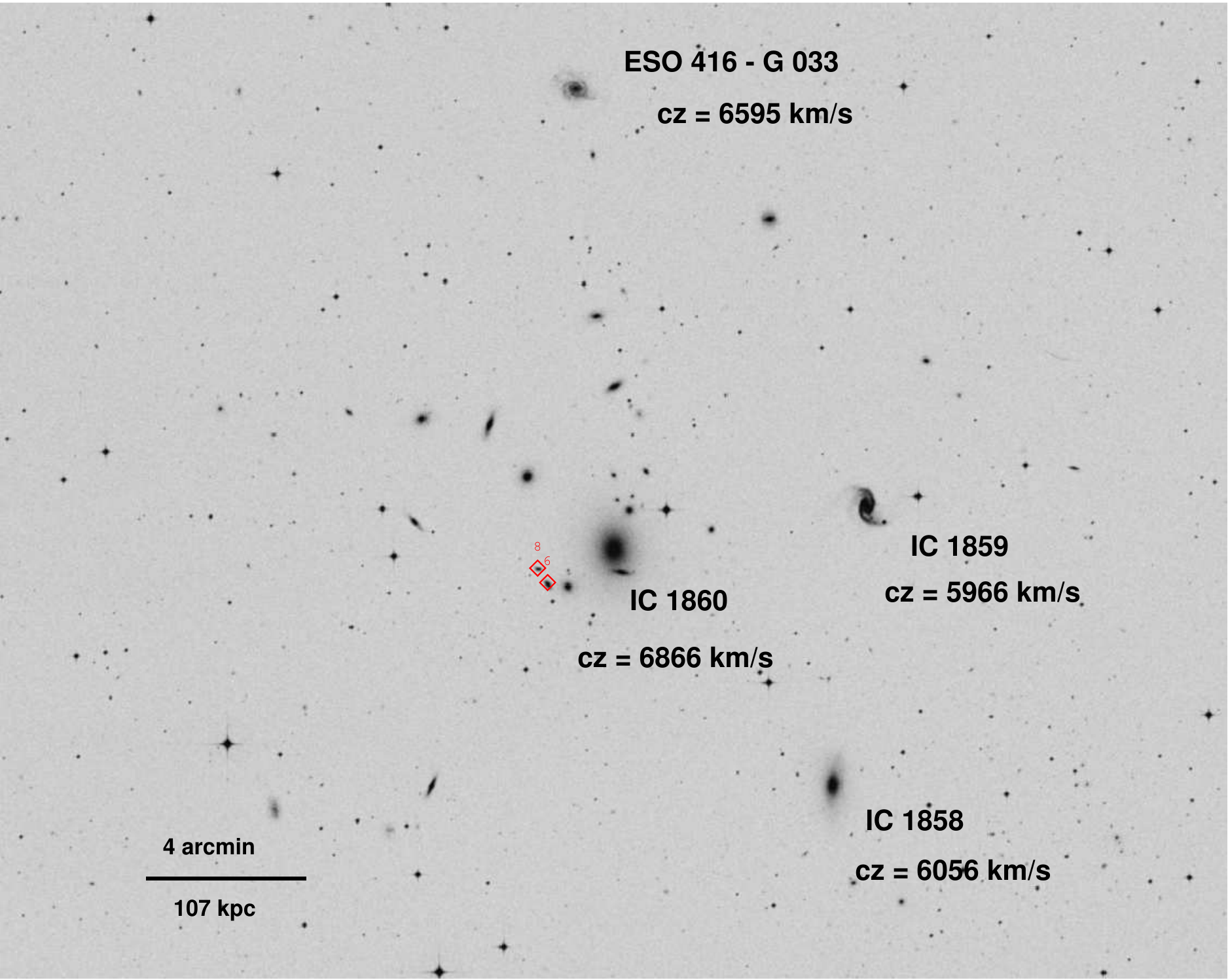}}
\caption{\label{fig.optical} \footnotesize DSS image of the central region of the IC 1860
group. IC 1860, IC 1859 and ESO 416 - G033 are indicated (see discussion in the text \ref{perturbers}). 
IC 1858 is also indicated (see discussion in Appendix \ref{sec.appendixic1858}). Red labels refer to galaxies rejected by the 
shifting gapper method (see \ref{membership}), with numbers refering to the number entry
in the catalogue of Table \ref{tab.members}.
 }
\end{figure*}
%%%%%%%%%%%%%%%%%%%%%%%%%%%%%%%%%%%%%%%%%%%%%%%%%%%%%%%%%%%%%%%%%%%%%%%%%%%%%%%%%%%%%
%
In this section we present our optical analysis that includes the determination 
of group membership and substructure detection. The Digital Sky survey (DSS) image of the
central region of the group is shown in Figure \ref{fig.optical}. The purpose of the optical
analysis is to search for substructures and to use them together with the X-ray observations
to investigate the dynamical state of the IC 1860 group.
\subsection{Sample selection and group membership}
\label{membership}

\begin{table*}[t] \footnotesize
\caption{IC1860 Group Galaxies
\label{tab.members}}
\begin{center} \vskip -0.4cm
\begin{tabular}{ccccccc}
\tableline\tableline\\[-7pt]
ID & Galaxy Identifier & RA & DEC & c$z$ & c$z$ Error & Source \\
\tableline \\[-7pt]
1 & IC 1860                 & 42.39047 & -31.1891  & 6866  & 20 &  \citet{Stein:96} \\  
2 & 2MASX J02493302-3111569 & 42.38674 & -31.19923 & 7532  & 33 &  \citet{Stein:96} \\  
3 & 2MASX J02493200-3110229 & 42.38304 & -31.17301 & 6865  & 64 &   2dF  \\  
4 & 2dFGRS S394Z186         & 42.38183 & -31.16742 & 7525  & 89 &   2dF  \\  
5 & 2MASX J02493933-3112169 & 42.41373 & -31.20474 & 7905  & 20 &  \citet{Stein:96} \\  
\textit{6}* & \textit{2MASX J02494174-3112139} & \textit{42.42371} & \textit{-31.20347} & \textit{8664} & \textit{64} & \textit{2dF} \\  
7 & 2dFGRS S394Z183         & 42.4035  & -31.2205  & 6416  & 123 &  2dF  \\  
\textit{8}* & \textit{2dFGRS S394Z176} & \textit{42.4285}  & \textit{-31.19736} & \textit{8784} &  \textit{64} &  \textit{2dF} \\  
9 & 2MASX J02492990-3109249 & 42.37436 & -31.15675 & 7945  & 64  &  2dF  \\  
10 & MCG -05-07-036          & 42.34263 & -31.18172 & 6835 & 64  &  2dF  \\  
11 & 2dFGRS S467Z174         & 42.43247 & -31.15795 & 6985 & 64  &  2dF  \\  
12 & 2MASX J02493349-3107159 & 42.38908 & -31.12111 & 6985 & 64  &  2dF  \\  
13 & 2dFGRS S467Z173         & 42.45072 & -31.13592 & 6835 & 64  &  2dF  \\  
14 & 2MASXi J0249572-311036  & 42.4886  & -31.17639 & 6086 & 64  &  2dF  \\  
15 & 2dFGRS S394Z181         & 42.398   & -31.09167 & 6266 & 64  &  2dF \\  
16 & 2MASX J02495627-3107599 & 42.48442 & -31.13339 & 7195 & 64  &  2dF  \\  
17 & IC 1859                 & 42.26635 & -31.17251 & 5966 & 64  &  2dF  \\  
18 & 2dFGRS S467Z168         & 42.52021 & -31.13019 & 7664 & 64  &  2dF  \\  
19 & 2MFGC 02270             & 42.48279 & -31.2875  & 6356 & 64  &  2dF  \\  
20 & IC 1858                 & 42.28504 & -31.28958 & 6056 & 89  &  2dF  \\  
21 & 2dFGRS S394Z189         & 42.30958 & -31.06367 & 6446 & 89  &  2dF \\  
22 & LSBG F416-020           & 42.50358 & -31.30597 & 6835 & 89  &  2dF \\  
23 & 2MASX J02485680-3106431 & 42.23634 & -31.11194 & 6835 & 64  &  2dF  \\  
24 & 2MASX J02491494-3103060 & 42.31196 & -31.05162 & 6760 & 23  &  \citet{Stein:96} \\  
25 & 2MASX J02493567-3101249 & 42.39858 & -31.02364 & 6955 & 64  &  2dF \\  
26 & 2dFGRS S394Z174         & 42.45371 & -31.03258 & 7345 & 123 &  2dF \\  
27 & 2dFGRS S467Z163         & 42.56    & -31.29644 & 7315 & 89  &  2dF \\  
28 & ESO 416- G 033          & 42.40642 & -30.99583 & 6595 & 123 &  2dF  \\  
29 & 2dFGRS S466Z088         & 42.23542 & -31.38433 & 6685 & 89  &  2dF \\  
30 & 2dFGRS S394Z172         & 42.47059 & -30.95476 & 7045 & 64  &  2dF \\  
31 & 2dFGRS S394Z162         & 42.68179 & -31.04658 & 7045 & 123 &  2dF \\  
32 & ESO 416- G 035          & 42.63233 & -31.39464 & 6535 & 64  &  2dF  \\  
33 & 2dFGRS S394Z161         & 42.68804 & -31.04619 & 6625 & 89  &  2dF  \\  
34 & DUKST 416-040           & 42.1205  & -31.37789 & 6026 & 123 &  2dF \\  
35 & 2dFGRS S394Z115         & 42.22129 & -30.88644 & 6505 & 123 &  2dF  \\  
36 & 2MASX J02511626-3119392 & 42.81783 & -31.32758 & 6559 & 45  &  6dF \\  
\textit{37}* & \textit{ESO 416- G 025} & \textit{42.16992} & \textit{-31.53622} & \textit{4977} & \textit{64} & \textit{2dF} \\  
38 & ESO 416- G 036          & 42.64808 & -31.52008 & 7015 & 64  &  2dF \\  
39 & 2MASX J02504193-3130145 & 42.67467 & -31.50403 & 7135 & 64  &  2dF \\  
40 & ESO 416- G 034          & 42.60833 & -31.55828 & 7045 & 64  &  2dF \\  
\tableline 
\vspace{0.5 cm}
\end{tabular}
\end{center}
\end{table*}
\addtocounter{table}{-1}
\begin{table*}[t] \footnotesize
\caption{Continued
\label{tab.members}}
\begin{center} \vskip -0.4cm
\begin{tabular}{ccccccc}
\tableline\tableline\\[-7pt]
ID & Galaxy Identifier & RA & DEC & c$z$ & c$z$ Error & Source \\
\tableline \\[-7pt]
41 & ESO 416- G 027          & 42.23883 & -30.79289 & 6835 & 64  &  2dF  \\  
42 & LCSB S0441P             & 41.89725 & -31.10831 & 6925 & 64  &  2dF  \\  
43 & 2dFGRS S467Z150         & 42.70867 & -31.53058 & 6026 & 89  &  2dF \\  
44 & 2dFGRS S467Z190         & 42.28596 & -31.68119 & 6745 & 89  &  2dF  \\  
\textit{45}* & \textit{2dFGRS S466Z100} & \textit{42.13721} & \textit{-31.64678} & \textit{5516} & \textit{89} & \textit{2dF} \\  
46 & 2dFGRS S394Z155         & 42.919   & -30.94231 & 6446 & 123 &  2dF  \\  
47 & 2dFGRS S467Z121         & 42.9955  & -31.09386 & 6386 & 64  &  2dF \\  
48 & 2dFGRS S467Z134         & 42.9345  & -31.48431 & 6865 & 64  &  2dF \\  
49 & LCSB L0144P             & 42.83546 & -30.78181 & 7010 & 31  &  6dF \\  
50 & 2MASX J02483150-3142206 & 42.13125 & -31.70575 & 6835 & 64  &  2dF \\  
51 & 2MASX J02493534-3145319 & 42.39725 & -31.75886 & 6955 & 89  &  2dF \\  
52 & ESO 416- G 021          & 41.78033 & -31.48347 & 7435 & 89  &  2dF \\  
53 & 2dFGRS S467Z114         & 43.11317 & -31.32844 & 6505 & 123 &  2dF \\  
54 & DUKST 416-030           & 43.07238 & -30.94333 & 6655 & 64  &  2dF \\  
55 & 2MASX J02483040-3147366 & 42.12675 & -31.79356 & 6356 & 64  &  2dF \\  
56 & 2dFGRS S466Z087         & 42.24479 & -31.83344 & 6955 & 89  &  2dF \\  
57 & 2dFGRS S466Z140         & 41.79083 & -31.60678 & 6595 & 89  &  2dF \\  
58 & 2MASXi J0250166-303222  & 42.5695  & -30.5395  & 7075 & 64  &  2dF \\  
59 & 2dFGRS S467Z135         & 42.91488 & -31.73853 & 6895 & 123 &  2dF  \\  
\textit{60}* & \textit{2dFGRS S393Z037} & \textit{41.68479} & \textit{-30.78836} & \textit{4707} & \textit{123} & \textit{2dF} \\  
61 & DUKST 416-043           & 41.63875 & -31.54614 & 7345 & 89  &  2dF \\  
62 & ESO 416- G 040          & 43.11304 & -30.77653 & 6775 & 89  &  2dF  \\  
63 & 2dFGRS S467Z137         & 42.854   & -31.82    & 6775 & 89  &  2dF  \\  
64 & 2dFGRS S394Z041         & 42.92617 & -30.5975  & 6745 & 89  &  2dF \\  
65 & 2dFGRS S466Z101         & 42.14604 & -31.91153 & 6865 & 89  &  2dF  \\  
66 & LCSB S0440P             & 41.86242 & -31.81589 & 6838 & 150 &  DUKST     \\  
67 & AM 0250-310             & 43.22633 & -30.85742 & 6745 & 64  &  2dF  \\  
\textit{68}* & \textit{2dFGRS S393Z025} &\textit{ 41.89079} & \textit{-30.50036} & \textit{5216} &  \textit{89} & \textit{2dF} \\  
69 & 2dFGRS S395Z027         & 43.00417 & -30.56647 & 6955 & 89  &  2dF \\  
70 & 2dFGRS S467Z136         & 42.87258 & -31.92583 & 6146 & 123 &  2dF \\  
71 & 2dFGRS S395Z037         & 43.33496 & -30.91456 & 6715 & 89  &  2dF \\  
72 & ESO 416- G 039          & 43.09304 & -31.81728 & 6625 & 64  &  2dF \\  
73 & 2dFGRS S394Z015         & 43.28067 & -30.77533 & 6416 & 64  &  2dF \\  
74 & 2dFGRS S393Z030         & 41.80092 & -30.47836 & 6805 & 89  &  2dF \\  
75 & 2dFGRS S395Z035         & 43.41779 & -31.16642 & 6386 & 64  &  2dF \\  
76 & ESO 416- G 041          & 43.37237 & -30.86036 & 6416 & 64  &  2dF \\  
\textit{77}* & \textit{2dFGRS S466Z122} & \textit{41.98921} & \textit{-32.05119} & \textit{4947} &  \textit{64} & \textit{2dF} \\  
78 & LSBG F416-018           & 43.264   & -31.75997 & 6296 & 123 &  2dF \\  
79 & 2dFGRS S466Z183         & 41.43917 & -31.69525 & 6595 & 89  &  2dF \\  
80 & 2dFGRS S466Z167         & 41.55225 & -31.85869 & 6446 & 123 &  2dF \\  
81 & 2dFGRS S466Z166         & 41.55658 & -31.88614 & 6985 & 89  &  2dF \\
\tableline 
\vspace{0.5 cm}
\end{tabular}
\tablecomments{Columns are as follows: (1) Galaxy Identifier (2) Right ascension (J2000), (3)
Declination (J2000), (4) Recession Velocity, (5) Error on the Recessio Velocity (6) Source of the measurement \\
The galaxies with entry in italics and marked with an asterisk are the galaxies rejected from memebership by the 
shifting gapper method.}
\end{center}
\end{table*}

%
%%%%%%%%%%%%%%%%%%%%%%%%%%%%%%%%%%%%%%%%%%%%%%%%%%%%%%%%%%%%%%%%%%%%%%%%%%%%%%%%
\begin{figure}[t]
\centerline
{\includegraphics[height=0.3\textheight,width=0.6\textwidth]{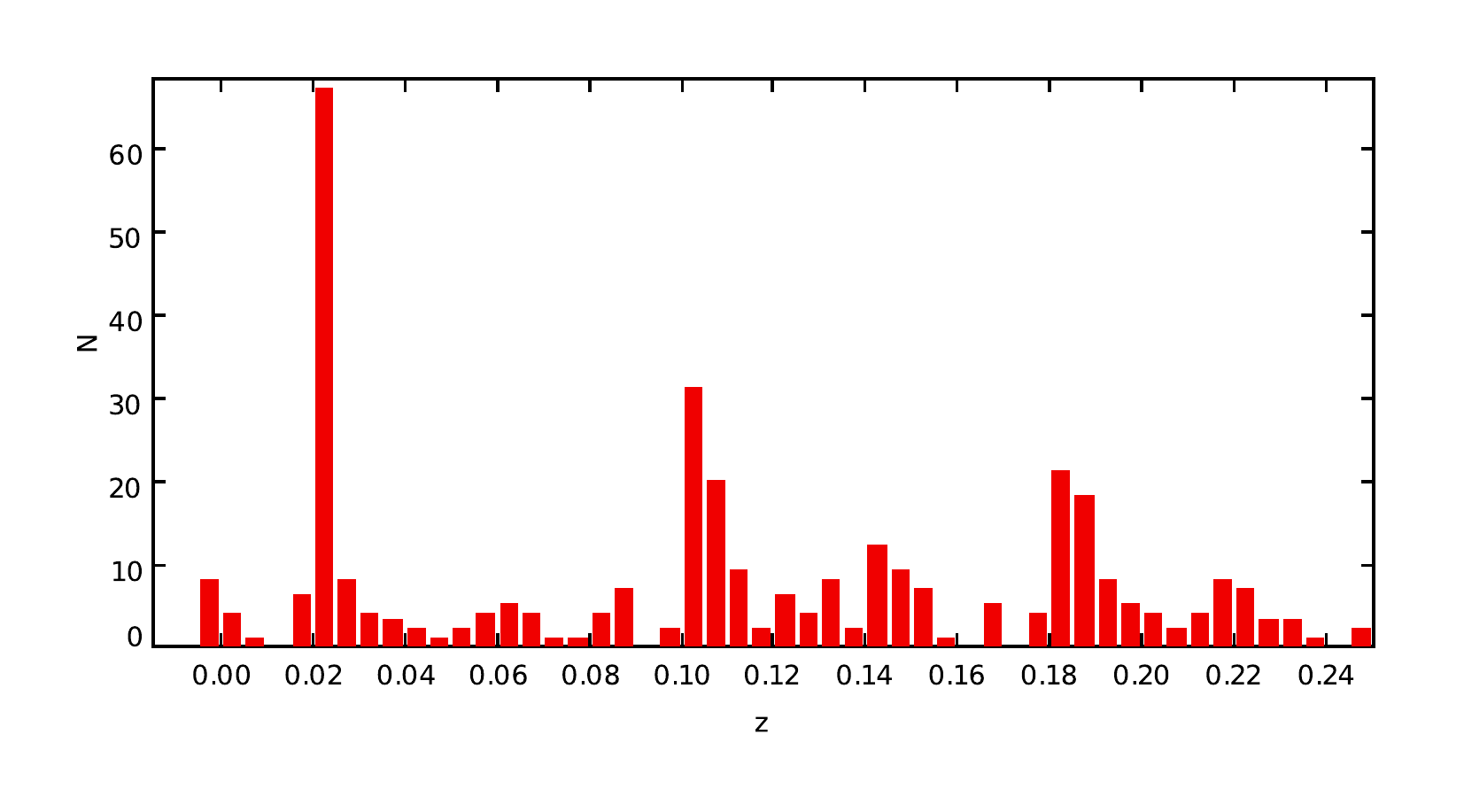}}
\caption{\label{fig.redshifts} \footnotesize Histogram of all the available redshifts in NED within
a 60\arcmin\ field around IC 1860 (galaxies with $z > 0.25$ are not shown for display purposes). 
The IC1860 group stands out as the redshift peak at $z=0.02$. 
 }
\end{figure}

%%%%%%%%%%%%%%%%%%%%%%%%%%%%%%%%%%%%%%%%%%%%%%%%%%%%%%%%%%%%%%%%%%%%%%%%%%%%%
%
%
We used the NASA/IPAC extragalactic database (NED) to collect galaxies with known recession
velocities within 60\arcmin\ (1.6 Mpc at the redshift of the source) of the optical
position of the galaxy IC 1860 which coincides with the peak of the diffuse X-ray emission of the group.
This corresponds to 1.7 times the virial radius ($r_{100}$) estimated from hydrostatic mass analysis of
the \xmm\ data in \citet{Gastaldello.ea:07*1}: this is justified by the search for a possible 
perturber following the scenario of \citet{Ascasibar.ea:06}. The primary sources of these velocities 
are: 2dFGRS \citep{Colless.ea:03}, the redshift catalogue of \citet{Dressler.ea:88}, the Southern Sky
Redshift Survey \citep{daCosta.ea:98}, the redshift survey using the ESO OPTOPUS multifiber instrument 
performed by \citet{Stein:96}, the 6dFGS \citep{JonesH.ea:09}, and the Durham/UKST (DUKST) redshift survey
\citep{Ratcliffe.ea:98}. 
Group members have been selected through the elimination of background and foreground galaxies along 
the line of sight to IC1860. An initial rejection was performed using the "velocity gap" method 
outlined by \citet{De-Propris.ea:02} where the galaxies are sorted in redshift space and their velocity
gap, defined for the $n$th galaxy as $\Delta v_n\;=\;cz_{n+1}-cz_{n}$, calculated.
Clusters and groups appears as well populated peaks in redshift space which are well separated by
velocity gaps from the nearest foreground and background galaxies: we uses 1000 \kms\ as velocity 
gap and Fig.\ref{fig.redshifts} shows that this procedure clearly detects IC 1860 as a peak at 
$z=0.02$ populated by 81 galaxies.
This initial 81 members list for the IC1860 group is shown in Table \ref{tab.members}, based mainly on 
the homogeneous 2dF sample which consists of 74 objects. Additional objects come from the other catalogues and 
we checked in Appendix \ref{app.redshifts} the consistency of the various measurements.

\begin{figure}[th]
\centerline
{\includegraphics[height=0.3\textheight]{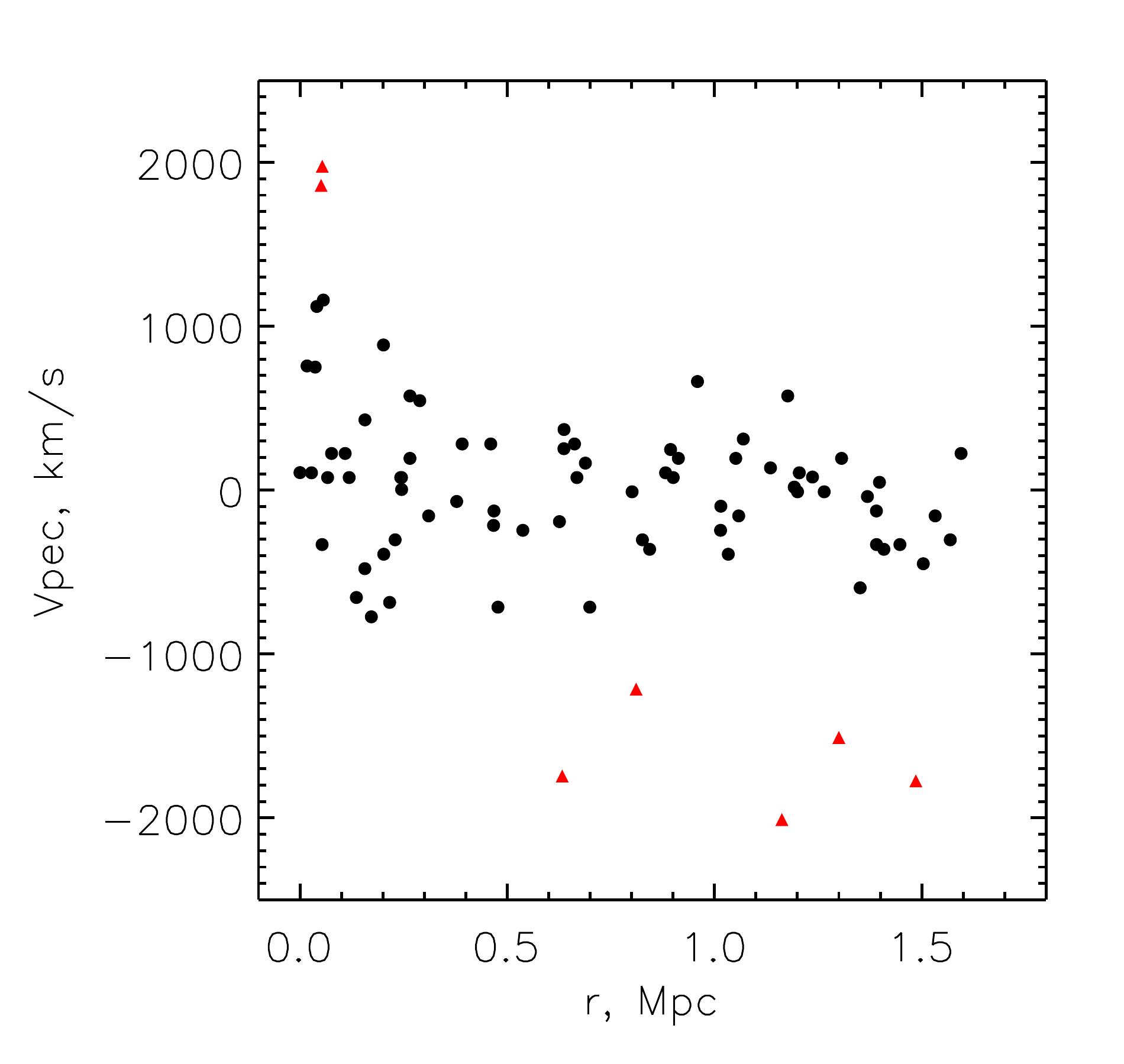}}
\caption{\label{fig.owers} \footnotesize Peculiar velocities as a function of the distance from the center of the
IC 1860 group, illustrating the procedure of refinement of the membership using the shifting gapper technique (see text). 
The black circles represent galaxies allocated as group members while the red triangles are rejected foreground 
and background galaxies lying close to the group in redshift space.
 }
\end{figure}

As a further refinement of the membership allocation process we used the slightly modified
version of the "shifting gapper", first employed by \citet{Fadda.ea:96}, as described by 
\citet{Owers.ea:09}. The method utilizes both radial and peculiar velocity information to separate
interlopers from members as a function of group-centric radius. The data are binned radially such that
each bin contains at least 20 objects. Within each bin galaxies are sorted by peculiar velocity with velocity
gaps determined in peculiar velocity. Peculiar velocities were determined by estimating the mean group velocity
using the biweight location estimator \citep{Beers.ea:90} which was assumed to represent the cosmological redshift
of the group, $z_{\rm{cos}}$. The peculiar redshift of the galaxy is 
$z_{\rm{pec}} = (z_{\rm{gal}} - z_{\rm{cos}})/(1 + z_{\rm{cos}})$ 
and the peculiar velocity is derived using the special re\-la\-ti\-vi\-stic formula 
$v_{\rm{pec}} = c((1 + z_{\rm{pec}})^2 - 1)/((1 + z_{\rm{pec}})^2 + 1)$, where $c$ is the speed of light.
The "f pseudosigma" \citep{Beers.ea:90} derived from the first and third quartiles of the peculiar velocity distribution
was used as the fixed gap to separate the group from the interlopers, because of its robustness to the presence of 
interlopers. The above procedure was iterated until the number of members was stable. The results are shown in 
Fig.\ref{fig.owers} and the rejected galaxies are highlighted in Table \ref{tab.members}.
%
%%%%%%%%%%%%%%%%%%%%%%%%%%%%%%%%%%%%%%%%%%%%%%%%%%%%%%%%%%%%%%%%%%%%%%%%%%%%%%%%
%
%%%%%%%%%%%%%%%%%%%%%%%%%%%%%%%%%%%%%%%%%%%%%%%%%%%%%%%%%%%%%%%%%%%%%%%%%%%%%
%
%%%%%%%%%%%%%%%%%%%%%%%%%%%%%%%%%%%%%%%%%%%%%%%%%%%%%%%%%%%%%%%%%%%%%%%%%%%%%%%%
\begin{figure}[th]
\centerline
{\includegraphics[height=0.3\textheight]{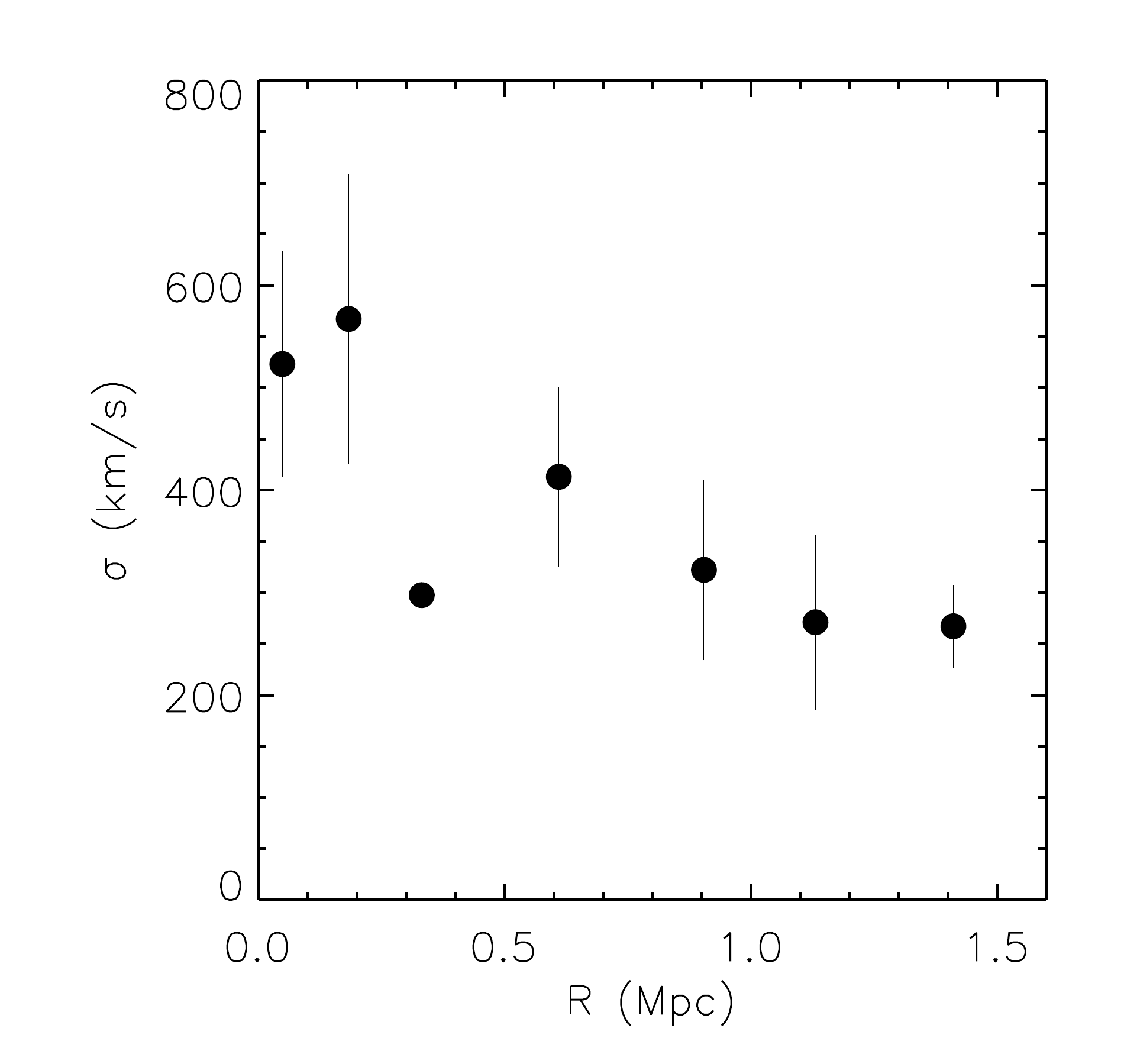}}
\caption{\label{fig.dispersion_profile} \footnotesize 
Velocity dispersion profile of the IC 1860 group, where IC 1860 is assumed to be the center of the group. 
Bins contain 10 galaxies each and 1$\sigma$ jackknife errors are shown.}
\end{figure}
%
%%%%%%%%%%%%%%%%%%%%%%%%%%%%%%%%%%%%%%%%%%%%%%%%%%%%%%%%%%%%%%%%%%%%%%%%%%%%%

The final group sample contains 74 members. The value for the biweight location estimator of the
mean group velocity is $6753\pm44$ \kms\ which corresponds to $z_{\rm{cos}}=0.02252\pm0.00015$.
We used the biweight scale estimator to estimate a velocity dispersion of $401\pm46$ \kms.
The errors for the redshift and velocity dispersion are 1$\sigma$ and are estimated using the jackknife resampling technique.

In Fig.\ref{fig.dispersion_profile} we show the velocity dispersion profile derived for the data. 
Bins contain 10 galaxies each and velocity dispersions have been calculated using the gapper algorithm 
\citep{Beers.ea:90}. The profile shows the typical falling behavior with radius expected for a relaxed object.

\subsection{Peculiar Velocity of the Central Galaxy}
\label{peculiarvelocity}

We calculated the peculiar velocity of the central galaxy IC 1860 which is $113\pm50$ \kms, just adding in quadrature the 
errors on the galaxy and group velocities. An offset velocity
must lie outside a range defined by the cluster velocity dispersion to be significant \citep{Gebhardt.ea:91}.
We therefore calculated the Z-score for the central galaxy \citep[eq. (7) of][]{Gebhardt.ea:91} which is 0.279 and its
90\% confidence limits are (0.036, 0.520). The confidence interval has been calculated using 10000 bootstrap resamplings and allowing 
for the measurement error of the central galaxy velocity by sampling from a Gaussian distribution with standard deviation set to the reported 
error as part of the bootstrap. The velocity offset is marginally significant at the 2.3$\sigma$ level.

\subsection{Substructure Detection}
\label{substructuretests}

Given the presence of sloshing cold fronts in the X-ray emitting medium
of the IC 1860 group it is relevant to look for the presence of substructure
and to possibly identify the merging sub-halo responsible for the onset of
sloshing. 

In general the higher the dimensionality of the substructure test, the more sensitive it is 
to substructure; however the sensitivity of individual diagnostics depends on the line of sight relative to 
the merger axis and therefore no single substructure test is the most sensitive in all the situations; it is essential 
to apply a full range of statistical tests \citep[e.g.,][]{Pinkney.ea:96}. In this section we apply a battery of
statistical methods to the detection of substructure in IC 1860.

\subsubsection{One-dimensional Tests for Substructure}

We analyze the velocity distribution to look for departures from Gaussianity which can be attributed to
dynamical activity. As a first test of Gaussianity, following \citet{Hou.ea:09}, we applied the Anderson-Darling (AD)
test as the more reliable and powerful test at detecting real departures from an underlying Gaussian distribution.  
We used the AD test as implemented in the task \emph{ad.test} of the package \emph{nortest} in version 2.10 of the R 
statistical software environment\footnote{http://www.r-project.org}\citep{R}. 
When applied to the final group sample of 74 objects (the galaxies not highlighted in Table \ref{tab.members}) 
the AD test returns a $A^{2}$ statistic \citep[following the notation of][just $A$ in the 
\emph{ad.test} documentation]{Hou.ea:09} of 0.6146 which corresponds to a $p$-value of $0.1059$ 
(computed using the modified statistic $A^{2*}$), therefore consistent with having a Gaussian distribution.

We estimated three shape indicators, the skewness, the kurtosis, and the scaled tail index, a robust indicator 
\citep{Bird.ea:93}. The values for the sample are 0.429 and 0.386 for skewness and kurtosis respectively and they
show no departure from a Gaussian distribution \citep[see Table 2 of][for 75 data-points]{Bird.ea:93}.
The scaled tail index is 1.14 again consistent with a Gaussian distribution.

We investigated the presence of significant gaps in the velocity distribution following the weighted gap analysis of 
\citet{Beers.ea:91} looking for normalized gaps larger than 2.25 since in random draws of a Gaussian distribution they 
arise at most in about 3\% of the cases, independent of the sample size. We did not find any significant gap in the
sample.

To detect subsets in the velocity distribution we also applied to the data the Kaye's mixture model 
(KMM) test \citep{Ashman.ea:94}.
The KMM algorithm fits a user-specified number of Gaussian distributions to a dataset and assesses the improvement of that
fit over a single Gaussian. In addition, it provides the an assignment of objects into groups. 
The KMM test is more appropriate in situations where theoretical and empirical arguments
indicate that a Gaussian model is reasonable, as for the line of sight velocities in a dynamically relaxed cluster.
We did not find again any significant partition of the sample.

To summarize the results of the 1D tests, we found no evidence of substructure in the velocity distribution in the
final group sample. Note that the same battery of tests applied to the unrefined sample 
would detect a significant departure from the Gaussianity. Thus,
the shifting gapper refinement of the membership allocation has been effective in removing
small foreground and background groups; in particular the central group of high velocity interlopers was already noticed by
\citet{Burgett.ea:04}.

%%%%%%%%%%%%%%%%%%%%%%%%%%%%%%%%%%%%%%%%%%%%%%%%%%%%%%%
\begin{figure}[t]
%%\vspace{-0.5cm}
\centerline{
\includegraphics[height=0.3\textheight]{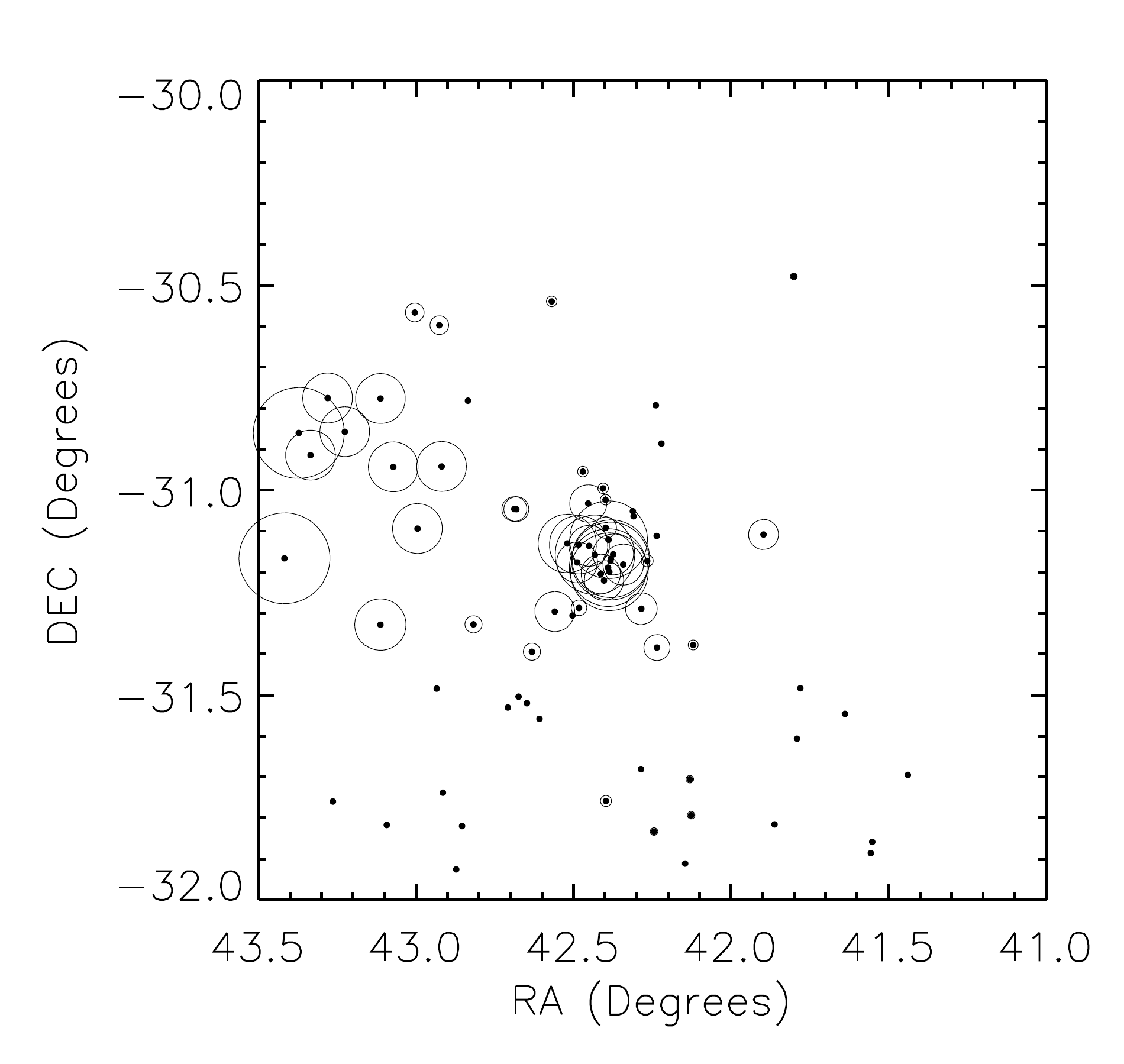}
}
\caption{\label{fig.ds} \footnotesize
Results for the $\kappa$ substructure test. Each dot represents a galaxy in the 
IC 1860 group and the size of the circle around the dot gives an indication of the difference of the local velocity distribution compared 
to the global group velocity distribution (the size is scaled by $\kappa_{i}$ where $\kappa_{i}$ is given by Equation \ref{eq.kappa}).
Clustering of large circles would indicate significant departures and the likely presence of a substructure.
}
\end{figure}
%%%%%%%%%%%%%%%%%%%%%%%%%%%%%%%%%%%%%%%%%%%%%%%%%%%%
%
%
\subsubsection{Three-dimensional Tests for Substructure}
\label{3Dtests}

The existence of correlations between positions and velocities of cluster galaxies is a signature of true
substructures.
As a first approach we adopt the Dressler-Shectman $\Delta$ statistic \citep{Dressler.ea:88} which tests
for differences in the local mean and dispersion compared to the global mean and dispersion and it is recommended by
\citet{Pinkney.ea:96} as the most sensitive 3D test.
Calculation of the $\Delta$ statistic involves the summation of local velocity anisotropy, $\delta$,
for each galaxy in the group, defined as

\begin{equation}
\label{eq.delta}
\delta^2 =
\left(\frac{N_{\mathrm{nn}}+1}{\sigma^2}\right)[(\bar{v}_{\mathrm{local}}-\bar{v}_{\mathrm{gro
up}})^2-(\sigma_{\mathrm{local}}-\sigma)^2],
\end{equation}

where $N_{\mathrm{nn}}$ is the number of nearest neighbours over which the local recession velocity 
($\bar{v}_{\mathrm{local}}$) and velocity dispersion ($\sigma_{\mathrm{local}}$) are calculated.
Here we adopt $N_{\mathrm{nn}}=\sqrt{N}$, following \citet{Pinkney.ea:96}. The significance of $\Delta$
is estimated using 10000 Montecarlo simulations randomly shuffling the galaxy velocities.
We find suggestions of substructure with the DS test at the marginal level 96.7\% level. 
A similar result (95.8\%) is obtained when using the $\kappa$ test of \citet{Colless.ea:96}. 
The $\kappa$-test searches for local departures from the global velocity distribution around each cluster member galaxy
by using the Kolmogorov-Smirnov (KS) test which determines the likelihood that the velocity distribution of the
$\sqrt{N}$ nearest neighbors around the galaxy of interest and the global cluster velocity distribution are drawn from the same parent distribution.
Calculation of the the $\kappa$ statistic involves the summation of the negative log likelihood for each galaxy defined as 

\begin{equation}
\label{eq.kappa}
\kappa_{i} = -log P_{KS} ( D > D_{obs,i}),
\end{equation}

over the entire sample.
By inspecting the bubble plot of Fig.\ref{fig.ds} the galaxies contributing to the signal of the $\kappa$ statistic 
are the galaxies with high velocities
in the center of the group (including the galaxy IC 1860 itself) and a subclump in the north-east outskirts of the group 
(the galaxies with ID \# 75, ID \# 76, ID \# 73, ID \# 81, ID \# 54, and ID \# 56).
Removing these galaxies and re-running the test results in a $\Delta$ value being significant at only the 71\% level. 
If on the contrary we remove the galaxies identified in the center with high values of  $\delta$ and we re-run the test we obtain 
a $\Delta$ value being significant at the 81\% level. It is important to bear in mind that the $\Delta$ statistic 
is particularly insensitive to superimposed substructures \citep{Dressler.ea:88,Pinkney.ea:96} and that it has 
the highest false positive detection of substructures due to velocity gradients across the group \citep{Pinkney.ea:96}.

We tested further the presence of these two candidates subgroups by applying the full 3D-KMM method.
The 3D-KMM test, starting from the above mentioned seeds, can allocate galaxies in subclusters. 
Using the NE subgroup and the rest of the galaxies as initial two-groups partition, we did not find 
any significant allocation.

On the contrary a two-group partition with as initial seeds the inner subgroup and the rest of 
the members we find that a partition of 60 and 14 galaxies is a more accurate description of the 3D distribution 
than a single Gaussian at more than the 99.999\% level. 

To summarize the results of the 3D tests we found marginal evidence of substructure related to a possible subgroup 
in the NE outskirts of the group and a central structure with relatively high peculiar velocities.

\subsubsection{Candidate Perturbers}
\label{perturbers}

The possible substructures identified by the statistical tests applied above do not seem 
likely to be the perturbers. The central substructure comprises the central galaxy, IC 1860, itself
and the north-east substructure is a loose aggregate of very faint and small galaxies.

A visual inspection (see Fig.\ref{fig.optical}) revealed two possible candidates: IC 1859 and ESO 416-G033.
IC 1859 is a Seyfert 2 galaxy with a peculiar optical morphology showing a pair of spurs in his eastern spiral arm;
it was already classified as peculiar in \citet{Kelm.ea:98}. It is at a projected distance of 172 kpc from IC 1860
\citep[corresponding to 0.18 $r_{\rm{vir}}$,][]{Gastaldello.ea:07*1} and it 
has a velocity difference of $-900\pm67$ \kms\ with respect to IC 1860 and of $-787\pm78$ \kms\ with respect to the group
mean velocity. ESO 416-G033 has a mildly disturbed optical morphology, in particular if compared to IC 1859, and it is at a 
projected distance of 310 kpc (0.33 $r_{\rm{vir}}$) from IC 1860 and it has a velocity difference of $-271\pm124$ \kms\ with respect to IC 1860 
and of $-158\pm130$ \kms\ with respect to the group mean velocity. A third galaxy SW of IC 1858, the S0 IC 1858, is a striking 
narrow-angle tail radio galaxy (NAT) as shown by the \gmrt\ data and discussed in Appendix \ref{sec.appendixic1858}. Thanks also
to the radio tail constraining the motion in the plane of the sky we are able to rule out this galaxy as the perturber 
(see \S\ref{simulations}).
%
%=====================================================================
% RADIO ANALYSIS
%=====================================================================
%
%
\section{Radio analysis}
\label{radio}
%
%%%%%%%%%%%%%%%%%%%%%%%%%%%%%%%%%%%%%%%%%%%%%%%%%%%%%%%%%%%%%%%%%%%%%%%%%%%%%%%%
\begin{figure}[th]
%%\vspace{-0.5cm}
\centerline{
{\includegraphics[height=0.25\textheight]{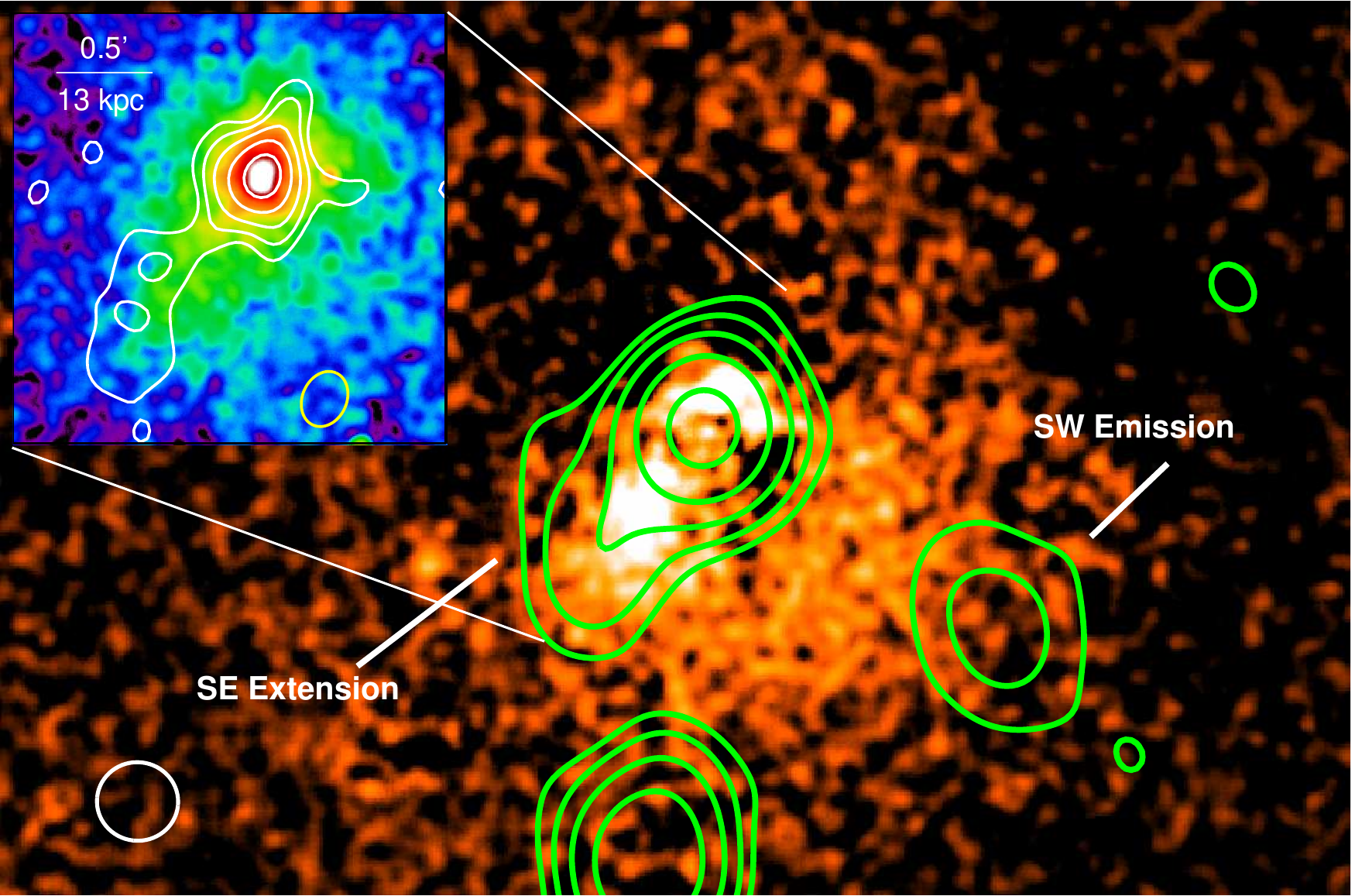}}
}
\caption{\label{fig.radioic1860} \footnotesize
{\em GMRT} radio contours at 325 MHz overlaid on the map of \xmm\ residuals (Fig.\ref{fig.residuals}). 
The restoring beam is $30^{\prime \prime}\times29^{\prime \prime}$, in p.a. $6^{\circ}$ (white ellipse). 
The rms noise level is 0.6 mJy beam$^{-1}$. Contours are spaced by a factor of 2 starting from 
2 mJy beam$^{-1}$. The emission which can be seen in the South is coincident with an X-ray point source.
The inset shows the 325 MHZ image at higher resolution 
($18^{\prime\prime} \times 14^{\prime\prime}$, in p.a. $-24$ $^{\circ}$; black ellipse), 
overlaid on the \chandra\ image.  The rms noise level is 0.5 mJy beam$^{-1}$. Contours start at 1.5 mJy beam$^{-1}$ and then scale by a 
factor of 2. No negative levels, corresponding to $-3\sigma$, are present in the portion of the image shown.
%Dashed contours correspond to the $-2$ mJy beam$^{-1}$ level.
}
\end{figure}
%%%%%%%%%%%%%%%%%%%%%%%%%%%%%%%%%%%%%%%%%%%%%%%%%%%%%%%%%%%%%%%%%%%%%%%%%%%%%%%%%%%%%%%%%%%%%%%%%%%%%
%
A radio image of IC1860 has been presented by \citet{Dunn.ea:10}, who 
analyzed {\em Very Large Array} ({\em VLA}) A-configuration data at
1.4 GHz with angular resolution of $\sim 3^{\prime \prime}$. A point
source with two faint, small-scale extensions was found associated
with the galaxy, but no correlation between these radio features and 
the X-ray emission was observed.

In search for possible extended radio emission, undetected in the 
current radio images, we analyzed low radio-frequency observations of 
IC1860 extracted from the {\em Giant Metrewave Radio Telescope} ({\em GMRT}) 
archive (project 06EFA01). The source was observed at 325 MHz in September 
2004 for approximately 8 hours on source, using only the upper side band (USB) 
with 16 MHz bandwidth. We calibrated and reduced the data set using the 
NRAO Astronomical Image Processing System package (AIPS), following 
the procedure described in \citet{Giacintucci.ea:08*1}. We applied phase-only 
self-calibration to reduce residual phase variations and improve the quality 
of the final images. The scale of \citet{Baars.ea:77} was adopted for the 
flux density calibration. Residual amplitude errors are within $8\%$ 
\citep[e.g.,][]{Chandra.ea:04}.

Fig.\ref{fig.radioic1860} shows the {\em GMRT} contours at 325 MHz overlaid on the map
of \xmm\ residuals shown in Fig.\ref{fig.residuals}. To highlight the extended emission, the radio image
has been produced with a resolution of $\sim 30^{\prime \prime}$, 
obtained by applying a taper to the $uv$-data through the parameters 
ROBUST and UVTAPER in the task IMAGR. The radio image of the central region is also shown at higher resolution (18"), 
overlaid on the \chandra\ image. The image shows an
unresolved central component, spatially coincident with the emission
detected at higher frequency and higher resolution by \citet{Dunn.ea:10}. 
Its flux density at 325 MHz is $60.3\pm4.8$ mJy, as
measured on the full resolution image ($14^{\prime\prime}\times9^{\prime
\prime}$; not shown here), and a total flux density of 13.9 mJy is 
reported at 1.4 GHz by \citet{Dunn.ea:10}, implying a spectral index\footnote{We adopt the 
convention $S_{\nu} \propto \nu^{-\alpha}$ for the synchrotron spectrum, 
where $S_{\nu}$ is the flux density at the frequency $\nu$.} 
$\alpha=1.0$.

In addition to the central point source, the {\em GMRT} image reveals
an extended feature with very low surface brightness which originates at the IC1860 center and 
extends $\sim 100^{\prime \prime}$ (45 kpc) from the center toward south-east. This radio emission
is contained within the inner cold front and traces the ``mushroom head'' feature
interpreted as the tip of the sloshing spiral in \S \ref{subsection.spiral}.
A flux density of $\sim$22 mJy is associated with this component. 
A second patch of faint radio emission is visible $\sim2^{\prime}$ 
south-west of the central source, with a flux density of $\sim$20 mJy. 
Neither of these extended features are visible in the image
by \citet{Dunn.ea:10}, due to the lack of short spacings in the $uv$ coverage 
of their data set (the largest angular scale that can be imaged 
with the {\em VLA} in A configuration at 1.4 GHz 
is $\sim 38^{\prime\prime}$). We thus inspected the NRAO 
{\em VLA} Sky Survey \citep[NVSS,][]{Condon.ea:98} image at 1.4 
GHz, but found no evidence of significant extended emission over 
the two extended structures detected by the {\em GMRT}. Given the
sensitivity level of the NVSS image ($1\sigma=0.45$ mJy beam$^{-1}$, 
one beam corresponding to $45^{\prime\prime}$), we placed a lower limit of 
$\alpha > 1.9 $ for the spectral index of the SE extension and $\alpha
> 1.8$ for the SW emission. Both spectral index limits are very steep
and suggest that the extended emission originates from aged 
radio plasma, perhaps associated with a past episode of activity of
the central radio galaxy.
%
%=====================================================================
% SIMULATIONS SIMULATIONS SIMULATIONS
%=====================================================================
%%
\section{Comparison with simulations}
\label{simulations}

The analysis of the available \chandra\ and \xmm\ data for the IC 1860
group has revealed a pair of surface brightness discontinuities 
at 45 kpc and 67 kpc that the spectral analysis indicated to be cold fronts.
We also detected a spiral-shaped excess in the core with the outer edges 
of this spiral being the cold fronts. These features are all indicators of
ongoing sloshing in the scenario put forward to explain these phenomena 
\citep{Ascasibar.ea:06,Roediger.ea:11}.

\begin{figure*}
%\centerline{
\hfill rotate box around x-axis  \hfill rotate box around y-axis \phantom{x}\hfill\hfill
\newline
\rotatebox{90}{orbital plane face-on}
\hspace{2.0cm}
\includegraphics[width=0.32\textwidth]{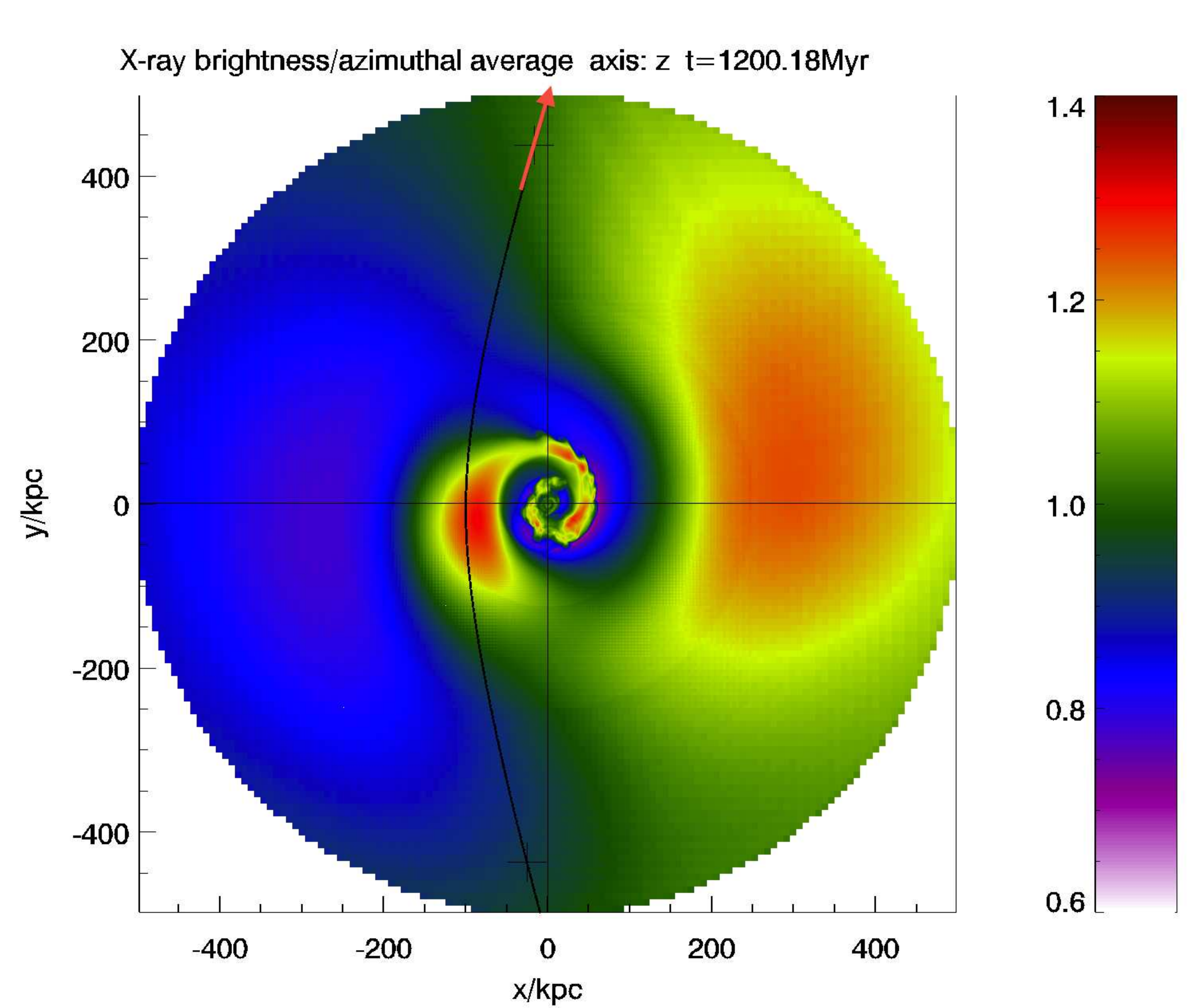}  
 \hspace{2.0cm}
\includegraphics[width=0.32\textwidth]{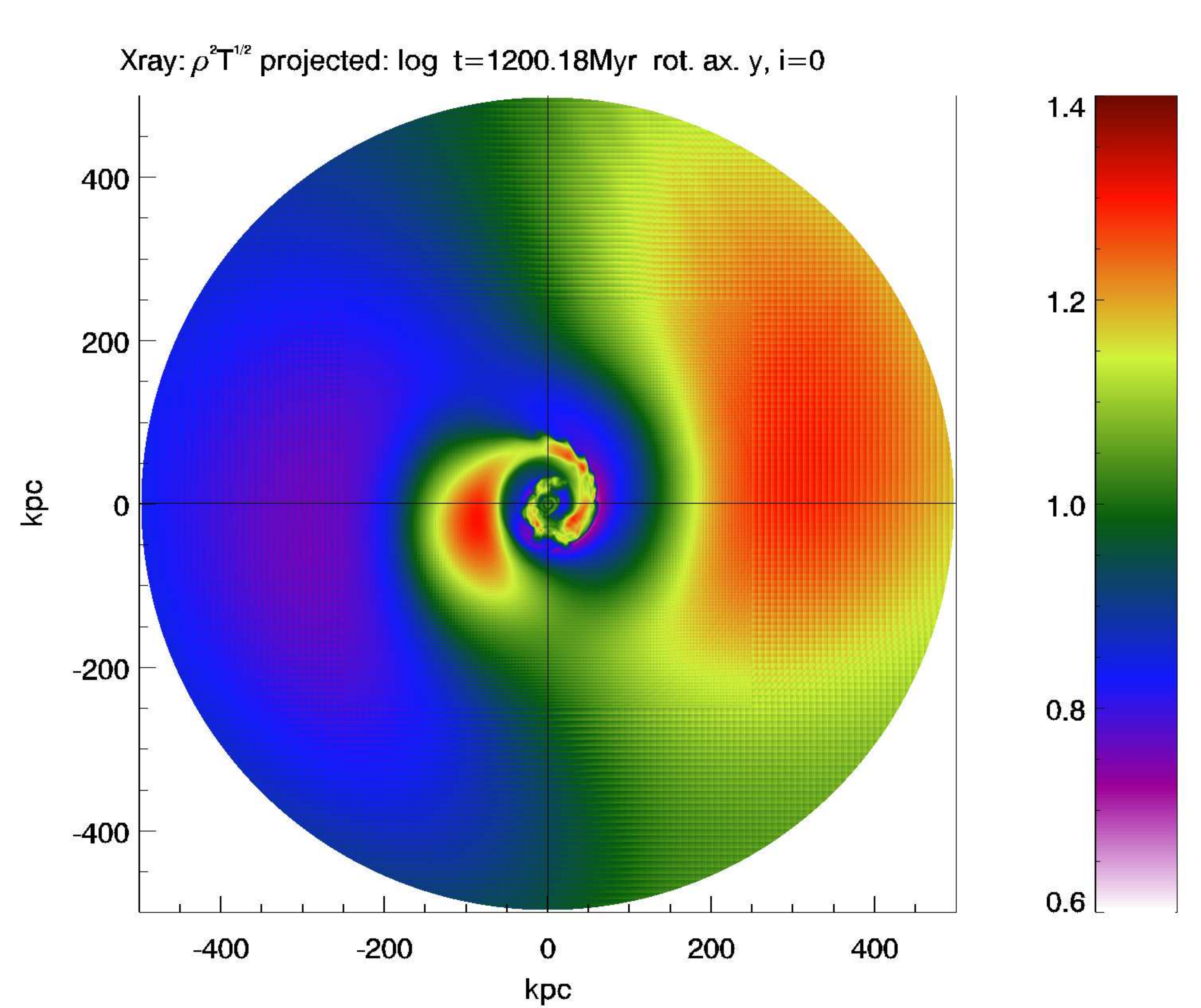}
\newline
\rotatebox{90}{22 degrees}
\hspace{2.0cm}
\includegraphics[width=0.32\textwidth]{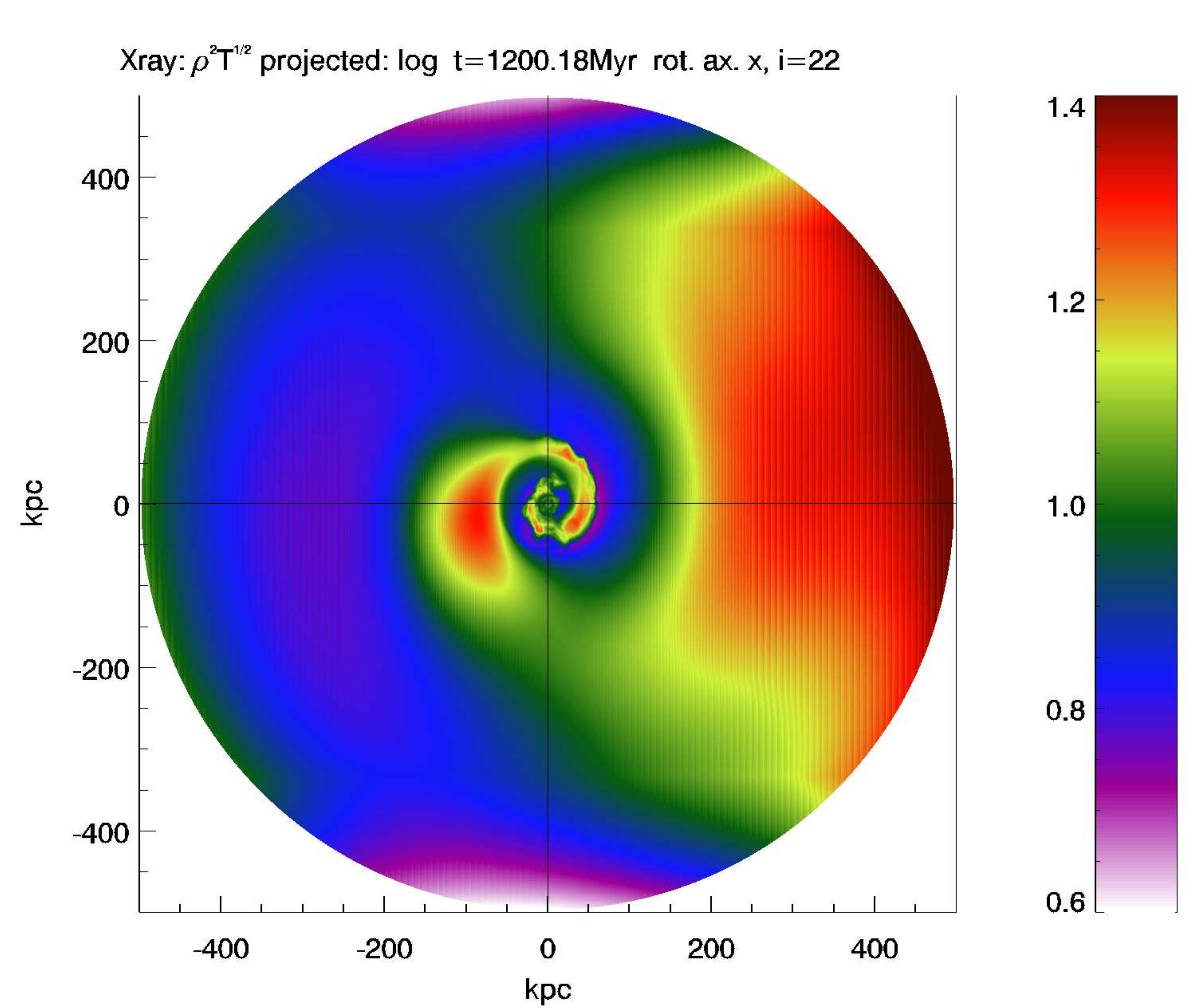}  
 \hspace{2.0cm}
\includegraphics[width=0.32\textwidth]{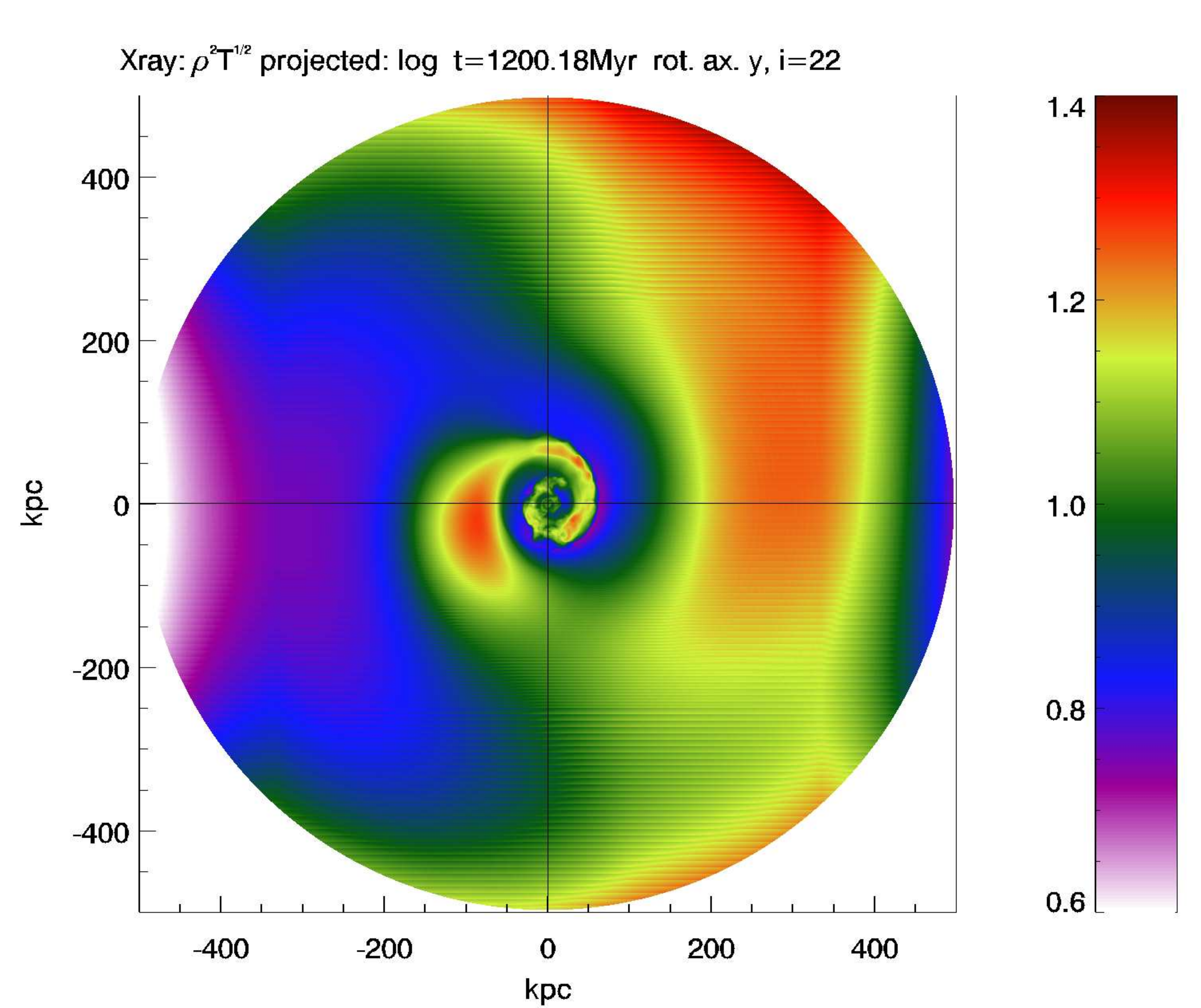}
\newline
\rotatebox{90}{45 degrees}
\hspace{2.0cm}
\includegraphics[width=0.32\textwidth]{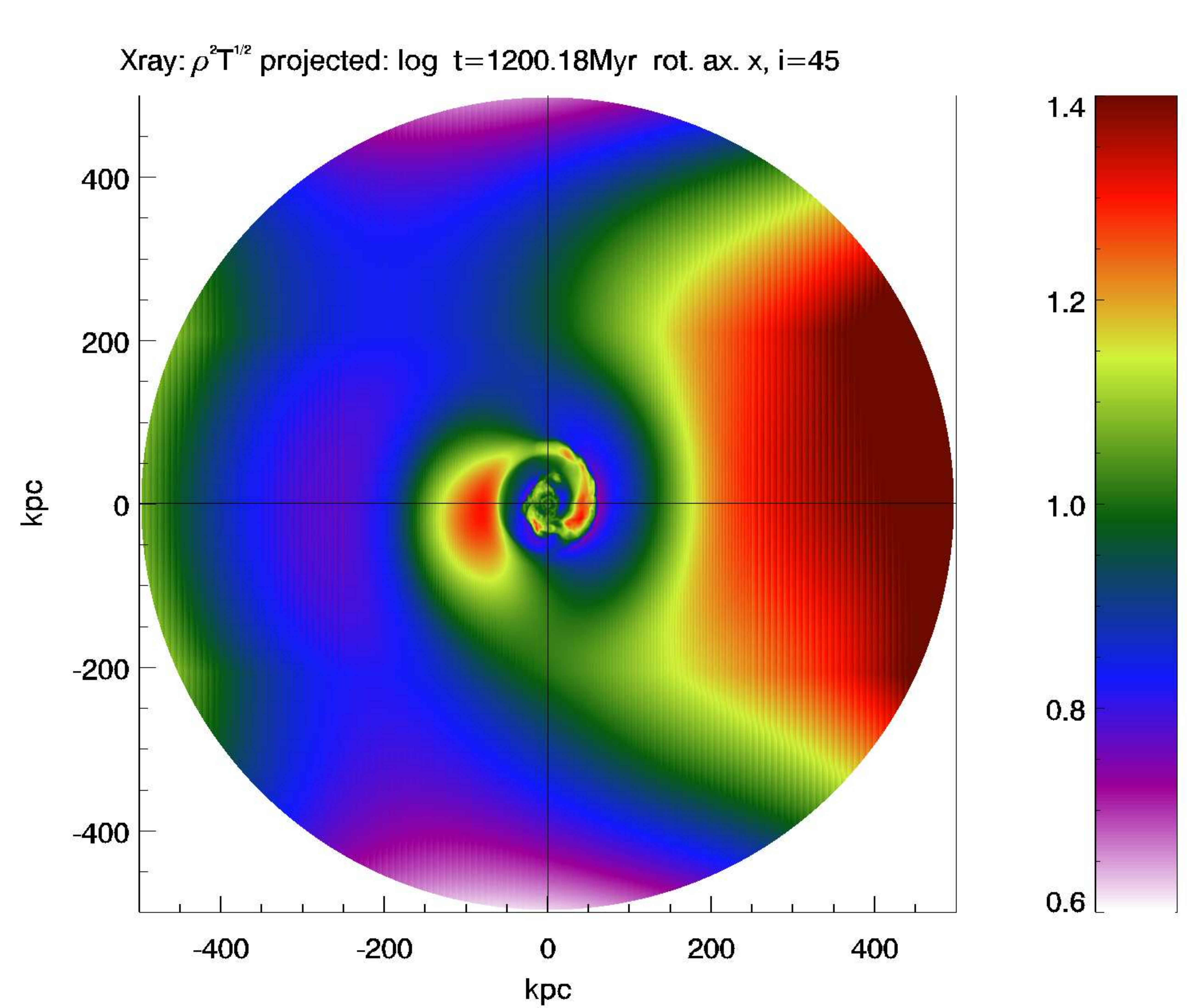} 
 \hspace{2.0cm}
\includegraphics[width=0.32\textwidth]{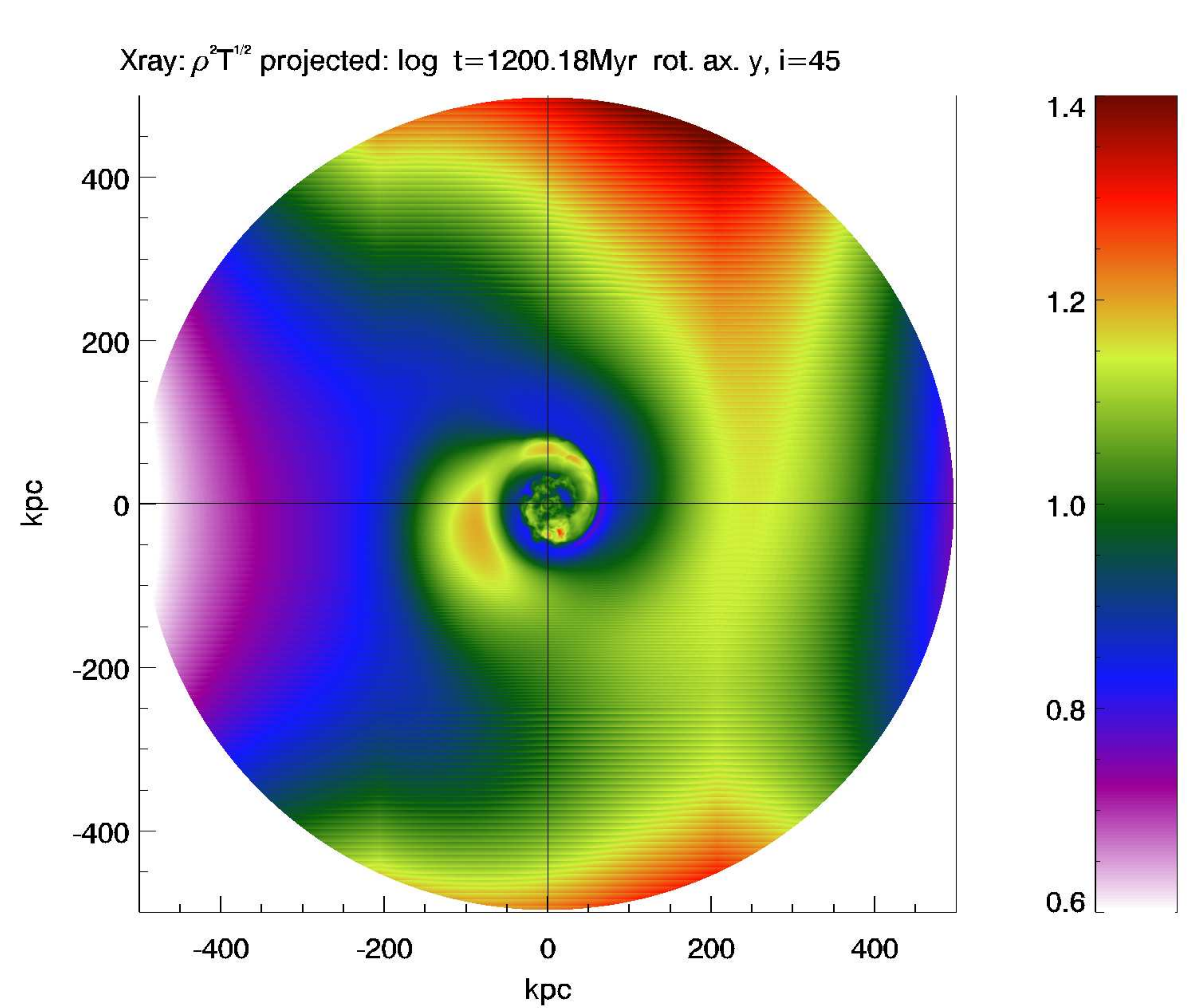}
\newline
\rotatebox{90}{67 degrees}
\hspace{2.0cm}
\includegraphics[width=0.32\textwidth]{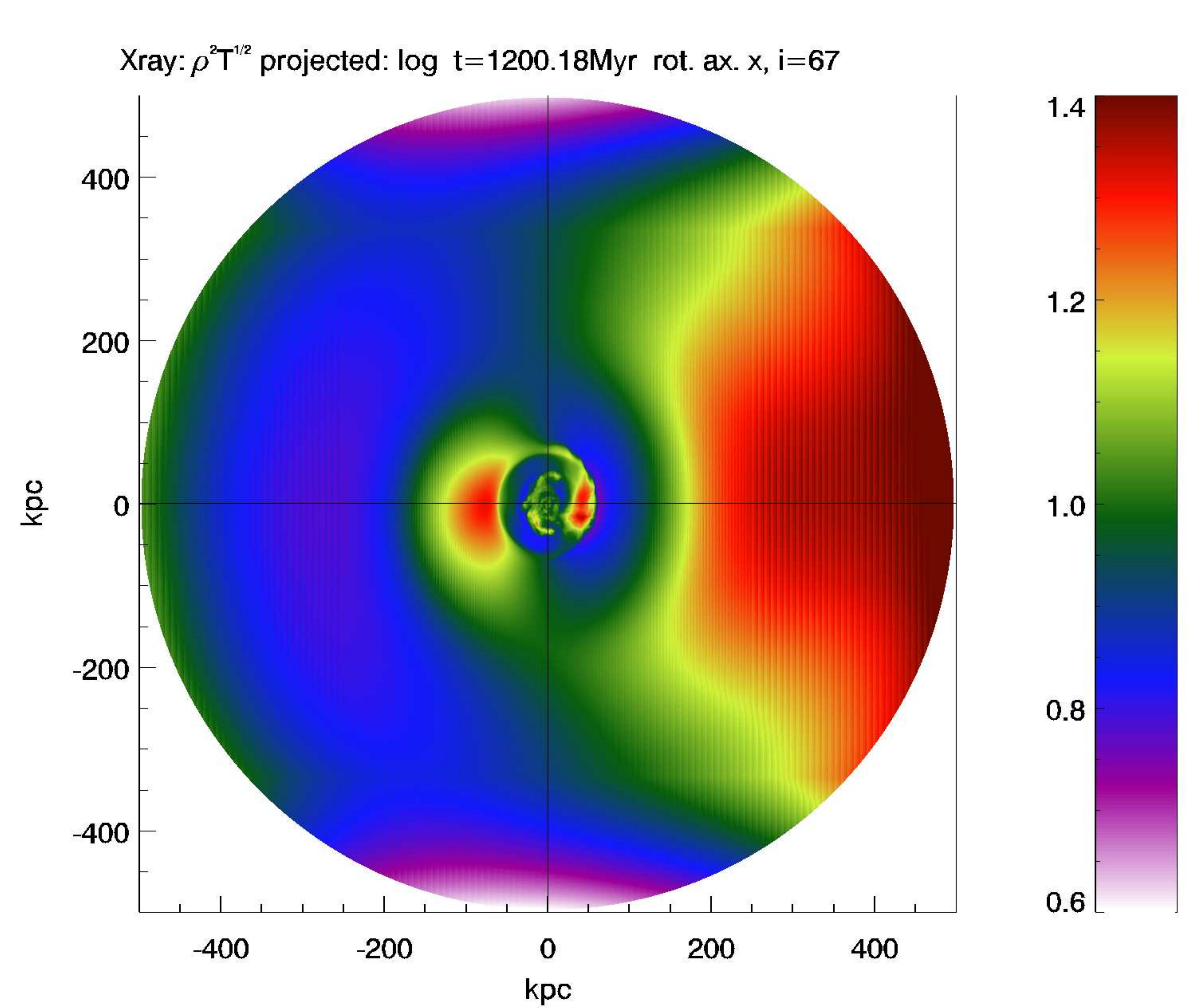}
\hspace{2.0cm}
\includegraphics[width=0.32\textwidth]{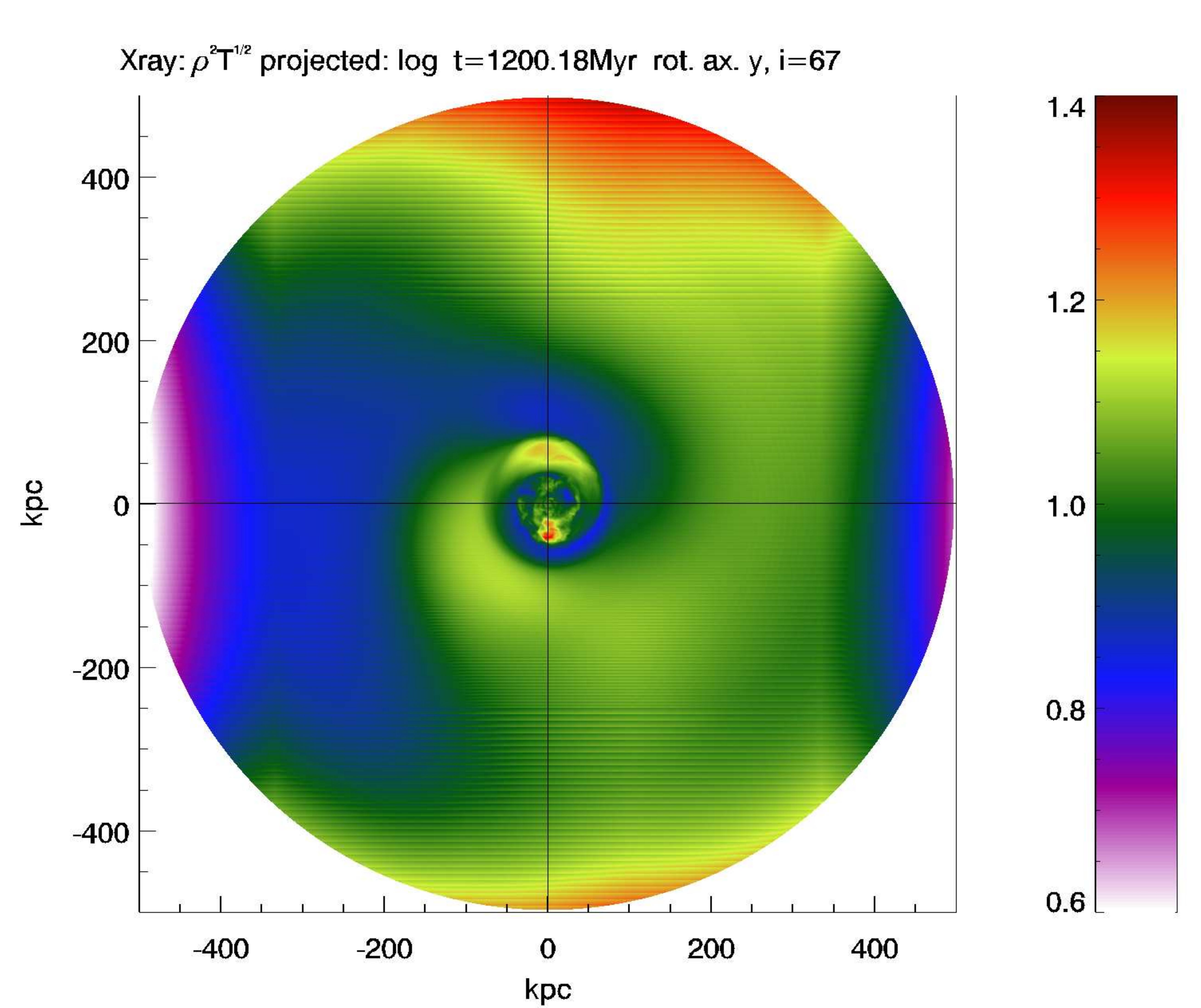}
\newline
\rotatebox{90}{90 degrees}
\hspace{2.0cm}
\includegraphics[width=0.32\textwidth]{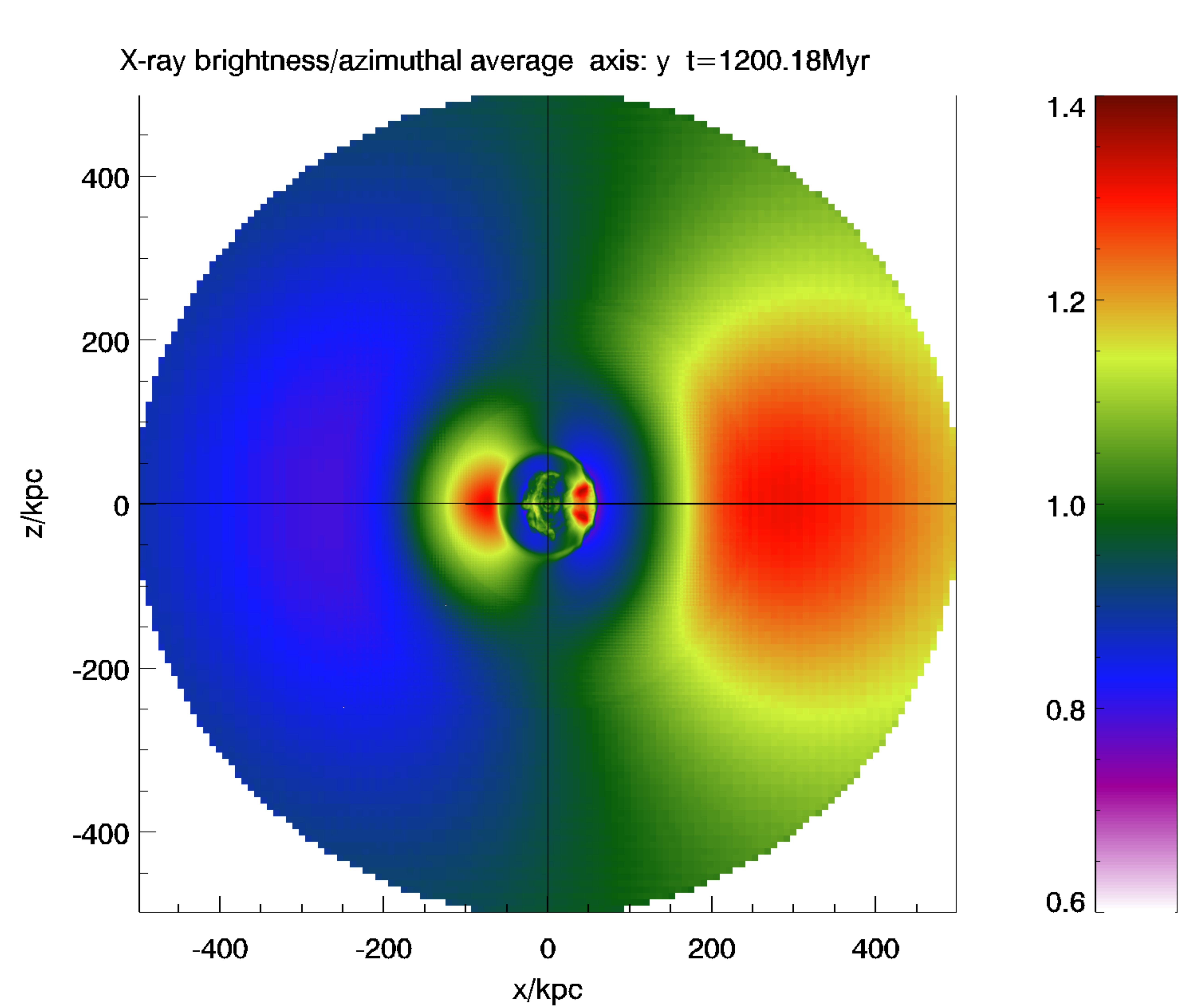}
\hspace{2.0cm}
\includegraphics[width=0.32\textwidth]{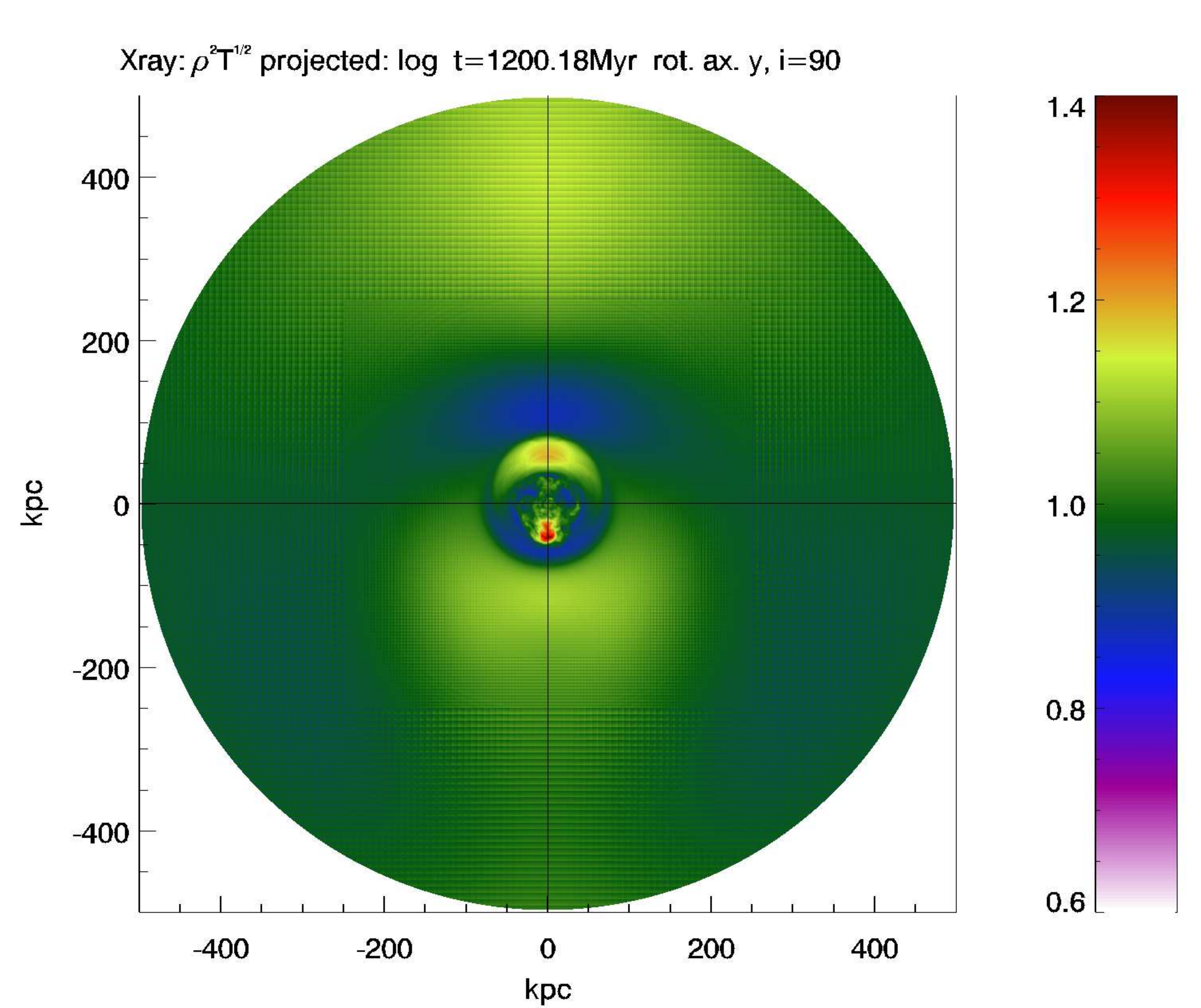}
\newline
\hfill\hfill\hfill \phantom{xxxxxxxxx}LOS along orbit \hfill  LOS perp. to orbit, orbital plane edge-on \hfill
%}
\caption{Brightness residuals with respect to azimuthally averaged brightness. Original image divided by averaged. Only the inner 500 kpc of the cluster have been used to create these images. This slightly affects the quantitative values of the brightness residuals at periphery of the images, but they are qualitatively correct. In the top left panel we mark the orbit of the perturber.}
\label{fig:residuals}
\end{figure*}

Following \citet{Roediger.ea:11, Roediger.ea:12}, we use the sloshing signatures to derive the  merger history. 
We perform a qualitative comparison to the sloshing simulations \citet{Roediger.ea:11} performed for the Virgo cluster. The sloshing structure appears 
clearest in brightness residual maps and we show such maps derived from the simulations for different lines of sight (LOSs) in Figure~\ref{fig:residuals}. 
The top row shows the residual map for the LOS perpendicular to the orbital plane. In the left-hand-side column the cluster is rotated around the horizontal 
axis, in the right-hand side column around the vertical axis. 
%
%%FFFFFFFFFFFFFFFFFFFFFF
%%FFFFFFFFFFFFFFFFFFFFFF
%
When the LOS is perpendicular to the orbital plane, the residual map shows the well-known spiral morphology. This morphology can be clearly 
recognized for inclinations of the orbital plane up to 45\deg away from this configuration. Beyond 70\deg the spiral morphology becomes weak. 
The residual map of the IC 1860 displays a  spiral-like morphology, therefore we conclude that our LOS is within 70\deg of the face-on view. 
We note that only the rotation around the horizontal axis in Fig.~\ref{fig:residuals} (i.e. the cases in the left-hand-side column) leads to a substantial 
radial component for the orbit of the perturber, i.e.~a LOS velocity component. The configurations in the right-hand-side column lead 
to a progressively smaller angle between the LOS and the orbital plane, but the LOS is always perpendicular to the orbit itself.  A characteristic 
feature of all cases in the left-hand-side column and the top panels in the right-hand-side column is the pronounced large-scale asymmetry, with a 
clear brightness excess in the right, i.e. at the side of the cluster center opposite the perturber's pericentre.  In all these cases, also in 
projection the pericentre of the perturber orbit is left of the cluster center. Only in the 2 or 3 bottom panels of the right-hand side column, where 
the pericentre is seen more and more behind/in front of the cluster center, this large-scale asymmetry is weak or absent. 
In IC 1860, the brightness outside the radius of the western cold front does not seem strongly asymmetrical.

The simulations \citep[e.g.,][]{Ascasibar.ea:06} show that following the spiral structure from the outside inward indicates the direction of the 
infall of the perturbing object. Given that the direction of the spiral inflow is clockwise the perturber also has a clockwise orbit.
We find two configurations of the orbit projected onto the sky fit the observation best.

\begin{description}
\item[E-N-W]  The perturber approaches approximately from the east, passes the group center in the North and departs approximately towards the West 
(case E-N-W).  The IC 1860 observation matches this configuration if IC 1860's western cold front corresponds to the excess region to the left in 
the top panels in Fig.~\ref{fig:residuals} (what was called cold fan by \citealt{Roediger.ea:11}), and IC 1860's south-eastern cold front to the 
outermost cold front to the right of the core in the top panels in Fig.~\ref{fig:residuals}, which was called major cold front in  \citet{Roediger.ea:11}. 
In the Virgo simulations, this cold front moved outwards with a velocity of $\sim$55 kpc per Gyr. Assuming that in the IC 1860 group the cold font moves 
at comparable velocity, the group-centric distance of 46 kpc of this cold front indicates that the pericenter passage of the perturber took place about 1 Gyr ago. 
This configuration predicts a large-scale brightness excess to the south. 
\item[S-E-N]  The perturber approaches approximately from the south, passes the group center in the east and departs approximately towards the 
north (case S-E-N). The IC 1860 observation matches this configuration if IC 1860's western cold front corresponds to the major cold front 
(outermost to the right in top panels of Fig.~\ref{fig:residuals}), and the SE front to the secondary front just left of the simulated cluster core. 
A close inspection of IC 1860's brightness residual map suggests that the spiral-shaped brightness excess may continue  beyond the excess just inside 
the cold front in the west, it may continue outwards anti-clockwise over north to the north-east, maybe even east. The faint outer eastern end of the 
brightness excess spiral would correspond to the brightness excess fan in the left of the simulated images. If the western, now identified as major, 
cold front expands at the same velocity as the major cold front in the Virgo simulations, its group-centric distance of 76 kpc indicates that the 
pericenter passage of the perturber took place about 1.5 Gyr ago. This configuration predicts a large-scale brightness excess to the west.
\end{description}
Correspondingly, orbits W-S-E and N-W-S do not match the  structure of the brightness residual map of the ICM 1860 group well, and we regard them as 
unlikely. Consequently, we expect the perturber in the approximately in the north or the west or in between, and the pericentre passage has taken place 
1 to 1.5 Gyr ago. 
The discussion above is based on the LOS being perpendicular to the orbital plane, but is valid to configurations at least 45\deg away from this. 
Only beyond 70\deg the spiral morphology becomes weak.

The group gas temperature is approximately 1.5 keV. This corresponds to sound speed of approx. $600\Kms$. If we assume an average relative 
velocity of the perturber between Mach 1.5 and Mach 2, within 1 Gyr it can move around 1 Mpc away from the group center in real space. 

The optical image in Fig.~\ref{fig.optical} shows three giant galaxies as potential perturber candidates in the IC 1860 group. 
There are two big distorted spirals in N and W, which fit either orbit configurations E-N-W or S-E-N. IC 1859 in the W has 
strong radial velocity ($\sim 800\Kms$) with respect to the group, and is quite close to IC1860 in projection(176 kpc). In the north, 
ESO 416 has only mild radial velocity (160 and 270 with respect to IC1860 or group mean, respectively), and is also slightly further away (310 kpc). 
If we assume both galaxies are about 1 Mpc from the group center in real space, it means the orbit of ESO 416 makes an angle of 70\deg with sky, 
the orbit of IC 1859 an angle of 80\deg to achieve correct projected distance. According to the simulated brightness residual maps, this is just 
within the range of inclinations where the central brightness excess takes on a clear spiral structure. If the galaxies were passing the group at 
high velocity, their real-space distance and thus the inclination angle between their orbit and the plane of the sky may be somewhat smaller. On the basis of 
the current X-ray data, either of these galaxies may have been the perturber. A clear detection of the large-scale brightness asymmetry would be helpful 
to distinguish both cases. The S0 galaxy in the south is a NAT with the tail pointing southwards, indicating a projected motion towards the group 
center (see Appendix \ref{sec.appendixic1858}). 
Therefore, both, its position and its possible orbit disfavor this galaxy as the perturber.

%=====================================================================
% DISCUSSION  DISCUSSION  DISCUSSION  DISCUSSION  DISCUSSION   
%=====================================================================
\section{Discussion}
\label{discussion}

\subsection{The sloshing scenario at the group scale: cold fronts, peculiar velocities of BCGs and perturbers}

The evidence presented in this paper indicates the presence of cold fronts in the galaxy group IC 1860
associated with a spiral excess in surface brightness in its core. These features fit nicely in the 
sloshing scenario put forward by \citet{Ascasibar.ea:06}. Furthermore, as qualitatively shown in \S \ref{simulations},
the observed geometry of the sloshing features (cold fronts and surface brightness excess) can be reproduced
by simulations and can constrain the recent merger history, as done quantitatively for the Virgo cluster and A496 \citep{Roediger.ea:11,Roediger.ea:12}.
Simulations tailored specifically to IC 1860 and in general to the groups showing sloshing will allow another step forward
in interpreting the data.

Another tell-tale sign of the off-axis merger with a satellite is the presence of a peculiar velocity
of the central galaxy, as also suggested by \citet{Ascasibar.ea:06}. The elliptical galaxy IC 1860, the BCG of the group, has a 
peculiar velocity of 113 \kms\ with respect to the group velocity distribution, a difference significant
at the $2.3\sigma$ level. If we identify the BCG position with the bottom of the potential
well (i.e. the dark matter peak), the order of magnitude of the measured velocity is consistent with Fig.4 of \citet{Ascasibar.ea:06} where one can roughly estimate 
a DM peak velocity of $\sim$100kpc/Gyr i.e $\sim 100$\kms. It would be interesting to estimate this parameter from simulations also for a group size main halo and 
for different merger geometries. The presence of BCG peculiar velocities is another piece of evidence in favor of the sloshing scenario: for example an alternative 
to displacing the ICM by gravitational interaction with a subcluster is the passage of a shock as proposed by \citet{Churazov.ea:03}. \citet{Roediger.ea:11}
showed in the case of Virgo that the bow shock of a fast moving galaxy causes sloshing and leads to very similar cold front
structures as observed. However in this scenario no BCG peculiar velocity should be observed.
The same minor mergers which are a common phenomena in the growth of galaxy groups and clusters are therefore the origin of both the sloshing features in the 
X-ray emitting gas and the peculiar velocities of the central galaxies, now consistently found in massive halos ($> 10^{13}$ \msun) as for example in 
\citet{vandenBosch.ea:05}. 
\citet{vandenBosch.ea:05} in-fact suggest as the preferred explanation for the latter finding the scenario of a non-relaxed halo, meaning 
that the BCG is located at rest with respect to the minimum of the dark matter potential, but the dark matter mass distribution is not fully relaxed
and the potential minimum does not coincide with the barycentre. This is the same scenario put forward by \citet{Ascasibar.ea:06} (and envisaged by 
\citet{Miralda-Escude:95}): therefore 
both sloshing and peculiar velocities of the BCGs are a natural outcome of a hierarchical picture of structure formation. 

The key ingredient in the sloshing scenario is the presence of a perturber. The identification of the perturber is straightforward 
in case of double gaseous components as in Abell 1644 \citep{Johnson.ea:10} and in the groups NGC 7618 - UGC 12491 \citep{Roediger.ea:12*1}; it is 
also identified in a convincing manner by means of optical substructure investigations \citep[e.g.][]{Owers.ea:11*1}. However in some cases,
as for example A496, notwithstanding many observed X-ray features, in particular the key presence of a large scale asymmetry, consistent with a dedicated
hydrodynamical simulation, it was not possible to identify the perturber. It may be impossible to ever identify it because tidal forces can 
disperse the galaxies in the subcluster which might not be recognizable as a compact structure anymore \citep{Roediger.ea:12}. 
At the group scale we might have a cleaner test case in this respect for the 
sloshing scenario than in more massive clusters: in a group environment the halo of a single moderately massive spiral galaxy can be the responsible perturber. 
In \S\ref{simulations} we explored two possible scenarios consistent with the identification of two big distorted spirals, ESO 416 and IC 1859.
Given the optical appearance of tidal features pointing in the projected direction of the central galaxy IC 1860, a preference
for the most likely candidate can be given to IC 1859.

This interpretation is strengthened by the analogy with two other groups reported to host sloshing cold fronts, NGC 5044
and NGC 5846. These two groups also have an optically disturbed spiral galaxy which can 
be identified as the possible perturber. For the NGC 5044 group we highlighted in \S\ref{ngc5044} a key feature not reported in previous analysis: the presence
of a large scale asymmetry in terms of a brightness excess in the east (see Fig.\ref{fig.ngc5044}). From their sloshing 
simulations for the Virgo cluster, \citet{Roediger.ea:11} predicted the existence of a large scale asymmetry. In Virgo 
the asymmetry could not be confirmed observationally, due to incomplete data coverage and to the fact that Virgo is still a very dynamic 
cluster where sloshing-related asymmetry may be difficult to be disentangled form other perturbations. In the cluster 
A 496 for the first time the connection between gas sloshing and the presence of a large scale asymmetry was confirmed: the
asymmetry was found in simulations and observations at the same position \citep{Roediger.ea:12}. The galaxy group NGC 5044
is therefore another example which confirms the picture of gas sloshing. Given the presence of almost concentric arcs
on opposite sides of the cluster core (the identified cold fronts) and the orientation of the large scale 
asymmetry, it will be possible to constrain quantitatively the merger scenario. Already qualitatively the LOS can be constrained
to be almost parallel to the orbital plane of the perturber and consistent with a position East of the center of the group. All these 
constraints are compatible with the identification of the perturber as the tidally disturbed spiral galaxy NGC 5054 
(see Fig.\ref{fig.n5054}), as already suggested by \citet{David.ea:09}.
For the NGC 5846 group the characterization of the cold fronts, the possible merger geometry and the identification with the 
optically \citep[and HI,][]{Higdon.ea:98} disturbed ringed spiral galaxy NGC 5850 (see Fig.\ref{fig.ngc5846_opt}) have already 
been discussed in \citet{Machacek.ea:11}. In \S\ref{ngc5846} we reported the \xmm\ MOS surface brightness residual map which is 
consistent with the \chandra\ one of \citet{Machacek.ea:11} and reported a possible evidence of a peculiar velocity
of the BCG NGC 5846 with respect to the mean group velocity, as observed in other cases of sloshing systems. The analysis and interpretation
may be challenging even if more data are collected given the possible identification of the NGC 5846 group as a subgroup of a more extended system
comprising also the NGC 5813 group \citep{Mahdavi.ea:05}.

X-ray sloshing features are witnessing the interaction of the perturber with the group halo. If the perturber as in the case of the three groups
discussed above is a optically disturbed spiral galaxy, further constraints might in principle be obtained by comparisons of the optical distorsion 
of the spiral arms produced in simulations following the evolution of disc galaxies within the global tidal field of the group \citep{Villalobos.ea:12}.

\subsection{The low frequency radio emission in IC 1860}

Mini radio halos, faint diffuse radio emission with radius of 100-300 kpc and a steep spectrum ($\alpha \sim 1.3$), are present in a number of relaxed
cool-core clusters. The central active galaxies of these clusters are not directly powering the diffuse radio emission as the 
radiative time scale of the electrons responsible of the observed emission is much shorter than the time required for these electrons
to diffuse across the cooling region \citep{Brunetti:03}. Reacceleration of pre-existing, low-energy electrons electrons in the ICM by turbulence
in the core region is a possible mechanism responsible for the radio emission. Sloshing motions can produce significant turbulence in the 
cluster core \citep{Fujita.ea:04,ZuHone.ea:13} and this solution is strengthened by the spatial correlation between radio mini-halo emission
and cold fronts in some clusters hosting mini-radio halos \citep{Mazzotta.ea:08}.

An ongoing statistical study of the properties of clusters with and without a minihalo shows that minihalos tend to occur only in very massive, cool-core clusters 
\citep{Giacintucci.ea:13}. Indeed, no clear detection of a minihalo at the group scale has been reported so far. In IC 1860 we have indication of a spatial 
coincidence between a cold front and extended radio emission, similar to what is seen in some minihalo clusters.
The \gmrt\ radio emission at 325 MHz is much more extended than the 1.4 GHz emission
presented in \citet{Dunn.ea:10} which shows a clear point source with faint extensions in the north-east and south-west directions 
with no clear correlation with the X-ray extension towards the south-east.
Instead one component of the 325 MHz emission traces the bright tip of the excess 
spiral feature and is confined within the inner cold front, very much like the spatial coincidence found
in clusters with mini radio-halos. Another component is more detached lying in projection at the
edge of the spiral. This latter feature resembles the detached components found at 235 MHz in NGC 5044 \citep{Giacintucci.ea:11,David.ea:09}.\\
However the extension (only 45 kpc) and the ultra-steep spectral index ($\alpha > 1.9$) of the radio emission can be explained without the need for in situ 
re-acceleration as for radio halos and radio mini-halos in galaxy clusters. 
Indeed, as discussed in \citet{Jaffe:77}, a possibility for the transport of the electrons responsible for the
radio emission is large scale motions of the gas carrying the electrons along. \citet{ZuHone.ea:13} showed how relativistic electrons preferentially
located in AGN-blown bubbles are redistributed by the sloshing motions so that most of them end up within the spiral shape traced out by the cold fronts.
Observations at different radio frequencies and simulations tailored for the groups mass scale are needed to investigate the possibility that 
re-acceleration might take place.

\subsection{Reconstructing the mass distribution in a sloshing group}

Since the introduction of the sloshing paradigm in the cluster Abell 1795, it was pointed out the the presence of cold front discontinuities
in the ICM density and temperature profiles can caused flawed results in cluster mass estimates \citep{Markevitch.ea:01}. Unphysical discontinuities
arise at the cold front in the derived cluster mass profile assuming hydrostatic equilibrium. This is true for most of the published cold front profiles
and this fact seems consistent with tangential flows associated with sloshing contributing an additional centrifugal acceleration \citep{Keshet.ea:10}.
\citet{Roediger.ea:12} address the issue to what extent the underlying mass distribution
can be inferred from fully azimuthally averaged temperature and density profiles with the presence of cold fronts in their simulations
for the case of the Abell 496 cluster. They found that sloshing modifies the azimuthally averaged profiles
by less than 10\% and concluded that given the quality of the current available data, the distortion of the profiles due to
sloshing does not seem to introduce an observable difference and will not be a major error source for cluster mass profile.
This is true also at the group mass scale: in \citet{Gastaldello.ea:07*1} this issue was addressed directly in IC 1860 by comparing the
mass profile reconstructed in two sectors with PA 15\deg-95\deg\ and 195\deg-15\deg\ finding results on the derived total mass (and NFW concentration parameter)
within their 2$\sigma$ errors. Therefore the low-level disturbances due to sloshing do not indicate a significant violation of
hydrostatic equilibrium.
%
%
%=====================================================================
% CONCLUSION  CONCLUSION  CONCLUSION  CONCLUSION  CONCLUSION  
%=====================================================================
\section{Summary and conclusions}
\label{summary}
We have presented results of a combined analysis of the currently available
\xmm\ and \chandra\ X-ray data, optical spectroscopy and radio \gmrt\ for the bright nearby galaxy group 
IC1860. 
The results can be summarized as follows:
\begin{enumerate}

\item the X-ray imaging and spectroscopy reveal the presence of a set of two cold fronts
at 45 and 76 kpc and of a spiral of excess surface brightness. 
Those features are consistent with the sloshing scenario put forward to explain these features,
as further supported by the presence of a peculiar velocity of the central galaxy IC 1860 and 
the identification of a possible perturber in the optically disturbed spiral galaxy IC 1859;
\item the analogy of the IC 1860 group with two other X-ray groups showing cold fronts, NGC 5044 and NGC 5846
is tantalizing in terms of the identification of the possible perturber with an optically
disturbed spiral galaxy and peculiar velocity of the bright central galaxy. We reported further
evidence for the sloshing scenario in those two systems. We detected a large scale surface brightness 
excess in NGC 5044 which strengthen earlier suggestions by \citet{David.ea:09} of the identification of the 
perturber with the disturbed spiral NGC 5054. We detected hint of a peculiar velocity of the central galaxy NGC 5846. 
\item The presence of X-ray sloshing features, peculiar velocity of the BCG and optically disturbed features
in the perturbing spiral galaxies are signatures of the interaction of the group tidal field with a disk galaxy
and indication that also this process is actively participating in the transformation of late type 
galaxies in the group environment. 
\item The low frequency \gmrt\ radio data show extended emission with a component confined by the inner cold front
as seen in some clusters with mini-halos. Unlike minihalos that require in situ particle reacceleration, the  
extended emission in IC 1860 is consistent with aged radio plasma redistributed by the sloshing motions.
\end{enumerate} 

The case studies presented in this paper show the rich phenomenology and the constraining power that
X-ray, optical and radio data can provide in assessing the dynamical state and past merging history of galaxy groups and their 
galaxies. We have provided the strongest evidence to date that cold fronts in groups exist and have similar properties to those in massive clusters. 
This is critical confirmation of the currently favored model of cold-fronts formation in relaxed objects.

%
%=====================================================================
% ACKNOWLEDGEMENTS  ACKNOWLEDGEMENTS  ACKNOWLEDGEMENTS
%=====================================================================
\begin{acknowledgements}
We thank M. Markevitch and T. Venturi for useful discussions and the anonymous referee 
for suggestions that improved the work presented here.
FG acknowledges financial contributions by the Italian Space Agency through ASI/INAF
agreements I/023/05/0 and I/088/06/0 for the data analysis and I/032/10/0 for the
XMM-Newton operations. SGi acknowledges the support of NASA through Einstein Postdoctoral
Fellowship PF0-110071 awarded by the {\em Chandra} X-ray Center (CXC), which is operated by the Smithsonian Astrophysical
Observatory (SAO).
We are grateful to the ACE SWEPAM instrument team and the ACE Science center for
providing the ACE data and to the 2dF team for the data of the 2dF survey. 
This research has made use of the NASA/IPAC Extragalactic Database (NED)
which is operated by the Jet Propulsion Laboratory, California Institute of Technology, under contract with the
National Aeronautics and Space Administration.
\end{acknowledgements}
%
%%=====================================================================
%% REFERENCES  REFERENCES  REFERENCES  REFERENCES  REFERENCES  
%%=====================================================================
%
\bibliographystyle{apj}
\bibliography{gasta}
\begin{appendices}

\section{The NGC 5044 and NGC 5846 groups}\label{comparison}

In this section we report additional X-ray and optical analysis for the NGC 5044
and NGC 5846 groups, which show sloshing cold fronts. The analogy of IC 1860 group 
with these systems will reinforce the suggestion of an optically disturbed spiral galaxy
as the candidate perturber and provide further evidence of the sloshing scenario at the 
group scale.

\subsection{NGC 5044}
\label{ngc5044}

%%%%%%%%%%%%%%%%%%%%%%%%%%%%%%%%%%%%%%%%%%%%%%%%%%%%%%%%%%%%%%%%%%%%%%%%%%%%%%%%
\begin{figure*}[th]
%%\vspace{-0.5cm}
\centerline{
\parbox{0.5\textwidth}{
\includegraphics[height=0.28\textheight]{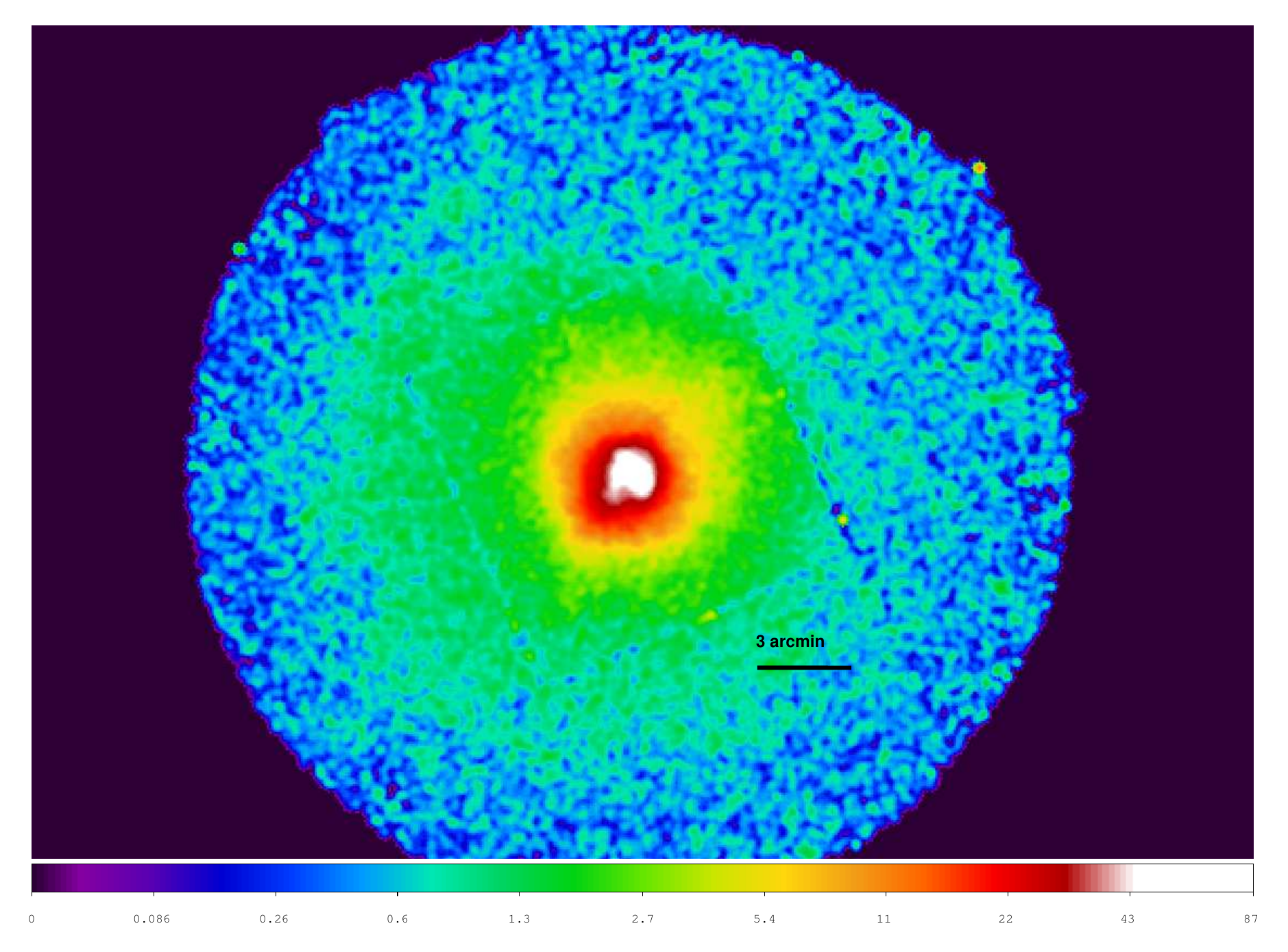}}
\parbox{0.5\textwidth}{
\includegraphics[height=0.28\textheight]{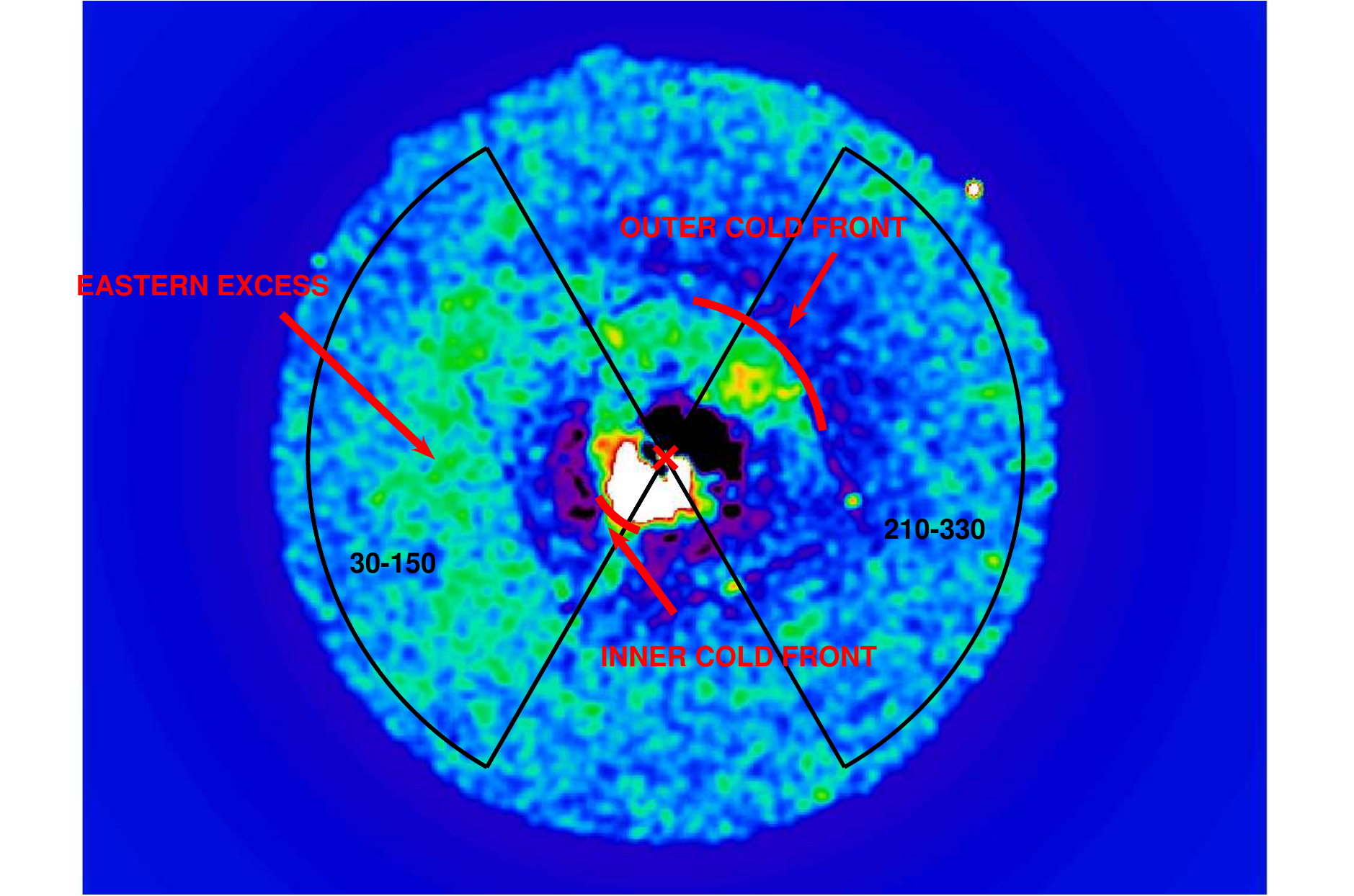}}
}
\caption{\label{fig.ngc5044} \footnotesize
\emph{Left panel:} 0.5-2 keV \xmm\ MOS image of the galaxy group NGC 5044. Point sources have been removed using the \ciao\ task {\tt dmfilth}. 
The excess of emission in the East direction is clearly seen. 
\emph{Right panel: } Map of the residuals between the 0.5-2 keV MOS image of the left panel and the best fit 
beta model for the data smoothed on a 20\arcsec\ scale. The center of NGC 5044 is shown by the red cross and the position
and extent of the cold fronts discussed in  \citet{Gastaldello.ea:09} by the red arcs. Another relevant feature, the Eastern excess, 
and the sectors used for extraction of the surface brightness profiles discussed in the text are also shown.
}
\end{figure*}
%%%%%%%%%%%%%%%%%%%%%%%%%%%%%%%%%%%%%%%%%%%%%%%%%%%%%%%%%%%%%%%%%%%%%%%%%%%%%%%%%%%%%%%%%%%%%%%%%%%%%

%%%%%%%%%%%%%%%%%%%%%%%%%%%%%%%%%%%%%%%%%%%%%%%%%%%%%%%%%%%%%%%%%%%%%%%%%%%%%%%%
\begin{figure}[th]
%%\vspace{-0.5cm}
\centerline{
\includegraphics[height=0.3\textheight]{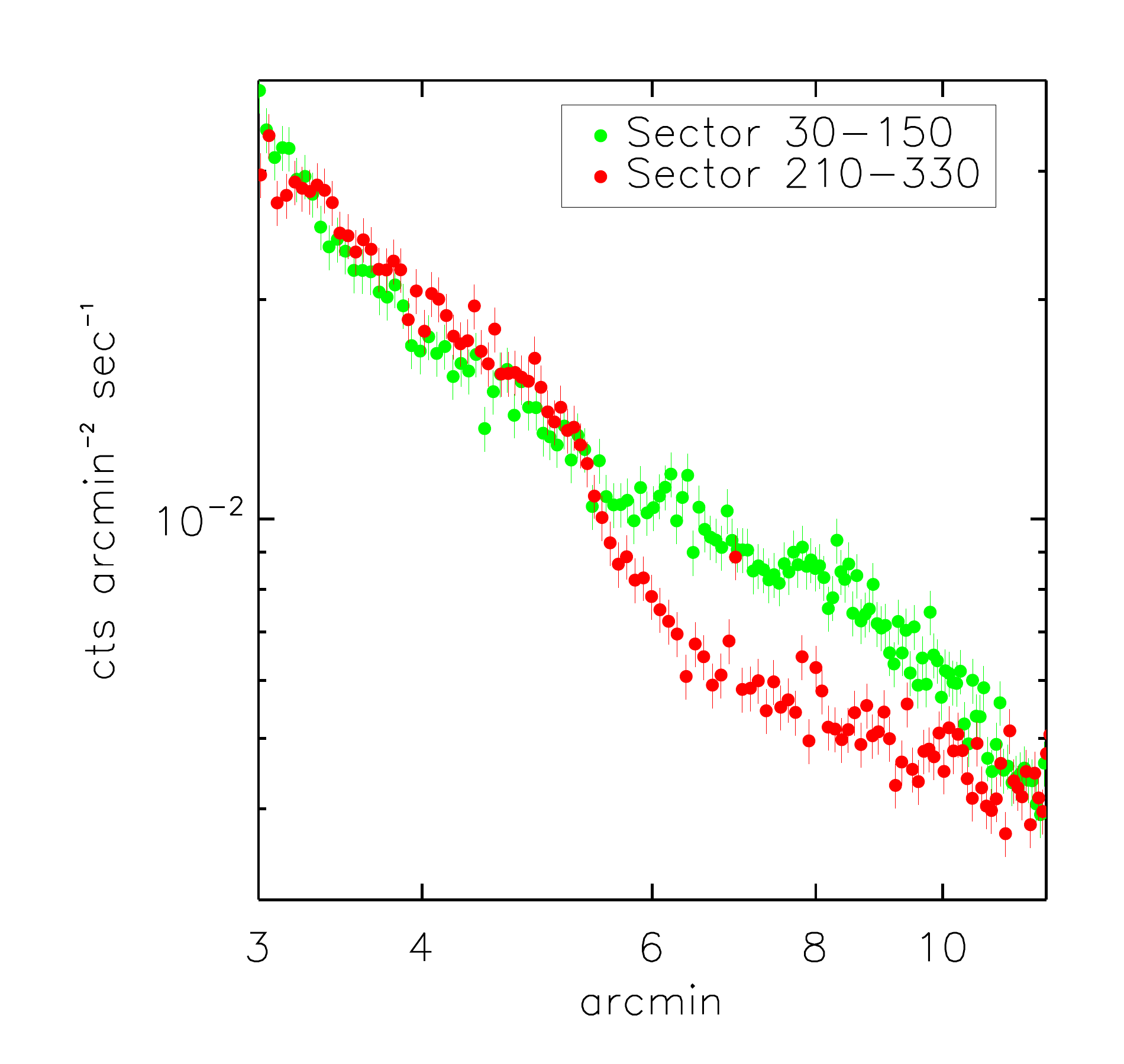}
}
\caption{\label{fig.sbngc5044} \footnotesize
Surface brightness profiles in the angular sectors shown in Fig.\ref{fig.ngc5044}, with position angle measured counterclockwise 
from the North direction. The surface brightness excess in the East can be seen in the 6\arcmin-10\arcmin\ radial range.
}
\end{figure}
%%%%%%%%%%%%%%%%%%%%%%%%%%%%%%%%%%%%%%%%%%%%%%%%%%%%%%%%%%%%%%%%%%%%%%%%%%%%%%%%%%%%%%%%%%%%%%%%%%%%%
%
%%%%%%%%%%%%%%%%%%%%%%%%%%%%%%%%%%%%%%%%%%%%%%%%%%%%%%%%%%%%%%%%%%%%%%%%%%%%%%%%
\begin{figure}[th]
%%\vspace{-0.5cm}
\centerline{
\includegraphics[height=0.24\textheight]{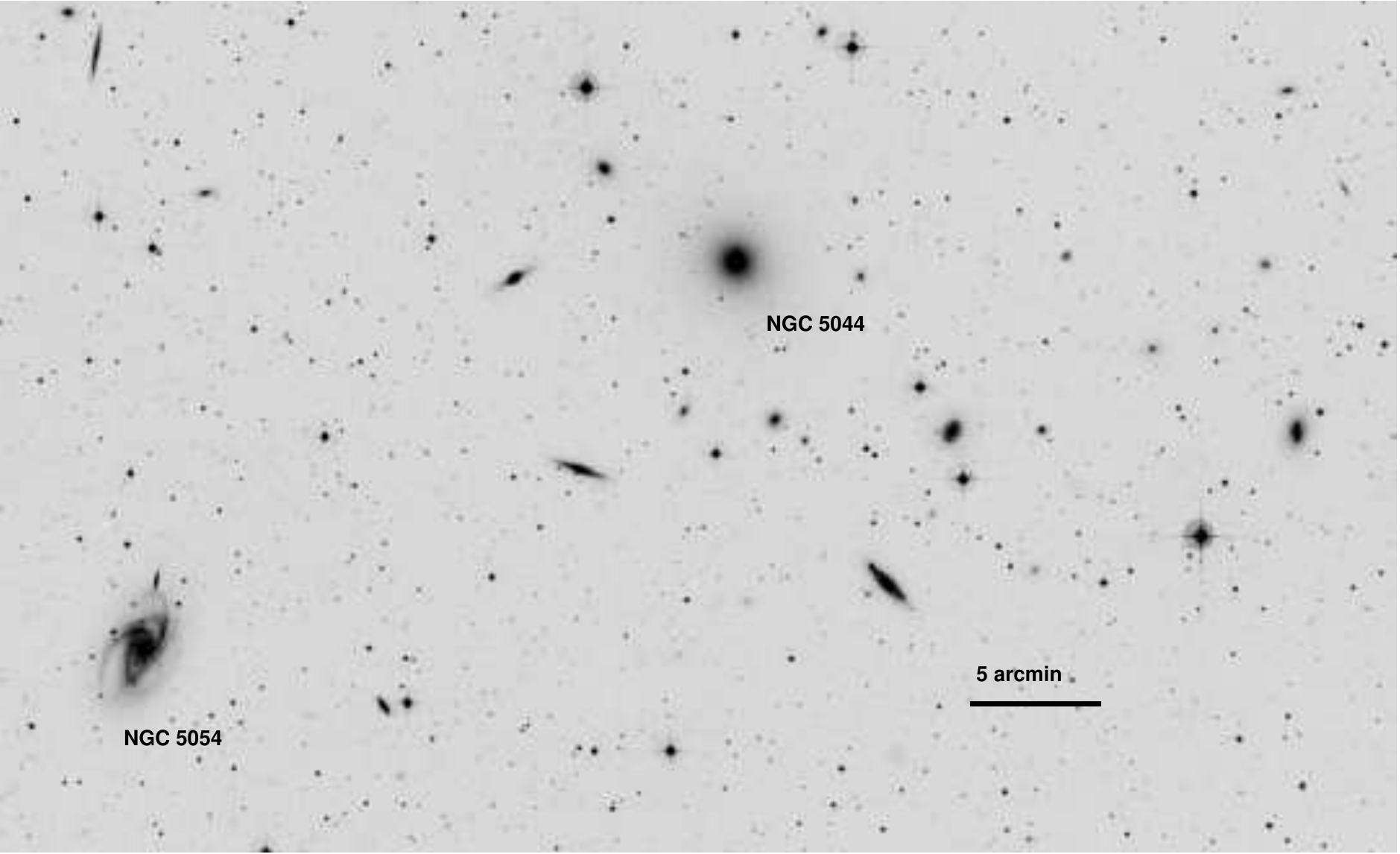}
}
\caption{\label{fig.ngc5044_opt} \footnotesize
DSS image of the central region of the NGC 5044 group. NGC 5044 and NGC 5054 are indicated.
}
\end{figure}
%%%%%%%%%%%%%%%%%%%%%%%%%%%%%%%%%%%%%%%%%%%%%%%%%%%%%%%%%%%%%%%%%%%%%%%%%%%%%%%%%%%%%%%%%%%%%%%%%%%%%
%
%%%%%%%%%%%%%%%%%%%%%%%%%%%%%%%%%%%%%%%%%%%%%%%%%%%%%%%%%%%%%%%%%%%%%%%%%%%%%%%%
\begin{figure}[th]
%%\vspace{-0.5cm}
\centerline{
\includegraphics[height=0.24\textheight]{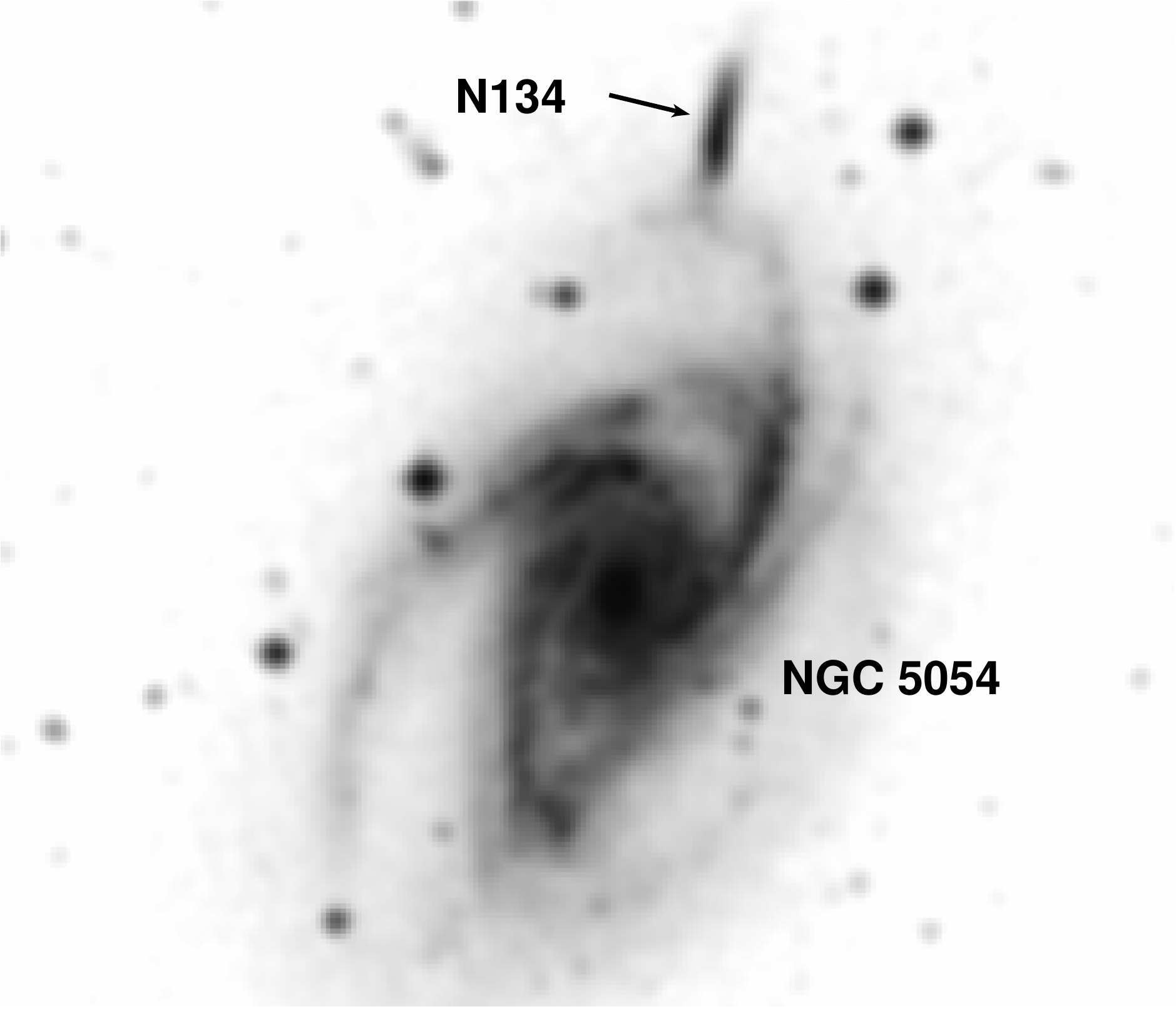}
}
\caption{\label{fig.n5054} \footnotesize
DSS image of the galaxy NGC 5054, together with the companion N134.
}
\end{figure}
%%%%%%%%%%%%%%%%%%%%%%%%%%%%%%%%%%%%%%%%%%%%%%%%%%%%%%%%%%%%%%%%%%%%%%%%%%%%%%%%%%%%%%%%%%%%%%%%%%%%%
%
The dynamical status of the NGC 5044 group (z=0.009) as revealed by the presence of cold
fronts and sloshing features in its X-ray emitting gas has been discussed extensively
using \xmm\ and \chandra\ data \citep[][and references therein]{Gastaldello.ea:09,David.ea:09}.
Here we report additional features in the X-ray morphology in this group, one of the
brightest in the X-ray, which further support the interpretation of the sloshing scenario
and the identification of the perturber.

In the left panel of Fig.\ref{fig.ngc5044} we report the MOS image of the group already reported
in \citet{Gastaldello.ea:09} with a better color coding which highlights the overall extension towards
the East direction of the X-ray emission in the 6\arcmin-10\arcmin\ (66-110 kpc at the redshift
of NGC 5044 for the cosmology adopted in this paper) radial range. In the right panel of Fig.\ref{fig.ngc5044}
we present the surface brightness residual map between the \xmm\ data and the best fit one-dimensional $\beta$-model describing 
the surface brightness radial profile. The cold fronts discovered and discussed in \citet{Gastaldello.ea:09}
are shown and the Eastern excess clearly stands out in this residual map. The surface brightness profile in 
Fig.\ref{fig.sbngc5044} show a factor of $\sim 4$ enhancement compared to the symmetrical sector in the West.
The presence of almost concentric arcs on opposite sides of the group core and the large scale asymmetry points towards
the orbital plane of the perturber almost parallel to the LOS \citep[e.g.][and references therein]{Roediger.ea:12}.
The SE (small scale) spiral feature extending from the SE cold front discussed in \citet{David.ea:09} seems a brightness enhancement
which connects the inner surface brightness excess marked by the cold front to the large scale SE excess. The same argument can be made for the
NW small spiral, also discussed in \citet{David.ea:09}, which seems to connect the core with the NW concentric arc marked by the outer cold front at
67 kpc. These small scale features are qualitatively seen in simulations and can further constrain the geometry of the merging.

NGC 5044 is also from an optical point of view an outstanding object in the local Universe for its
notable population of dwarf galaxies \citep[e.g.][]{Ferguson.ea:90,Mathews.ea:04,Faltenbacher.ea:05*1,Cellone.ea:05,Mendel.ea:08,Mendel.ea:09,Buzzoni.ea:12}.
\citet{Mendel.ea:08} using new, deep spectroscopic observations in conjunction with archival velocity information 
provided a sample of 111 spectroscopically confirmed members. We used that sample to gain further information on the
dynamical status of the NGC 5044 group. As already reported by \citet{Mendel.ea:08} the biweight location and scale estimators
give a mean group velocity of $2577\pm33$ \kms\ and a velocity dispersion of $331\pm26$ \kms. Using a velocity
for the NGC 5044 galaxy of $2733\pm51$ \kms\ we calculate that its peculiar velocity is $156\pm61$ (just adding errors in quadrature) 
corresponds to a Z-score of $0.473\pm0.098$ therefore the velocity offset is significant at the $4.8\sigma$ level.

As for the candidate perturber, \citet{Mendel.ea:08} reported a low-mass substructure detected by means of the 
$\Delta$ test with 6 members $\sim$ 1.4 Mpc from the group center to the NE. However as suggested by \citet{David.ea:09}
this subgroup is too far away in the outskirts of the group to be a likely candidate, also because it has a very small 
component of the velocity in the LOS (the biweight mean velocity of the subgroup is 2819 \kms).
A more likely candidate suggested by \citet{David.ea:09} is the giant perturbed spiral NGC 5054 (see Fig.\ref{fig.n5054}): it is 
one of the brightest members of the group and it has a stellar mass of $2.5\times10^{11}$ \msun\ comprising 11\% of the total 
stellar mass of the group \citep{Buzzoni.ea:12}. It is at a projected distance of 300 kpc from NGC 5054 and it has a velocity of 
1658 \kms\ \citep{Mendel.ea:08} corresponding to a peculiar velocity of $-919\pm61$ \kms\ with respect to the group velocity.
At a projected distance of 2.6 \arcmin\ (29 kpc) from NGC 5054 lies the Sm/Im galaxy N134, one of the bluest dwarf galaxies of the
group, partially overlapping with the north arm; their interaction is likely the cause of the intense star formation 
activity in N134 and whether or not the interaction is also responsible for the peculiar arm morphology of NGC 5054 is a matter
of debate \citep[][and references therein]{Buzzoni.ea:12}. However we reinforce the \citet{David.ea:09} suggestion that NGC 5054 is the likely
perturber given also the X-ray evidence presented in Fig.\ref{fig.ngc5044} and suggest that the peculiar morphology of NGC 5054
is mainly caused by interaction with the global tidal field of NGC 5044 group halo.

\subsection{NGC 5846}
\label{ngc5846}

%%%%%%%%%%%%%%%%%%%%%%%%%%%%%%%%%%%%%%%%%%%%%%%%%%%%%%%%%%%%%%%%%%%%%%%%%%%%%%%%
\begin{figure*}[th]
%%\vspace{-0.5cm}
\centerline{
\parbox{0.5\textwidth}{
\includegraphics[height=0.28\textheight]{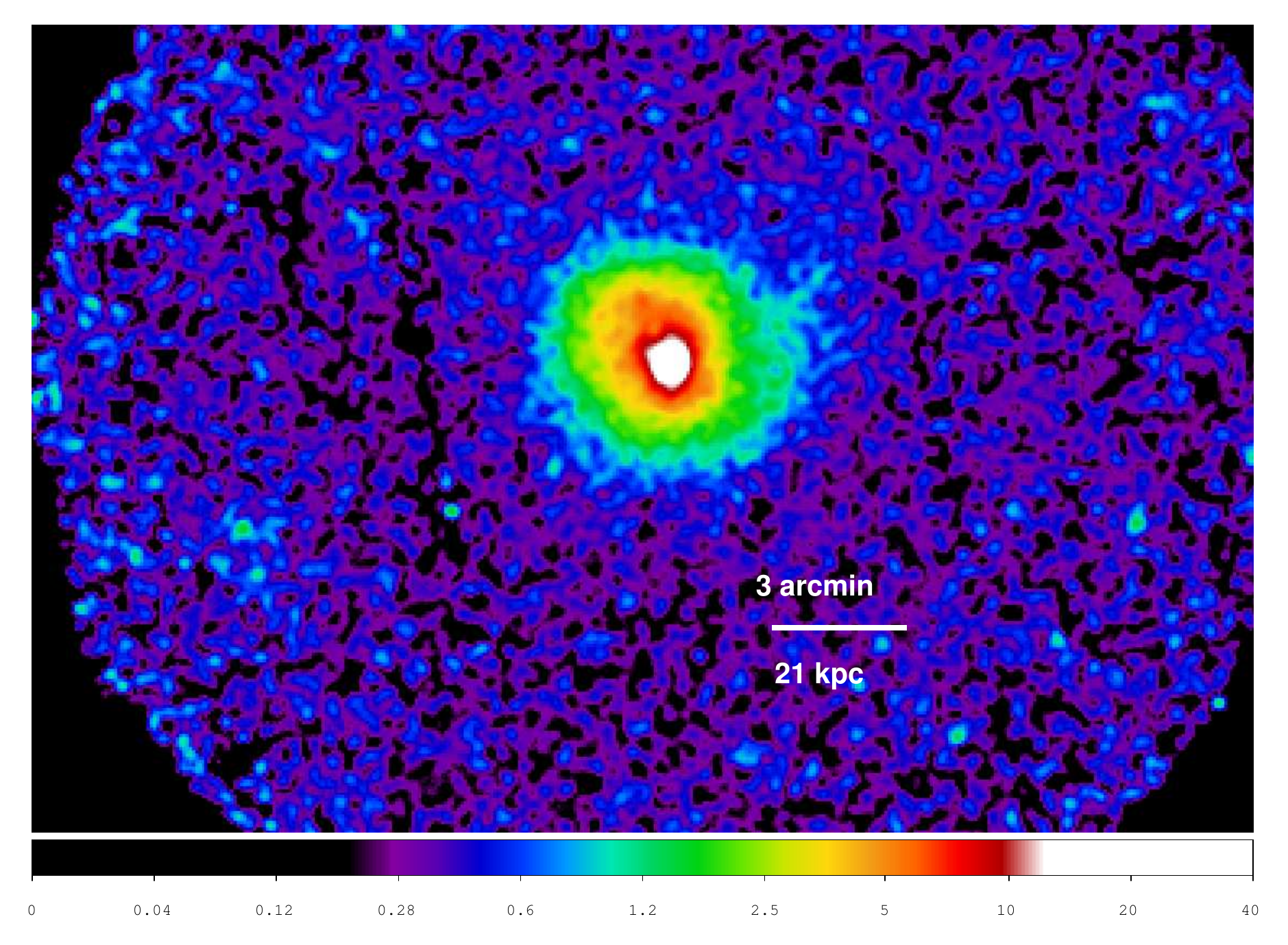}}
\parbox{0.5\textwidth}{
\includegraphics[height=0.28\textheight]{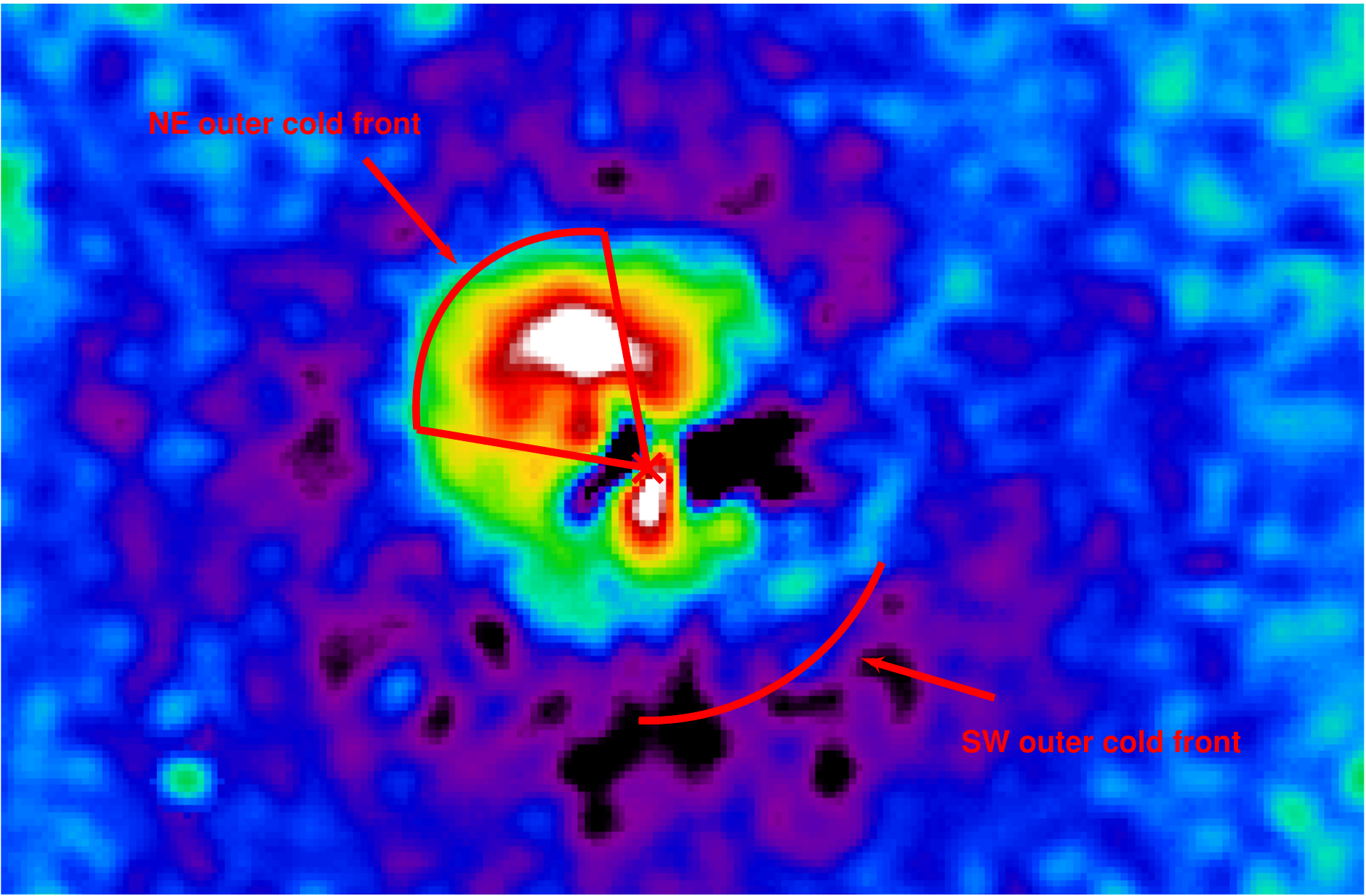}}
}
\caption{\label{fig.ngc5846} \footnotesize
\emph{Left panel:} 0.5-2 keV \xmm\ MOS image of the galaxy group NGC 5846. Point sources have been removed using the \ciao\ task {\tt dmfilth}. 
The excess of emission in the East direction is clearly seen 
\emph{Right panel: } Map of the residuals between the 0.5-2 keV MOS image of the left panel and the best fit 
beta model for the data smoothed on a 20\arcsec\ scale. The center of NGC 5846 is shown by the red cross and the position
and extent of the outer cold fronts discussed in  \citet{Machacek.ea:11} by the red arcs.
}
\end{figure*}
%%%%%%%%%%%%%%%%%%%%%%%%%%%%%%%%%%%%%%%%%%%%%%%%%%%%%%%%%%%%%%%%%%%%%%%%%%%%%%%%%%%%%%%%%%%%%%%%%%%%%

%%%%%%%%%%%%%%%%%%%%%%%%%%%%%%%%%%%%%%%%%%%%%%%%%%%%%%%%%%%%%%%%%%%%%%%%%%%%%%%%
\begin{figure}[th]
%%\vspace{-0.5cm}
\centerline{
\includegraphics[height=0.24\textheight]{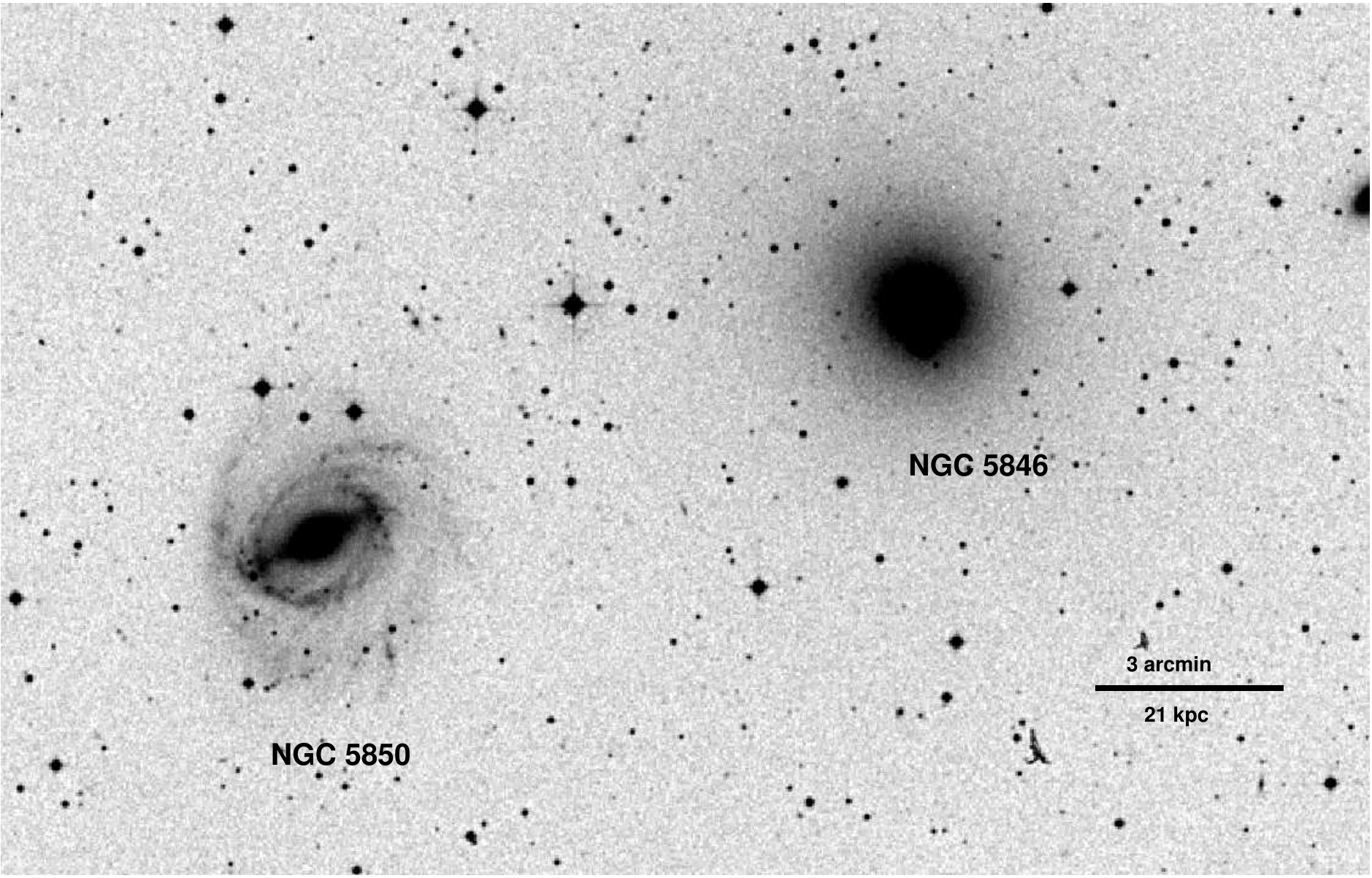}
}
\caption{\label{fig.ngc5846_opt} \footnotesize
DSS image of the central region of the NGC 5846 group. NGC 5846 and NGC 5850 are indicated.
}
\end{figure}
%%%%%%%%%%%%%%%%%%%%%%%%%%%%%%%%%%%%%%%%%%%%%%%%%%%%%%%%%%%%%%%%%%%%%%%%%%%%%%%%%%%%%%%%%%%%%%%%%%%%%

NGC 5846 is a nearby group (z=0.0056) where evidence for gas sloshing with multiple cold fronts has been presented
using a deep (120 ks) \chandra\ exposure in \citet{Machacek.ea:11}. Numerous ring-like structures are present, in particular the
SW feature was already pointed out in the \xmm\ observation presented in \citet{Finoguenov.ea:06}.
In the left panel of Fig.\ref{fig.ngc5846} we report the MOS image of the group (taken from the \xmm\ observation 
obsID 0021540501) and in the right panel the surface brightness residual map between the \xmm\ data and the best fit one-dimensional 
$\beta$-model describing the surface brightness radial profile. The outer cold fronts discussed in \citet{Machacek.ea:11} are shown:
The residual map is in agreement with the \chandra\ one reported in \citet[][see their Fig.4]{Machacek.ea:11} with a less pronounced tail 
like feature given the shorter (15 ks) \xmm\ exposure. 

\citet{Mahdavi.ea:05} performed a photometric and spectroscopic survey of a 10 deg$^2$ surrounding NGC 5846 and together with \rosat, \xmm, 
and \asca\ data they suggested that NGC 5846 together with the other giant elliptical NGC 5813 are the dominant components of two distinct
subgroups separated by 600 kpc in projection and embedded in a larger 1.6 Mpc diameter dark matter halo. Both the NGC 5813 and NGC 5846 
galaxy groups are in the ROSAT-ESO Flux-Limited X-Ray (REFLEX) catalog \citep{Bohringer.ea:04}. Given the presence of the subgroup 
associated with NGC 5813 we restricted the spectroscopically confirmed sample of 83 galaxies of \citet{Mahdavi.ea:05} to the members 
within 400 kpc (0.9 deg) of NGC 5846 resulting in a sample of 39 objects. The biweight location and scale estimators
give a mean subgroup velocity of $1833\pm78$ \kms\ and a velocity dispersion of $386\pm41$ \kms. 
Using a velocity for the NGC 5846 galaxy of $1714\pm5$ \kms\ \citep{Trager.ea:00} we calculate that its peculiar velocity is $-119\pm78$ 
corresponding to a Z-score of $-0.339\pm0.206$ therefore the velocity offset is marginally significant at the $1.6\sigma$ level.

As suggested by \citet{Machacek.ea:11} a likely perturber is the big spiral galaxy NGC 5850 (see Fig.\ref{fig.ngc5846_opt}) which is located
10\arcmin\ (71 kpc) in projection to the East of NGC 5846 and it has a velocity of $2556\pm4$ \kms\ \citep{rc3}. It therefore has a relative
velocity of $842\pm6$ \kms\ with respect to NGC 5846 and of $723\pm78$ \kms\ with respect to the mean group velocity. 
The optical image reveal numerous morphological peculiarities, particularly in the
spiral arms and structural asymmetries in the HI gas in the direction of NGC 5846 are even more pronounced \citep{Higdon.ea:98}. 
These features already suggested a recent ($\lesssim 200$ Myr) tidal interaction between the two galaxies \citep{Higdon.ea:98}: the new 
element suggested by the X-ray evidence is that the interaction has been between NGC 5850 and the overall mass of the group halo 
interior to the orbit of NGC 5850.

\end{appendices}

\begin{appendices}

\section{Redshift uncertainties in the catalogue of members of the IC 1860 group}\label{app.redshifts}

%%%%%%%%%%%%%%%%%%%%%%%%%%%%%%%%%%%%%%%%%%%%%%%%%%%%%%%%%%%%%%%%%%%%%%%%%%%%%%%%%%%%%%%%%%%%%%
%
\begin{table*}[h]
\caption{Comparison of 2dF redshift measurements with literature 
values.\label{tab_vel_comparison}}
\begin{center} \vskip -0.4cm
\begin{tabular}{ccccccc}
\tableline\tableline\\[-7pt]
Literature source & $N_{match}$ & $\overline{\Delta cz}$ & $\sigma (\Delta{cz})$ & Literature redshift & Quadrature difference 
& Ratio \\
&   &  \kms && uncertainty \kms &  \kms &  \\
\tableline \\[-7pt]
6dF & 10 & $-17\pm20$ & $52\pm11$ & 31 & $42\pm14$  &  0.35 \\
\citet{Stein:96} & 9 & $-7\pm18$ & $60\pm13$ & 23 & $55\pm14$ & 0.26 \\
\citet{Dressler.ea:88} & 23 & $-61\pm22$ & $90\pm13$ & 45 & $78\pm15$ & 0.50 \\
RC3  & 11 & $-30\pm29$ & $75\pm22$ & 43 & $61\pm27$ & 0.48 \\
\citet{daCosta.ea:98}  & 14 & $-46\pm19$ & $56\pm15$ & 37 & $42\pm18$ & 0.41 \\
\tableline 
\vspace{0.5 cm}
\end{tabular}
\tablecomments{Columns are as follows: (1) Origin of external redshift catalog for comparison; 
(2) number of redshifts in common with the 2dF, $N_{match}$; (3) the mean redshift difference, 
$\overline{\Delta cz}=\overline{cz_{2dF}-cz_{ext.}}$, as determined by the 
biweight estimator; (4) dispersion in the $\Delta cz$ distribution, 
$\sigma ({\Delta cz})$, determined with the biweight estimator; (5) the 
redshift uncertainty for the external redshift measurements; (6) the 
quadrature difference between $\sigma ({\Delta cz})$ and the external redshift 
error which gives an external estimate of the uncertainty on the 2dF
measurements; (7) the ratio of the external uncertainty measurement to 
the median uncertainty measurement of the 2dF sample which is 89\kms.
The errors for $\overline{\Delta cz}$ and $\sigma (\Delta{cz})$ are obtained as one standard deviation of the
distribution obtained by 10,000 bootstrap resamplings of the data.}
\end{center}
\end{table*}

%%%%%%%%%%%%%%%%%%%%%%%%%%%%%%%%%%%%%%%%%%%%%%%%%%%%%%%%%%%%%%%%%%%%%%%%%%%%%%%%%%%%%%%%%%%%%%%%%%%%%%%

Following \citet{Owers.ea:11} we performed an independent check of the measured velocities and their
uncertainties for the galaxies in the IC 1860 group by cross-correlating the catalogues from different sources. We used as reference the 
more homogeneous 2dF sample which consists of 74 objects and correlated it  with
the 6dFGS, the \citet{Stein:96}, the \citet{Dressler.ea:88}, the \citet{daCosta.ea:98} and the 
RC3 \citep{rc3} redshifts catalogues.
The results of the comparisons are shown in Table~\ref{tab_vel_comparison}. 

As can be seen there are no significant offsets from a zero mean for all the redshift differences
between the various catalogues, ensuring that also the few external redshifts are consistent with the 2dF 
ones.

\end{appendices}

%\newpage
\begin{appendices}

\section{The narrow angle tail radio galaxy IC 1858}\label{sec.appendixic1858}

%%%%%%%%%%%%%%%%%%%%%%%%%%%%%%%%%%%%%%%%%%%%%%%%%%%%%%%%%%%%%%%%%%%%%%%%%%%%%%%%
\begin{figure*}[th]
%%\vspace{-0.5cm}
\centerline{
\includegraphics[height=0.28\textheight]{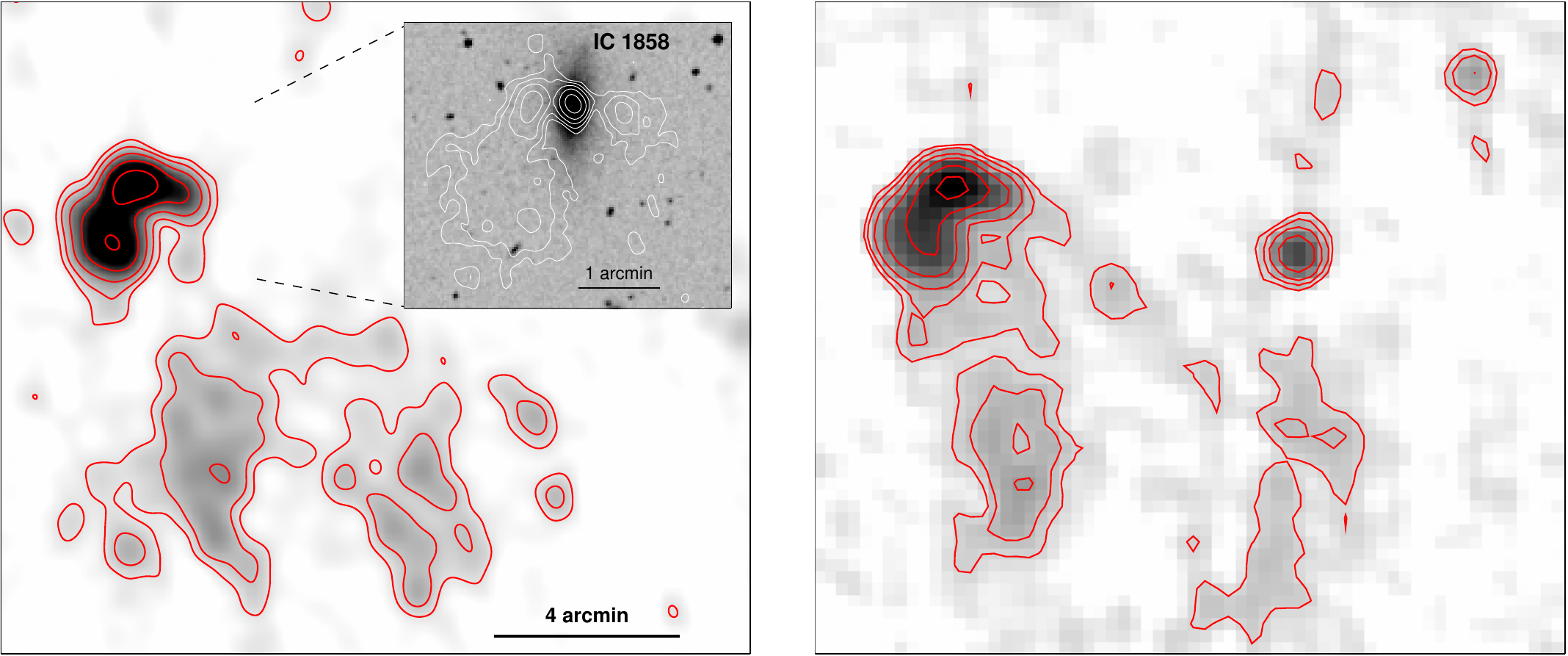}
}
\caption{\label{fig.ic1858} \footnotesize
{\em Left panel:} \gmrt\ image (contours and gray scale) at 325 MHz of the NAT associated with IC 1858. The restoring beam is
$44^{\prime \prime}\times40^{\prime \prime}$, in p.a. $0^{\circ}$. The rms noise level is 1 mJy beam$^{-1}$. Contours are spaced 
by a factor of 2 starting from 3 mJy beam$^{-1}$. The inset shows the \gmrt\ full-resolution 
($14^{\prime \prime}\times9^{\prime \prime}$, $-22^{\circ}$) contours overlaid on the DSS image of IC 1858. Contours are spaced 
by a factor of 2 starting from 1.2 mJy beam$^{-1}$. The rms noise level is 0.36 mJy beam$^{-1}$. {\em Right panel:} 1.4 NVSS image
of IC 1858. The restoring beam is $45^{\prime \prime}\times45^{\prime \prime}$, in p.a. $0^{\circ}$. The rms noise level is 0.5 mJy beam$^{-1}$. 
Contours are spaced by a factor of 2 starting from 1 mJy beam$^{-1}$.}
\end{figure*}
%%%%%%%%%%%%%%%%%%%%%%%%%%%%%%%%%%%%%%%%%%%%%%%%%%%%%%%%%%%%%%%%%%%%%%%%%%%%%%%%%%%%%%%%%%%%%%%%%%%%%

Radio galaxies located in dense environments often show complex and prominently distorted radio
structures. A common morphology is represented by tailed radio
galaxies, i.e., double sources whose jets and lobes are bent. Sources whose jets and lobes form a small angle
are referred to as narrow-angle-tail radio galaxies \citep[NAT, or
U-shaped;][and references therein]{Feretti.ea:02}. Many NAT galaxies reside in clusters, however an increasing number
are found in lower mass environments like group of galaxies \citep[][and references therein]{Freeland.ea:08}. Here we report
on the \gmrt\ data showing a typical NAT morphology for the IC 1858 galaxy (see left panel of Fig.\ref{fig.ic1858}).
The total flux at 325 MHz is $880\pm70$ mJy and it has a largest angular size of $\sim 11$\arcmin\ corresponding at $\sim 300$ kpc at the
redshift of the source. It is also detected in the NVSS with a flux of $\sim 180$ mJy in the same region of the 325 MHz emission (see 
right panel of Fig.\ref{fig.ic1858}) implying a spectral index of $\alpha \sim 1.1$ typical for such extended sources. Its radio power at 
1.4 GHz is $2.0\times10^{23}$ W Hz$^{-1}$, again typical for NAT sources. It is at a projected distance of 216 kpc from IC 1860 and it has a 
velocity difference of $-697\pm99$\kms\ with respect to the mean group velocity. Hence there is a substantial component of the velocity of the 
galaxy along the LOS; the overall bending of the radio jets points for a component in the plane of the sky pointing in the N-NE direction. The galaxy
could therefore just falling in for the first time but also, given the it probably has already time to experience the action of the group environment 
transforming its morphology into S0, having already experience its first core passage.

\end{appendices}

\end{document}